\PassOptionsToPackage{unicode}{hyperref}
\PassOptionsToPackage{hyphens}{url}
\PassOptionsToPackage{dvipsnames,svgnames,x11names}{xcolor}
\documentclass[
  10pt,
]{article}
\usepackage{xcolor}
\usepackage[margin=2.2cm]{geometry}
\usepackage{amsmath,amssymb}
\usepackage{iftex}
\ifPDFTeX
  \usepackage[T1]{fontenc}
  \usepackage[utf8]{inputenc}
  \usepackage{textcomp} 
\else 
  \usepackage{unicode-math} 
  \defaultfontfeatures{Scale=MatchLowercase}
  \defaultfontfeatures[\rmfamily]{Ligatures=TeX,Scale=1}
\fi
\usepackage{lmodern}
\ifPDFTeX\else
\fi
\IfFileExists{upquote.sty}{\usepackage{upquote}}{}
\IfFileExists{microtype.sty}{
  \usepackage[]{microtype}
  \UseMicrotypeSet[protrusion]{basicmath} 
}{}
\makeatletter
\@ifundefined{KOMAClassName}{
  \IfFileExists{parskip.sty}{%
    \usepackage{parskip}
  }{
    \setlength{\parindent}{0pt}
    \setlength{\parskip}{6pt plus 2pt minus 1pt}}
}{
  \KOMAoptions{parskip=half}}
\makeatother
\usepackage{longtable,booktabs,array}
\usepackage{caption}
\usepackage{calc} 
\usepackage{etoolbox}
\makeatletter
\patchcmd\longtable{\par}{\if@noskipsec\mbox{}\fi\par}{}{}
\makeatother
\usepackage{graphicx}
\makeatletter
\newsavebox\pandoc@box
\newcommand*\pandocbounded[1]{
  \sbox\pandoc@box{#1}%
  \Gscale@div\@tempa{\textheight}{\dimexpr\ht\pandoc@box+\dp\pandoc@box\relax}%
  \Gscale@div\@tempb{\linewidth}{\wd\pandoc@box}%
  \ifdim\@tempb\p@<\@tempa\p@\let\@tempa\@tempb\fi
  \ifdim\@tempa\p@<\p@\scalebox{\@tempa}{\usebox\pandoc@box}%
  \else\usebox{\pandoc@box}%
  \fi%
}
\def\fps@figure{htbp}
\makeatother
\ifLuaTeX
  \usepackage{luacolor}
  \usepackage[soul]{lua-ul}
\else
  \usepackage{soul}
\fi
\providecommand{\tightlist}{%
  \setlength{\itemsep}{0pt}\setlength{\parskip}{0pt}}
\usepackage{newunicodechar}
\newunicodechar{α}{\ensuremath{\alpha}}
\newunicodechar{ε}{\ensuremath{\varepsilon}}
\newunicodechar{π}{\ensuremath{\pi}}
\newunicodechar{Σ}{\ensuremath{\Sigma}}
\newunicodechar{′}{\ensuremath{{}'}}
\newunicodechar{⁴}{\ensuremath{{}^{4}}}
\newunicodechar{⁶}{\ensuremath{{}^{6}}}
\newunicodechar{⁻}{\ensuremath{{}^{-}}}
\newunicodechar{→}{\ensuremath{\rightarrow}}
\newunicodechar{−}{\ensuremath{-}}
\newunicodechar{√}{\ensuremath{\surd}}
\newunicodechar{≈}{\ensuremath{\approx}}
\newunicodechar{≥}{\ensuremath{\geq}}
\newunicodechar{≲}{\ensuremath{\lesssim}}
\newunicodechar{⊥}{\ensuremath{\perp}}
\usepackage{bookmark}
\IfFileExists{xurl.sty}{\usepackage{xurl}}{} 
\makeatletter
\@ifundefined{xmpquote}{}{}
\makeatother
\hypersetup{
  pdftitle={Animarium: Technical Report Version 1},
  colorlinks=true,
  linkcolor={Maroon},
  filecolor={Maroon},
  citecolor={Blue},
  urlcolor={Blue},
  pdfcreator={LaTeX via pandoc}}

\title{Animarium: Technical Report Version 1}
\author{}
\date{2026-10-08 · f8c96d8}

\begin{document}
\maketitle

{
\setcounter{tocdepth}{2}
\tableofcontents
}
\newpage

\section{Front matter --- version v1 (26 August
2026)}\label{front-matter-version-v1-26-august-2026}

\begin{center}\rule{0.5\linewidth}{0.5pt}\end{center}

\section{Animarium: Technical Report Version
1}\label{animarium-technical-report-version-1}

\textbf{Technical report v1.0 --- Mirko Degli Esposti} Department of
Physics and Astronomy (DIFA), University of Bologna ORCID
0000-0003-0316-3449 · mirko.degliesposti@unibo.it

Viewer \url{https://animarium.it} · pipeline
\url{https://github.com/mirko-degli-esposti/gsp} · viewer code
\url{https://github.com/mirko-degli-esposti/animarium}

\subsection{Abstract}\label{abstract}

Synthetic populations of eleven Italian municipalities --- 1,814,317
individuals in 887,937 households --- generated from published
aggregates alone: ISTAT census and register tables, census-section
counts, the national civic-address register, public-use survey
microdata, and six municipal open-data portals, every source certified
in a registry with licence, fingerprint and declared affordances. Four
rings give every attribute a declared place: a maximum-entropy joint
model of up to nine demographic attributes; whole-vector donation of
twenty-three attitudinal and health variables from survey respondents;
placement to census section, single year of age and address; and
households constrained by the census size distribution per section.
Every downstream layer --- detailed titles, work, names, biographies ---
is a declared derivation adding no information. The pipeline is
deterministic to the byte: regenerating all eleven municipalities from
the tagged commit reproduces every file of every ring bit for bit, in 33
minutes on one workstation. Populations are released in a public regime
enforced in the data (no names, no addresses, coordinates randomised
within census section), browsable in Animarium --- a dependency-free web
viewer where every number carries its comparison and every view is a
citable URL --- and downloadable as an open dataset. The report
documents the architecture, the sources and their certification, the
reproducibility and quality measurements at the release tag, the viewer,
and the narrative layer that renders records into personas for
LLM-driven simulation --- with the platform's controllability
demonstrated in companion experiments, and validation explicitly out of
scope.

\subsection{Context: what this report
is}\label{context-what-this-report-is}

This is the technical report of Animarium (\url{https://animarium.it}).
The companion paper, which positions the work in the literature and
reports the measurements on the full regional release, is Degli Esposti
(2026c), arXiv:XXXX.XXXXX. This report is the detailed technical record
that the paper refers to.

A synthetic population is a set of simulated individuals whose
collective statistics match what is known about a real one. It is the
usual starting point for agent-based and microsimulation models, and
more recently for simulations whose agents are driven by large language
models (Larooij and Törnberg, 2025). Most methods reweight a survey
sample against small-area totals. Iterative proportional fitting (Deming
and Stephan, 1940), calibration (Deville and Särndal, 1992) and
combinatorial optimisation (Williamson et al., 1998) are the standard
techniques, reviewed by Hermes and Poulsen (2012), Tanton (2014) and
Chapuis et al.~(2022) and implemented in packages such as simPop (Templ
et al., 2017). Deep generative models learn the joint distribution from
the same individual records instead (Borysov et al., 2019; Kim and
Bansal, 2023; Tang et al., 2025). For Italy, Scarpa et al.~(2026)
compare reweighting approaches for every municipality of Emilia-Romagna.
All of these methods need a sample of individual records. A sample-free
route instead builds the population from published tables alone
(Lenormand and Deffuant, 2013). Animarium takes that route for its
demographic core: it draws individuals from the maximum-entropy
distribution consistent with published constraints, using the solver of
Degli Esposti (2026a).

Openly released national synthetic populations exist for Canada
(Prédhumeau and Manley, 2023), the United States (Jiang et al., 2024;
Rineer et al., 2025) and Ireland (Caulfield Curley et al., 2025). Italy
has none, although ISTAT publishes unusually rich sub-municipal
aggregates (ISTAT, 2026). The FOSSR cluster offers a population
generator as a service (FOSSR, 2026), and individual studies build
located populations for their own needs (Garrone, 2026). Animarium is
meant to fill that gap with released populations, resolved to the census
section. That resolution is also what urban digital twins need in order
to have inhabitants (Batty, 2018; Marçal Russo et al., 2026). What
simulations built on such populations can validly show is the subject of
Degli Esposti (2026b).

\begin{quote}
\mbox{}%
\subsection{Status of this report}\label{status-of-this-report}

This report documents release \texttt{report-v1.0} of the pipeline:
eleven municipalities, generated in August 2026, and every measurement
in Part III was taken at that tag. It remains the technical record of
the architecture, which has not changed since.

The released collection has. Release \texttt{release-v2.0} (September
2026) covers 245 municipalities --- every municipality of Emilia-Romagna
above the population threshold declared in its registry, plus Brescia
--- 4,433,976 individuals in 2,051,901 households, produced by the same
four rings from the same sources, and deposited at
\href{https://doi.org/10.5281/zenodo.22647404}{10.5281/zenodo.22647404}.
The fleet-wide checks on that release, and the constraint-set
construction in the form it took when the collection grew, are reported
in the companion paper; this report describes the system, and the twelve
municipalities it documents in full remain the subset measured in the
greatest detail.

Two corrections belong here rather than in an unrevised text.
\textbf{Sampling.} §III.4 explains a standard deviation of the per-cell
standardised errors above unity as autocorrelation of a Gibbs chain.
That explanation is withdrawn: the fit is exact on the dual and the
population is drawn i.i.d. from the fitted distribution over the
enumerated state space, so \texttt{sd(z)\ ≈\ 1} is what independent
draws predict, and the one municipality above it is an open item rather
than an explained one. \textbf{Ring 4.} The two post-tag patches
declared in §III.5 are in the code that produced \texttt{release-v2.0};
the household counts of this report are those of the tagged commit.
\end{quote}

\subsection{A note on tools}\label{a-note-on-tools}

Large language models --- principally Anthropic's Claude, with
occasional use of OpenAI's ChatGPT --- were used throughout this work:
as a writing partner for these notes and for this report, as a reviewer
and occasional author of code, and as an interlocutor while designing
and debugging the pipeline. Every design decision, every measurement,
and every claim in this document was made, checked and is owned by the
author; where a number appears, it was produced by code in the tagged
repository and verified against its source, not by a model. The errors
that remain are mine.

\subsection{Reading conventions}\label{reading-conventions}

Two markers appear throughout. \textbf{{[}m{]}} marks a figure measured
at the release tag, with the log or diagnostic that produced it
available in the repository; \textbf{{[}n{]}} marks a claim taken from a
working note, which is cited by name and travels with the code. Nothing
else in this report is a number: where a quantity appears without a
marker, it is a definition or a design parameter.

Claims marked \textbf{{[}n{]}} come from the project's working notes,
which live in \texttt{note/} of the GSP repository and are versioned
with the code: they are cited by file name, so
\texttt{nota\_nucleo\_familiare\_v3} is that file at that version. They
are in Italian, as the project's internal record is, and they are the
layer where a decision is argued at length before it becomes a paragraph
here.

\subsection{Version binding}\label{version-binding}

Every claim marked \textbf{{[}m{]}} in this report was measured at the
versions below; the hashes are the verification path.

{\def\LTcaptype{none} 
\begin{longtable}[]{@{}
  >{\raggedright\arraybackslash}p{(\linewidth - 4\tabcolsep) * \real{0.3333}}
  >{\raggedright\arraybackslash}p{(\linewidth - 4\tabcolsep) * \real{0.3333}}
  >{\raggedright\arraybackslash}p{(\linewidth - 4\tabcolsep) * \real{0.3333}}@{}}
\toprule\noalign{}
\begin{minipage}[b]{\linewidth}\raggedright
artefact
\end{minipage} & \begin{minipage}[b]{\linewidth}\raggedright
identifier
\end{minipage} & \begin{minipage}[b]{\linewidth}\raggedright
verify with
\end{minipage} \\
\midrule\noalign{}
\endhead
\bottomrule\noalign{}
\endlastfoot
this report & v1.0 (arXiv:2608.27111v2)--- & \\
GSP pipeline & tag \texttt{report-v1.0},
github.com/mirko-degli-esposti/gsp & \texttt{git\ rev-parse} \\
Animarium viewer & tag \texttt{report-v1.0},
github.com/mirko-degli-esposti/Animarium & idem \\
MaxEnt solver (\texttt{maxent-popsynth-pcd}) & commit \texttt{14f5bab}
(2026-08-03), github.com/mirko-degli-esposti/maxent-popsynth-pcd &
\texttt{git\ rev-parse\ -\/-short\ HEAD} in the clone \\
code snapshots & GSP:
\href{https://doi.org/10.5281/zenodo.22127410}{10.5281/zenodo.22127410}
· Animarium:
\href{https://doi.org/10.5281/zenodo.22127473}{10.5281/zenodo.22127473}
& checksum on Zenodo \\
open dataset &
\href{https://doi.org/10.5281/zenodo.22127581}{10.5281/zenodo.22127581},
CC-BY-4.0 & SHA-256 below \\
open dataset (v2.0) & Zenodo
\href{https://doi.org/10.5281/zenodo.22647404}{10.5281/zenodo.22647404}
· all versions: 10.5281/zenodo.22647403 & \texttt{SHA256SUMS.txt} in the
deposit \\
code snapshots (v2.0) & viewer 10.5281/zenodo.22646253 & checksum on
Zenodo \\
public bundle & eleven \texttt{pop.parquet}, SHA-256 in
\texttt{note/misure/rilancio\_report\_v1.0/hash\_parquet\_report\_v1.0.txt}
& \texttt{sha256sum} \\
source fingerprints & \texttt{fonti/registro.yaml} at the tag &
\texttt{python\ -m\ gsp.fonti\ -\/-verifica} \\
companion papers & arXiv:2603.27312 (solver) · arXiv:2607.00910 (SIVE) &
--- \\
\end{longtable}
}

One entry above deserves its warning. The fitting stage imports
\texttt{ConstraintSet} from the solver repository by \emph{filesystem
discovery} --- a glob over a sibling directory --- and does not verify
which commit it found. Until the solver is a pinned package dependency
(planned for v1.1), reproducing the fit requires checking out the commit
above by hand.

The tag \texttt{report-v1.0} names the same commit as
\texttt{report-v1.0-rc1}, under which the measurements of Part III were
taken: the release candidate was promoted rather than rebuilt, so every
\textbf{{[}m{]}} in this report refers to the artefact the tag points
at. (Two ring-4 patches were committed after that commit and are
declared in §III.5.)

\subsection{What, in a record, is actually
real}\label{what-in-a-record-is-actually-real}

The single table to read before reading anything else. A released record
looks like a person; its components have exactly three provenances.

{\def\LTcaptype{none} 
\begin{longtable}[]{@{}
  >{\raggedright\arraybackslash}p{(\linewidth - 4\tabcolsep) * \real{0.3333}}
  >{\raggedright\arraybackslash}p{(\linewidth - 4\tabcolsep) * \real{0.3333}}
  >{\raggedright\arraybackslash}p{(\linewidth - 4\tabcolsep) * \real{0.3333}}@{}}
\toprule\noalign{}
\begin{minipage}[b]{\linewidth}\raggedright
component
\end{minipage} & \begin{minipage}[b]{\linewidth}\raggedright
provenance
\end{minipage} & \begin{minipage}[b]{\linewidth}\raggedright
what is real
\end{minipage} \\
\midrule\noalign{}
\endhead
\bottomrule\noalign{}
\endlastfoot
demographic attributes (sex, age, marital status, citizenship,
education, condition, background, parents' origin, zone) & drawn from a
maximum-entropy distribution under census constraints & the
\emph{aggregates} are real; the individual is a sample of a
distribution, and no record corresponds to anyone \\
attitudes and health (23 AVQ variables) & the complete response vector
of one real survey respondent, copied whole & the vector is real ---
already protected at source by ISTAT's public-use release, and reused
across tens of synthetic individuals, so no combination is unique to one
record \\
section, exact age, address & allocated within the census section; the
address uniformly among its registered civic numbers & the \emph{counts}
per section are real; the assignment carries no information about anyone
(and the public regime randomises the coordinate within the section) \\
household id and role & assembled under the census size distribution per
section, composition from a survey-based repertoire & the size
distribution is real per section; the household is a model \\
names, detailed titles, sector, biography & deterministic rendering from
registered repertoires and census-conditioned structure & nothing:
plausible by construction, and therefore collident --- a name
individuates no one \\
\end{longtable}
}

\textbf{No component is personal data.} The population is simulated from
published aggregates, not anonymised from individual records; there is
no one to re-identify (§I.7).

\subsection{How to cite}\label{how-to-cite}

The report:

\begin{quote}
Degli Esposti, M. (2026). \emph{Animarium: Technical Report Version 1.}
arXiv:2608.27111.
\end{quote}

The dataset:
\href{https://doi.org/10.5281/zenodo.22127581}{10.5281/zenodo.22127581},
CC-BY-4.0, with the attributions in \texttt{fonti/ATTRIBUZIONI.md}. The
software: the Zenodo snapshot DOIs in the binding table, or
\texttt{CITATION.cff} in either repository.

\subsection{Licences}\label{licences}

Code MIT (both repositories). Released data and populations CC-BY-4.0,
inheriting the attributions listed in \texttt{fonti/ATTRIBUZIONI.md};
sources used for validation only are registered, fingerprinted and not
redistributed (§II.2). The viewer runs on any static server; no
component requires an account or a key.

\subsection{Scope}\label{scope}

This report is a release document and a reproducibility record: what the
pipeline does, on which sources it stands, what was measured at the tag,
and in which regimes the results leave the machine. It is not a methods
paper --- the maximum-entropy solver is arXiv:2603.27312 --- and it is
not a validation of LLM-driven synthetic populations, which it treats
strictly as evidence of use (Part V, first paragraph). Where the report
and the working notes disagree, the report is later and wins. Every
quantitative claim in Part III is measured at the release tag; where a
number comes from a working note instead, it carries the note's name.
Two boundaries are declared rather than hidden: the fit requires the
solver repository at the pinned commit above, and rings 2--4 require
survey microdata obtainable by anyone but through a manual request
(§III.1). Part V describes work in progress and is expected to change.

\newpage

\section{Part I --- Architecture: the four
rings}\label{part-i-architecture-the-four-rings}

\subsection{Version 1 --- §I.0--I.7 (26 August
2026)}\label{version-1-i.0i.7-26-august-2026}

\begin{quote}
\texttt{GSP\_popolazioni\_full\_riferimento\_v24} is cited throughout
this part as \emph{riferimento}.
\end{quote}

\begin{center}\rule{0.5\linewidth}{0.5pt}\end{center}

\subsubsection{I.0 The pipeline, as run}\label{i.0-the-pipeline-as-run}

Before principles, the machine. This section walks the chain exactly as
it executes for one municipality, with the real commands; everything
after it --- the design principles of §I.1, the rings of §I.2--I.5 ---
attaches to a step the reader has already seen run. The walkthrough of
every script against two municipalities added from scratch, with real
outputs and the findings it produced, is in
\texttt{note/collaudo\_acquisizione\_v0.2.md} (in Italian, as the
project's working notes are); this section is its distillation. Terms
are used here before they are defined --- \emph{ring} for the four
stages that give every attribute its place (§I.2--I.5), \emph{K6C/K9C}
for the two sizes of the joint model, without and with the zone
coordinate (§I.2) --- because the reader who sees the chain run once
will attach the definitions to steps already seen, not the other way
around. On first reading, the command names carry enough: fetch,
sections, zones, constraints, fit, then the three enrichment steps.

\begin{figure}
\centering
\pandocbounded{\includegraphics[keepaspectratio,alt={The pipeline as
run: three doors feeding one chain, ring 1 producing the population and
rings 2--4 enriching it, with the publication regimes as the single
exit. Source:
note/figure/fig\_I0\_pipeline.dot.}]{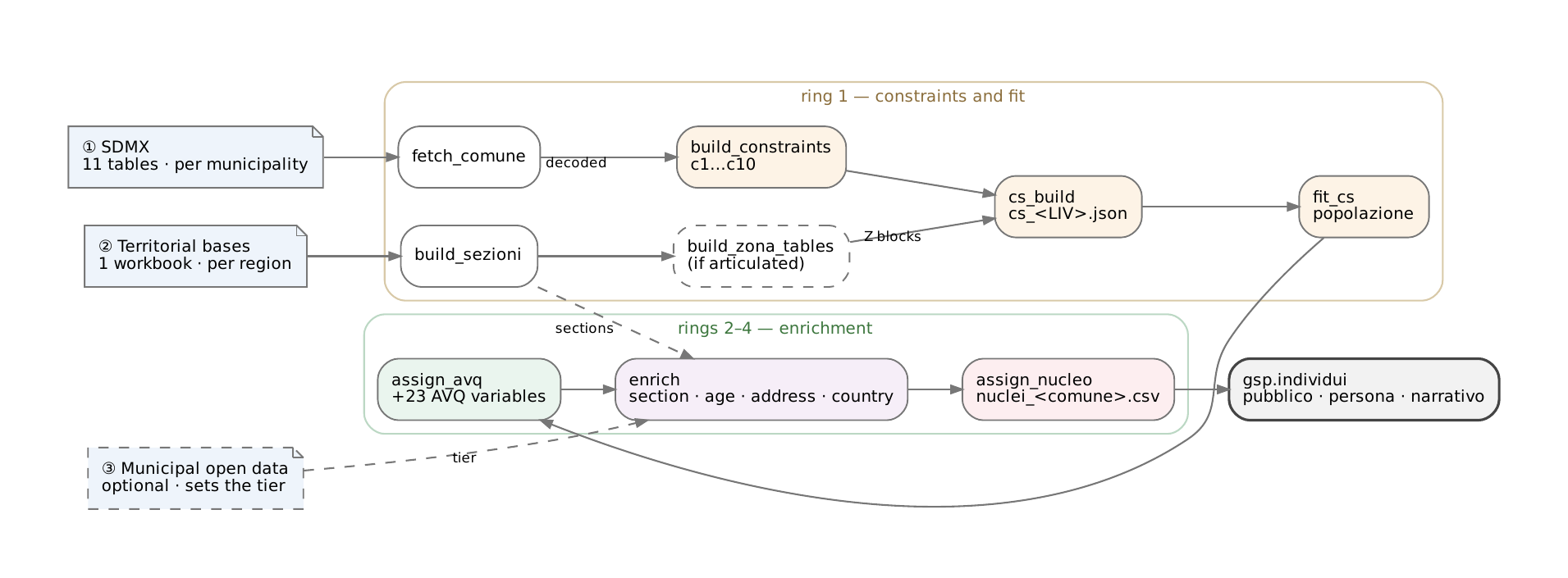}}
\caption{The pipeline as run: three doors feeding one chain, ring 1
producing the population and rings 2--4 enriching it, with the
publication regimes as the single exit. Source:
\protect\texttt{note/figure/fig\_I0\_pipeline.dot}.}
\end{figure}

\textbf{Three doors.} Every municipality receives data through three
doors at three granularities:

\begin{enumerate}
\def\labelenumi{\arabic{enumi}.}
\tightlist
\item
  \textbf{SDMX → the municipality as a whole.} Eleven ISTAT tables ---
  register and permanent census --- downloaded per municipality. Their
  row counts measure \emph{structure, not size}: Milano (1.4 M
  residents) and Mantova (49 k) produce the same 819 rows of education ×
  age, because the grid is fixed and only the counts change. Download
  time is dominated by the API's rate limit (\textasciitilde5
  queries/minute, enforced globally by \texttt{gsp.istat.sdmx};
  violations lead to multi-day IP blocks), not by bytes.
\item
  \textbf{Territorial bases → census sections.} One workbook \emph{per
  region}, one row per section, 138 columns (population by five-year
  class and sex, citizenship by three macro-classes, education at five
  levels, employment, migratory background, household sizes PF3--PF8).
  This is where a metropolis outweighs a town --- Milano has 6,059
  sections, Mantova 574 --- and where the sub-municipal articulation
  lives, as the \texttt{COM\_ASC*} columns.
\item
  \textbf{Municipal open data → what only the municipality publishes.}
  Citizenship by country at sub-municipal level, chiefly. This door is
  optional by design: the \emph{tier} system of §I.3 degrades gracefully
  to tier 0 (country conditioned on sex and geography from the census
  margin alone) when it is closed, and tier 0 was exercised end-to-end
  on both test municipalities.
\end{enumerate}

\textbf{The chain, per municipality.} With the region's one-off files in
place (door 2, plus the ANNCSU address extraction and the regional AVQ
pool), a new municipality is one entry in the \texttt{COMUNI} registry
of \texttt{gsp.common} --- name, slug, region, and the declared zone
level or the declared absence of one --- followed by:

\begin{verbatim}
python scripts/acquisizione/fetch_comune.py 020030          # door 1: 11 tables
python scripts/vincoli/build_sezioni.py     020030          # door 2: sections, ASC check
python scripts/vincoli/build_zona_tables.py 015146          # only if articulated
python scripts/vincoli/build_constraints.py 020030 --anno 2024
python scripts/vincoli/cs_build.py  020030 --anno 2024 --livello K6C
python scripts/fit/fit_cs.py        020030 --anno 2024 --livello K6C --pool 65000
python scripts/attributi/assign_avq.py 020030 --anno 2024 --pop-file popolazione_K6C.csv
python scripts/attributi/enrich.py     020030 --anno 2024 --pop-file popolazione_K6C_avq.csv
python scripts/attributi/assign_nucleo.py 020030 --anno 2024
\end{verbatim}

\texttt{rigenera.sh} runs this chain for the whole fleet --- eleven
municipalities, 44 output files, 33 minutes on one workstation,
byte-identical across runs (§III.2).

Two of these steps deserve a word here because their division of labour
is not guessable from the names. \texttt{build\_constraints} is the
\emph{municipal preparer}: it reads the eleven decoded tables, verifies
year coverage per table before building anything (a missing mandatory
table is a fatal error; a missing optional one is a declared skip), and
writes the municipal constraint blocks \texttt{c1..c10} together with a
manifest and a consistency report. \texttt{cs\_build} is the
\emph{assembler}: it takes those blocks, adds the zone blocks where the
municipality is articulated, applies the declared zeros and the ε floor,
audits the shared margins, and emits the constraint set the solver
reads. The two-stage relay means the municipal preparation is
inspectable on its own --- the \texttt{report.md} each run writes is
where the register--census identity of §I.2 was first noticed.

\textbf{What the test run established.} Mantova (K6C, no articulation)
and Milano (K9C, nine municipi) were added from scratch as test cases
--- they are not part of the released bundle (§III.5). Mantova's
constraint set has m = 263 constraints on \textbar X\textbar{} = 5,376
states and fits exactly in 0.17 s at MRE 3.4·10⁻⁴; Milano's has m =
1,037 on \textbar X\textbar{} = 1,451,520 and fits exactly in 54 s at
MRE 4.2·10⁻⁴ --- the same error as the fleet, because for the solver
what counts is the number of zones, not of residents: Milano, with nine
municipi, is a \emph{smaller} problem than Parma with thirteen
quartieri. The chain needed nothing beyond the registry entry; the two
limits it surfaced (a solver-repository dependency resolved by
filesystem discovery, and address coverage depending on each
municipality's ANNCSU georeferencing) are declared in the front matter
and §I.4 respectively.

\subsubsection{I.1 Design principles}\label{i.1-design-principles}

\begin{figure}
\centering
\pandocbounded{\includegraphics[keepaspectratio,alt={The four rings and
the derived layer: what each adds, from which source, by which method,
and what is real in it. Source: note
figure/fig\_I1\_anelli.dot.}]{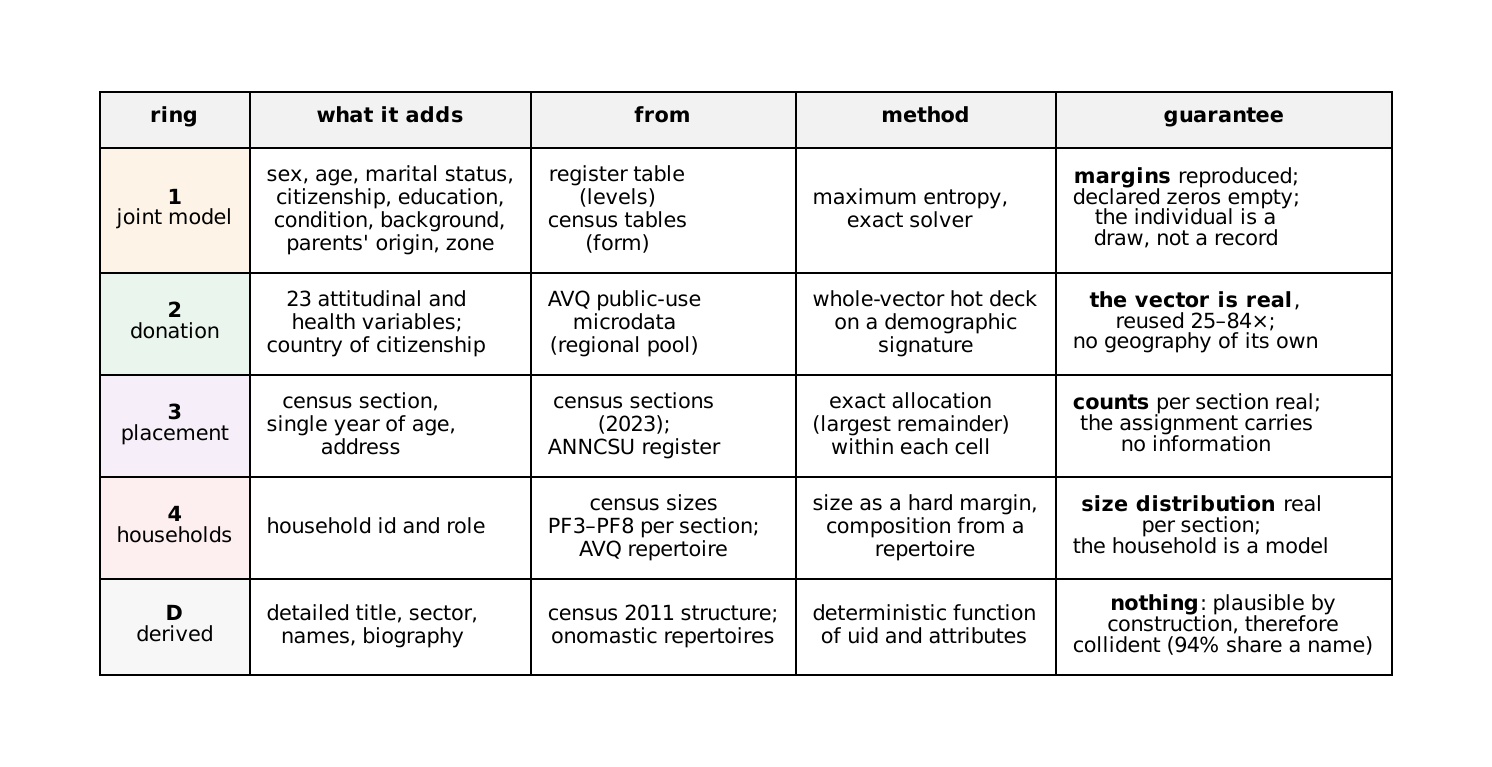}}
\caption{The four rings and the derived layer: what each adds, from
which source, by which method, and what is real in it. Source:
\protect\texttt{note\ figure/fig\_I1\_anelli.dot}.}
\end{figure}

\emph{Four principles run through the whole design; they are worth
stating before the mechanism they govern.}

\emph{Every attribute has a declared place}: education lives in the
joint model, because its association with age and citizenship must hold
under simultaneous constraints; the sector × position pair is derived
downstream, conditioned on sex and municipality, because that is where
the measurement put it --- a total-variation distance between
conditional and marginal compositions, defined below.

An attribute lives either in the joint model (ring 1), where the solver
generates it under simultaneous constraints, or in a downstream
derivation, conditioned on what is already there. The choice is not a
matter of taste, it is a measurement: for a candidate attribute X and
each available conditioning S, the total-variation distance d(S) =
TVD(P(X \textbar{} S), P(X)) --- the share of mass one would have to
move, readable without a scale (``one card in five would be wrong'').
Large d on a conditioning available downstream → derive; large d only on
a conditioning that is not available downstream, typically fine
geography → joint model; small everywhere → assign unconditionally. When
two derived variables are involved, a second measurement decides whether
they are drawn separately or jointly --- TVD(P(X, Y \textbar{} S), P(X
\textbar{} S) · P(Y \textbar{} S)) --- and here a secondary criterion
supplements the number: at equal distance, preserve the structure that
makes an error recognisable. A failed joint produces a manager in
agriculture, which anyone notices; a failed conditioning produces
plausible individuals in slightly wrong proportions, which nobody does.
It is the same reason the AVQ hot-deck copies whole vectors (§I.3). The
criterion is implemented as a self-contained library (\texttt{gsp.tvd}:
distance, per-conditioning profiles, support-partition checks); it knows
nothing of municipalities or populations, and it refuses to measure
across mismatched supports --- on real census tables, a distance
computed over different supports produced plausible and meaningless
values five times in two days before that refusal was built in.

The criterion has already reversed one decision. The economic sector
seemed to belong in the joint model, conditioned on sex --- the natural
pairing. Measured, the pairing is the wrong one: on the ATECO
composition, education dominates (TVD 0.149--0.512 across
municipalities, median 0.346), sex moves half as much (median 0.153),
and age sits below both. Education, however, is unusable as a
conditioner at municipal level --- the census does not publish the
sector × title cross outside agriculture --- while putting the sector in
the state space would have cost a much larger support and a conditioning
on the least informative of the three variables. The measurement settled
it: the pair sector × position is drawn downstream, jointly (the
joint-vs-product distance is 0.138--0.166 across five territories ---
structure, not noise), conditioned on sex and municipality, with
education's absence declared as a limit rather than silently absorbed
(§I.6, and the work module's own documentation).

\emph{Every metric is normalised against its null}. A raw error, a raw
distance, a raw correlation scales with the geometry of what it measures
--- how many cells, how small the sections, how fine the partition ---
before it says anything about quality. One example stands for all: the
municipality with the ``worst'' raw constraint error in this report (16
\%) is a town of sixteen thousand whose sampling floor --- the error a
perfect sampler would show, given its cell sizes --- is 24 \%; the
municipality with the finest zoning has a floor of 3,000 \% --- its
constraint set contains many cells whose expected count is a fraction of
an individual, and on a cell that expects 0.04 people, finding one is a
relative error of 2,400 \% while being off by one person. At that point
the relative scale has stopped measuring the fit and started measuring
the cell sizes. The comparison that still works is the z-score: each
cell's deviation divided by the fluctuation a perfect sampler would show
on that cell --- so every cell is judged against its own floor, and
cells of any size become comparable. On that scale the finest-zoned
municipality is unremarkable (§III.3). Comparing raw numbers across
configurations compares their geometry, not their quality. Every table
in Part III therefore carries the null beside the observation, and the
rule has retired at least four published-in-note numbers in this
project's history (§III.6).

\emph{The population is read-only.} Once generated, a population file is
never modified: derived attributes are computed at consumption time from
the \texttt{uid}, ring 4 writes its columns to a separate file joined on
\texttt{uid}, and every fix that would touch the file is instead queued
for a full regeneration cycle. The premise makes the regression test
trivial --- the file must never change --- and makes \texttt{diff} the
universal instrument.

\emph{Nothing is silently corrected.} When a claim in a working note
turns out to be wrong, the text is not fixed: the original stays where
it was, and a dated annotation beside it says what the correct number is
and why the first one was wrong. Predictions are written down before the
measurement runs, so that a falsified prediction cannot be quietly
rephrased as an expected result --- this report contains two, both
falsified, both kept (§III.6). And code is held to the same standard as
prose: a refactoring is accepted only when the regenerated populations
are byte-identical to the archived ones, so that ``nothing changed'' is
a measurement, not an assurance (§III.2). The cost is a paper trail of
one's own errors; the return is that every number still standing has
survived something.

\subsubsection{I.2 Ring 1 --- the joint
model}\label{i.2-ring-1-the-joint-model}

Ring 1 produces, for each municipality, the joint distribution of up to
nine demographic attributes --- zone, sex, age (eight bins), marital
status, citizenship (Italian/foreign), education (six classes),
occupational condition (seven), migratory background (six), parents'
origin (five) --- as a maximum-entropy distribution over the discrete
support X subject to the census constraint set, and then samples N
individuals from it.

{\def\LTcaptype{none} 
\begin{longtable}[]{@{}
  >{\raggedright\arraybackslash}p{(\linewidth - 4\tabcolsep) * \real{0.3333}}
  >{\raggedright\arraybackslash}p{(\linewidth - 4\tabcolsep) * \real{0.3333}}
  >{\raggedright\arraybackslash}p{(\linewidth - 4\tabcolsep) * \real{0.3333}}@{}}
\toprule\noalign{}
\begin{minipage}[b]{\linewidth}\raggedright
attribute
\end{minipage} & \begin{minipage}[b]{\linewidth}\raggedright
classes
\end{minipage} & \begin{minipage}[b]{\linewidth}\raggedright
values
\end{minipage} \\
\midrule\noalign{}
\endhead
\bottomrule\noalign{}
\endlastfoot
\texttt{zona} & 4--33 & the municipality's declared articulation
(statistical zones, quartieri, circoscrizioni, aree) \\
\texttt{sesso} & 2 & M, F \\
\texttt{eta} & 8 & 0--8, 9--14, 15--24, 25--34, 35--49, 50--64, 65--74,
75+ \\
\texttt{stato\_civile} & 4 & never married, married or in civil union,
divorced, widowed \\
\texttt{cittadinanza} & 2 & Italian, foreign \\
\texttt{istruzione} & 6 & no title, primary, lower secondary, upper
secondary diploma, tertiary (incl.~ITS), postgraduate \\
\texttt{condizione} & 7 & employed, seeking work, student, homemaker,
pension recipient, other, not applicable (under 15) \\
\texttt{background} & 6 & native Italian; returned Italian; naturalised,
born in Italy; naturalised immigrant; foreign, born in Italy; foreign
immigrant \\
\texttt{origine\_genitori} & 5 & both Italian; Italian mother, foreign
father; foreign mother, Italian father; both foreign; not applicable \\
\end{longtable}
}

(Class boundaries follow the census: \emph{upper secondary} is the
five-year diploma of any track (liceo, tecnico, professionale);
\emph{tertiary/ITS} pools bachelor's, master's degrees and ITS diplomas
without distinguishing them; \emph{postgraduate} includes doctorates and
specialisation degrees as a class, although the detailed-title rendering
(§I.6) can only produce master's-level titles, its 2011 source having no
doctoral entries --- a declared limit of the derived layer, not of the
class. Three- and four-year vocational qualifications are pooled by the
census itself into the \emph{upper secondary} class, together with IFTS
--- the source's aggregation, verified on the decoded table's own label
(\texttt{USE\_IF}: ``diploma di istruzione secondaria di II grado o di
qualifica professionale, compresi IFTS'').

The method --- exact solution of the dual where \textbar X\textbar{}
allows, persistent contrastive divergence where it does not --- is the
subject of arXiv:2603.27312 and is not repeated here; what this report
documents is the \emph{configuration}: what the constraints are, where
they come from (Part II), and what the fit achieves on the eleven
municipalities (Part III).

The constraint set is a \emph{template}, written once per level and
applied identically to every municipality at that level. Two levels are
in production. \textbf{K9C} --- nine constrained attributes, the full
table above --- applies where the municipality has a declared
sub-municipal articulation and the migratory-background tables to
populate it: the nine provincial capitals and Brescia. \textbf{K6C} ---
six attributes: sex, age, marital status, citizenship, education,
condition --- applies where it does not (Ferrara and Castenaso): no
\texttt{zona}, and no \texttt{background} / \texttt{origine\_genitori},
whose census tables have no sub-municipal counterpart there. The level
is a property of the available sources, not of the pipeline, and the
viewer inherits it (a K6C municipality shows six filterable attributes,
and its census-reference coverage has a different denominator --- 26 of
96 pairs against 67 of 333, which is why coverage \emph{counts} are
never compared across levels, only shares; §IV.1).

At K9C the template has sixteen blocks --- sub-tables of the census each
constraining a small set of attributes jointly --- of which eleven are
complete, their masses summing to one, and five are partial
\textbf{{[}n{]}} riferimento §14.1.

A partial block's unlisted cells are \emph{free}, not forbidden; the
distinction matters because its opposite also occurs: every constraint
set carries explicit zeros, of three provenances. Twenty-six are
\emph{imposed by construction}, identically in every municipality at
both levels: the logically impossible pairs of age × education (8) and
age × occupational condition (18). Six more appear identically in every
K9C municipality in citizenship × background: combinations the
variables' definitions exclude --- an Italian citizen cannot carry a
\emph{foreign} background, a foreign citizen cannot carry an Italian or
naturalised one. They are as structural as the twenty-six, and differ
only in provenance: the census table delivers them as observed zeros,
\texttt{cs\_build} does not need to impose them. The rest are the only
\emph{contingent} zeros --- demographically observed, in sex × age ×
marital status, varying with the municipality as data should: none in
Bologna, where even widowed 15-to-24-year-olds exist (two of them
costing the fit its largest z-score, §III.3a); two in most
municipalities --- the widowed young of both sexes; six in Castenaso,
whose sixteen thousand residents leave more demographic cells empty.
Mechanically all three kinds are the same object --- a constraint with
target zero, whose mass the solver suppresses --- and the generated
populations honour them exactly: no zero cell is realised in any
municipality (§III.3), no impossible combination in 1,814,317
individuals (§III.2). Absent and zero are opposites for maximum entropy
--- unconstrained versus forbidden. An unconstrained cell receives the
\emph{most} mass compatible with the remaining constraints, which is the
principle itself at work: where nothing is imposed, the distribution
spreads as evenly as it can. A zero-constrained cell receives none.
There is no middle ground, so misclassifying a cell does not produce a
small error but the largest possible one on that cell --- in either
direction. Read a missing row as ``no constraint'' when it is an
observed zero, and the model manufactures widowed teenagers in quantity;
read it as ``zero'' when the cell is merely outside the table's
universe, and education becomes forbidden for children under nine ---
for whom the census publishes nothing, but who exist and carry
\texttt{nessun\_titolo} --- and the population is distorted or cannot be
generated at all. The source offers no help: a downloaded table shows
the same missing value for both, and the difference lives only in the
table's declared universe. Translating tables into constraints therefore
requires knowing, cell by cell, \emph{why} a value is missing ---
reading the universe, not just the numbers --- and doing it implicitly
is the characteristic error of the exercise \textbf{{[}n{]}} riferimento
§14.2. The partial blocks encode exactly this reading: out-of-universe
cells stay free, observed zeros enter at α = 0.

Readers from the population-synthesis literature will look here for the
\emph{zero-cell problem} --- the impossibility, for seed-based methods
like IPF, of restoring mass to a combination the sample seed happens to
miss. The problem does not arise in this construction, because there is
no seed: the maximum-entropy distribution is determined by the
constraints alone, and an unconstrained cell receives the most even mass
compatible with them rather than inheriting a sampling accident. What
the literature treats as one pathological category is here three
deliberate ones --- observed zero (forbidden by a constraint),
impossible pair (excluded from the support, α = 0), and uncovered cell
(free) --- and the residue of the classical problem survives only as the
sampling floor on small cells (§III.3).

Two margins have distinct roles, because they describe different
instants: the population-register table (1 January of year N) fixes the
municipal totals \emph{exactly} --- the hard margin --- while the census
tables (31 December of year N−1) enter as soft constraints. A word on
what \emph{hard} and \emph{soft} mean here, because the natural reading
is wrong. They do not rank two competing measurements by trust. Since
2018 the official resident population is produced \emph{by} the
permanent census, so the register table and the census tables publish
the same demographic base: on sex × single year of age the two flows
agree exactly --- 2,821 cells across fourteen municipalities, zero
discrepant, maximum absolute difference 0 \textbf{{[}m{]}} (§III.3). The
identity extends down one more layer: the census-section tables,
aggregated over a municipality, reproduce the same base --- at Brescia,
the thirty-two five-year cells (sixteen classes × two sexes) summed from
6,000-odd sections match the register with zero discrepancy
\textbf{{[}m{]}}. Register, municipal census and section tables are one
dataset published at three granularities. The few cells present in one
flow only are empty ones in the tail of the age distribution (a
99-year-old male in a town of 16,000), not disagreements.

What distinguishes the two families is therefore not accuracy but
\emph{extension}: the register table extends the common base along
marital status --- the one axis no census table carries --- while the
census tables extend it along citizenship, education, occupational
condition and migratory background. The register block enters as exact
counts; the census blocks enter as conditional distributions applied to
those counts (\texttt{share\ ×\ count}, per group), which is why the
demographic margins remain exact by construction rather than by
reconciliation. Two costs come with the conditional form and are
declared: census quotas defined on wide age classes (9--24, 15--24) are
applied to finer groups under an assumption of homogeneity within the
class, and census values are rounded, with a per-table rounding sigma
recorded in the constraint-set manifest.

The conditional form is not a device for the census tables alone: it is
\emph{the} architectural rule, applied three times. Levels come from the
spine; everything else contributes form. The census blocks contribute
the socio-economic form (\texttt{share\ ×\ count} per demographic group,
above); the zone blocks contribute the geographic form ---
\texttt{P(zone\ \textbar{}\ group)\ ×\ municipal\ counts}, with IPF
closing the double margin --- each at the resolution its section columns
carry: five-year classes for age, three macro-classes for citizenship,
five education levels against the population's six, and the employed
side only for occupational condition (the geography of unemployment is
constrained by no observed datum, §I.4); and the country of citizenship
repeats the same pattern downstream, in ring 3's tier system (§I.3). One
principle, three applications --- which is also why the zone-margin
audits printed by \texttt{cs\_build}
(\texttt{Z1\ vs\ A:\ max\textbar{}diff\textbar{}\ =\ 0.000000}) are
algebraic identities verifying the implementation, not facts about the
sources; the fact about the sources is the register--census identity
above.

At the release tag every municipality was solved \emph{exactly}: the
eleven fits converge with MRE between 2.4·10⁻⁴ and 5.0·10⁻⁴, in 0.17 s
to 182 s \textbf{{[}m{]}} (generation logs at the tag). No municipality
in the fleet required PCD, which the solver provides for state spaces
this release does not contain --- Brescia at K10C, or the 88-NIL fit of
Milano registered as an experiment (§III.5). Two regularities are worth
noting, because they are counter-intuitive: fitting time follows
\textbar X\textbar{} = 161,280 × zones and not population --- Brescia,
198k residents in 33 quartieri, takes 182 s against Bologna's 79 s for
390k residents in 18 zones --- and the fit error does not follow either,
staying inside a narrow band across two orders of magnitude of
population, which makes it a property of the stopping criterion rather
than of the problem's difficulty.

\subsubsection{I.3 Ring 2 --- donated attributes and the country of
citizenship}\label{i.3-ring-2-donated-attributes-and-the-country-of-citizenship}

The population's attitudinal and health layer --- twenty-three AVQ
variables: self-rated health, chronic conditions, smoking, BMI, the
mental-health index, environmental satisfaction, interpersonal trust,
and the institutional-trust battery on a 0--10 scale --- is not
modelled. It is \emph{donated}: each synthetic individual receives the
complete response vector of one respondent of ISTAT's \emph{Aspetti
della vita quotidiana} public-use microdata (pools: Emilia-Romagna 4,629
donors, Lombardy 8,111; survey years 2023--24 stacked, weights
renormalised within year --- the 2022 wave is declared in the
acquisition list but excluded, because it lacks the chronic-conditions
item with no equivalent: an absent variable excludes the wave rather
than entering as a silent gap), drawn from the conditioning cell sex ×
macro-age × education-4 with a declared hierarchical collapse when a
cell holds fewer than twenty donors.

Copying the whole vector rather than sampling variable by variable is
the decision that preserves the inter-variable correlations by
construction --- and the same recognisability criterion as §I.1: a
mis-sampled joint produces visibly impossible respondents, a mis-sampled
marginal produces invisible bias \textbf{{[}n{]}} riferimento §2.2.

The price is stated, measured, and carried through every product. The
donated vector holds no geography below the region (assumption 6): if a
neighbourhood shows lower institutional trust, it is because people of
the groups that everywhere express lower trust live there in greater
numbers --- composition --- not because the neighbourhood adds anything
of its own, which the data cannot contain. The viewer says this on every
spatially filtered panel (§IV.1). The honest sample size behind any AVQ
mean is not n but Kish's effective size on the variable's own universe.
If each donor i is reused w\_i times in the population, the synthetic
sample of Σw\_i individuals is worth n\_eff = (Σ w\_i)² / Σ w\_i²
independent respondents ---the count the weights would give if they were
all equal, and less than the number of distinct donors whenever they are
not --- which is why counting distinct donors, the obvious shortcut, is
not enough. Two limits make it readable: with every donor used the same
number of times, n\_eff is exactly the number of donors; with one donor
carrying most of the population, n\_eff approaches one. Computed per
variable, on the universe that variable actually has --- the trust
battery is asked of adults, the school questions of a narrower group ---
the confidence bands it implies are 2 to 20 times wider than the naive
ones, the factor growing with the municipality's population and varying
with the universe (§III.3b). The hierarchical collapse of the
conditioning cell touches 1.5--3.1 \% of individuals --- but not at
random: the cells that fall below the twenty-donor threshold are all
low-education cells, so the collapse concentrates precisely where
educational conditioning would matter \textbf{{[}n{]}} riferimento
§13.2. The donor's identity survives as an equivalence class --- the
donated tuple \emph{is} the signature --- which is what the viewer
exposes on every individual card and what §III.3b counts.

Three counts appear in this report and they are not in conflict:
twenty-three variables are donated and all twenty-three form the donor
signature; the viewer's institutional-trust panel displays the fifteen
of them that share a 0--10 trust scale (the two health-service
judgements, the armed forces, eleven institutions of the PUNTIFI battery
and the local-health-authority rating), leaving out health, chronic
conditions, smoking, mental health, environment, BMI and weight ---
donated all the same, and present in the record.

The second half of ring 2 is the country of citizenship. The census
gives the municipal margin --- country × sex --- and, per census
section, the count of foreign residents; where a municipality publishes
a sub-municipal table of residents by country, that table is register
data of a different date, and its \emph{levels} are incompatible with
the census by amounts that matter (Brescia's register counts 40,090
foreigners against 37,478 census). The construction therefore uses the
local source's \emph{shape} only --- the third application of the rule
stated in §I.2: levels from the base, form from the source that has it.
It seeds an iterative proportional fit whose two margins are both census
--- the municipal country × sex table, and the census count of
foreigners of each geography --- so the system is consistent by
construction and converges in 9--88 iterations to \textasciitilde10⁻¹¹
\textbf{{[}n{]}} riferimento §6. Coverage is complete for the same
reason: a section's weight for foreigners is proportional to its census
foreign count, so no foreigner can land where the census counts none.

The result is a \emph{tier} per municipality --- the resolution of the
best available local source, a property of the municipality's open data,
not of the pipeline: tier 0, census only, seed = the national
composition replicated (five municipalities, and every new one); tier
1--2 where a neighbourhood- or zone-level table exists (Brescia, Forlì,
Ravenna, Reggio; Bologna); tier 3, section level, in Parma, from the
municipal register extract. Tier 0 is the default branch and was
commissioned as such: on Modena it reproduces the no-tier behaviour to a
total-variation distance of 0.0023 --- and that cross-check between two
branches that must agree surfaced two latent classification bugs that no
single-branch test had caught (§I.1) \textbf{{[}n{]}} riferimento §6.
Measured against its null (two independent permutations), the geographic
conditional carries 2.1--2.6 times the mass that chance would move at
neighbourhood level, in every municipality that has one --- and a poor
but resolved source is worth a rich one: Brescia's nineteen countries
without sex achieve 2.08 against 2.57 for Bologna's one hundred
fifty-five with sex. For a future municipality, a truncated country ×
neighbourhood table is enough \textbf{{[}n{]}} riferimento §6. The
default branch is not a theoretical fallback: both municipalities added
during testing ran at tier 0 end to end, including Milano, where the
branch had never been exercised on an \emph{articulated} municipality
--- 171 countries × 2 sexes × 9 municipi, IPF converging in one
iteration to 5·10⁻¹⁶.

\subsubsection{I.4 Ring 3 --- sub-municipal
placement}\label{i.4-ring-3-sub-municipal-placement}

This is the ring where the pipeline goes \emph{below} the level at which
its constraints are defined. Rings 1 and 2 work where tables exist; ring
3 places individuals inside a zone, at a census section, a single year
of age, and an address --- none of which the constraint set knows about.
Everything here is therefore allocation, not estimation, and that is why
the assumptions accumulate in this section rather than elsewhere.

\emph{Section.} Within a zone, individuals are allocated to census
sections by \emph{largest remainder} rather than by multinomial draw:
each section receives the integer part of its quota, and the remaining
places go to the sections with the largest fractional remainders. The
method is deterministic and bounded --- at most one individual of error
per allocation --- and the residue that remains comes from allocating
within each demographic group separately, so that the ±1 of several
groups accumulate on a section. The measured mean absolute error per
section is 0.72--1.57 individuals against ≈ 9.6 for a multinomial draw
\textbf{{[}n{]}} riferimento §5, on sections averaging 66--175
residents, with section totals matching the census exactly (§III.3).

The zone blocks of ring 1 already carry geography for age, citizenship,
education and employment, each at the resolution its section columns
have (§I.2). Assumption (8) concerns the step \emph{below} that one ---
zone to section --- where no table conditions education, occupational
condition or migratory background at all: within a zone, those three are
spread independently of the section, given sex, age-3 and citizenship.
It is a concession of the ring, not a claim about the world, and its
cost is measured twice over. The compositional analysis that motivated
the ring found 80--98 \% of the compositional signal living \emph{below}
the zone --- which is why placement cannot stop there \textbf{{[}n{]}}
\texttt{nota\_segnale\_compositivo\_v3}; and the M-EM measures, run
against the census's own section-level migratory-background columns,
find a real residual on all eleven municipalities (median net
\textasciitilde0.022 for Italians, \textasciitilde0.018 for foreigners)
--- assumption (8) discards section-level structure that exists. The
refinement through the census EM columns is designed and queued with the
next regeneration cycle (§III.5) \textbf{{[}n{]}}
\texttt{nota\_background\_sezione\_v1}.

\emph{Single-year age.} Here, unlike education or condition, the section
\emph{does} have something to say, and the exact age is drawn in two
stages: section → five-year class → single year. The five-year class
comes from the section's own census columns (sixteen five-year counts
per sex), so the age structure of every section is respected at
five-year resolution; where a bin boundary cuts a five-year class, the
class is split under uniformity: the 0--8 bin takes the whole of 0--4
and four fifths of the 5--9 class (ages 5 to 8), the remaining fifth ---
the nine-year-olds --- going to the 9--14 bin. The single year
\emph{within} the class then follows the municipality's register
distribution by single year of age: this is assumption (9) --- below
five-year resolution, every section inherits the municipal shape. The
ex-post diagnostic measures both artefacts of the construction: the seam
left at the split classes (mean residual 2.3--5.7 individuals per
section, §III.3), and a systematic within-bin lean --- too few
nine-year-olds where the 4/5--1/5 split operates, and a young lean in
the adult bins, ten concordant signs over two cities, p ≈ 0.002
(§III.4). Both measured, both queued with the next regeneration cycle
(§III.5).

\emph{Address.} A civic number drawn uniformly among the section's
ANNCSU entries (assumption 10), with its coordinates. In production the
section supplies the address directly for ≥ 99.5 \% of individuals
\textbf{{[}m{]}} (the \texttt{{[}3e{]}} line of every generation log);
the residue is sections whose registered civic numbers cannot be joined
--- either because the section has none, or because the ones it has
carry no coordinates. In the fleet it is the first case, and it is
distributional, not one of coverage: Modena has the largest share (0.15
\%) while holding \emph{more} civic numbers than Brescia for fewer
residents. Those individuals fall back to the nearest declared level.
One case is handled apart, and declared: individuals in collective
households sit in a fictitious section and carry \emph{no} address at
all, their coordinates being the zone centroid --- an institution is not
a home, and inventing a civic number for it would manufacture exactly
the kind of false precision the regime exists to prevent. The uniformity
of the draw is the load-bearing fact of the disclosure argument (§I.7):
every address exists, and the assignment carries no information about
anyone --- which is also why the public regime can randomise the
coordinate within the section at zero analytic cost (§IV.2).

The second case is rare in the fleet and dominant outside it, and was
found by adding a municipality outside it. Address coverage is a
property of each municipality's ANNCSU \emph{georeferencing}: the
fleet's ≥ 99.5 \% reflects Emilia-Romagna's near-complete coordinate
coverage, while Mantova --- added as a test case --- has certified
17,009 accesses in ANNCSU and georeferenced 240 of them (1.4 \%)
\textbf{{[}m{]}}, \texttt{collaudo\_acquisizione\_v0.2}. The addresses
exist as text, not as coordinates, and the spatial join legitimately
finds nothing to attach. The consequences split exactly along the regime
boundary: the public regime is unaffected, since its coordinates are
drawn within the census section, which every individual has; what
degrades is the textual address of the persona and narrative regimes,
which falls back to the zone level. A municipality's ANNCSU completeness
is therefore part of its declared source profile, on the same footing as
its open-data tier (§I.3).

\subsubsection{I.5 Ring 4 --- households}\label{i.5-ring-4-households}

Household structure cannot live in the joint model:
\texttt{\textbar{}X\textbar{}} would explode and the role variable would
introduce structural zeros that make the Gibbs chain reducible. Nor can
it be ignored --- assumption (11) of the earlier releases. The
resolution, taken with the TVD criterion on the Parma microdata, is a
two-part design whose asymmetry is \emph{measured, not assumed}:
household \textbf{size} has geographic structure down to the section and
is therefore constrained by the census size distribution PF3--PF8 per
section (a hard margin: household totals match the census to the unit,
§III.3); internal \textbf{composition} has weak structure below the
quartiere and comes from a repertoire of configurations built on the AVQ
microdata (8,443 nuclei, 19,003 components \textbf{{[}m{]}}; the size-6+
tail from the Parma microdata, a mixed source, declared), conditioned
demographically \textbf{{[}n{]}} nota\_nucleo §6, riferimento §16.

\texttt{assign\_nucleo.py} writes \texttt{uid,\ id\_nucleo,\ ruolo} to a
separate file --- the population stays read-only --- with individuals in
collective households carried with an empty \texttt{id\_nucleo} rather
than dropped. These are the same individuals who carry no address in
ring 3: the fictitious sections of §I.4 and the empty
\texttt{id\_nucleo} here identify one population, those living in
collective households --- barracks, student halls, care homes, reception
centres --- for which neither a civic number nor a family structure is
meaningful. In Milano they are 9,992 people, 0.73 \% of the
municipality; in Mantova, 34. The friction of meeting the size
constraint is itself reported (which sections required truncating the
open 6+ class, which betray a collective household) \textbf{{[}n{]}}
riferimento §16.4, and the headline anomaly --- 18--25 \% of married
individuals not in a married couple --- decomposes cleanly. Three
quarters of it is structural: 17.6 \% of the married carry a role the
repertoire never pairs (13.9 \% are children living with their parents,
the rest heads of complex or non-family arrangements), and these are
incoherent by construction. The pairing itself is accurate: of those in
a pairable role, 4.6 \% of partners and 8.8 \% of household heads have a
spouse who is not married --- together six points of the 23.6 \%
measured on Modena \textbf{{[}m{]}}. The constraint set never required
that people marry in pairs, so ring 4 reveals an incoherence already
present in the ring-1 population rather than creating it (§III.3). The
rate falls where single-person households are more common, as the
mechanism predicts: 17.7 \% in Milano --- the lowest measured, below the
fleet's minimum --- against 23.0 \% in Mantova, whose age structure is
older and whose households are larger. Same-sex couples are absent
because they are absent in the donor data (0 of 4,525 partner pairs), an
inherited limit declared with the civil-union statistics registered for
the next repertoire \textbf{{[}n{]}} nota\_nucleo, assign\_nucleo
header.

Ring 4 is the youngest ring, and this report freezes it as it stands
rather than as it will be. Three declared immaturities. \emph{Names are
still assigned per individual, not per household} (§I.6): two spouses do
not share a surname, and no onomastic rule links parent to child --- the
rendering layer treats each \texttt{uid} alone, and household-aware
naming is designed but not built. \emph{The address is per individual}
(assumption 11's residue, §III.4): spouses can carry different civic
numbers, and household-level address assignment is the stated
prerequisite of any building-level work.And \emph{the assembly still
produces rare demographic oddities at the margins} --- a handful of
married fifteen-year-olds have been observed: an artefact of the
assembly's age conventions, not a demographic claim, since the register
table itself has no married cell below sixteen (§I.2, the sparsity that
encodes legal ages), so the population's own source excludes what ring 4
produces. Rare enough to survive the current diagnostics, declared here,
on the list for the next repertoire iteration. The 17-to-24 widowed of
§I.2 were an observed zero the constraints enforce; the married
fifteen-year-old is the same kind of cell one ring further out, where no
constraint yet reaches.

\subsubsection{I.6 Derived layers --- no new information, declared as
such}\label{i.6-derived-layers-no-new-information-declared-as-such}

Everything else an individual can carry --- the detailed education title
(«diploma di istituto tecnico industriale» rather than «diploma»,
«laurea magistrale in ingegneria civile» rather than «tertiary»), the
sector × position pair, first name and surname, the rendered biography
--- is a deterministic function of the \texttt{uid} and the attributes
already generated:

{\def\LTcaptype{none} 
\begin{longtable}[]{@{}
  >{\raggedright\arraybackslash}p{(\linewidth - 6\tabcolsep) * \real{0.2500}}
  >{\raggedright\arraybackslash}p{(\linewidth - 6\tabcolsep) * \real{0.2500}}
  >{\raggedright\arraybackslash}p{(\linewidth - 6\tabcolsep) * \real{0.2500}}
  >{\raggedright\arraybackslash}p{(\linewidth - 6\tabcolsep) * \real{0.2500}}@{}}
\toprule\noalign{}
\begin{minipage}[b]{\linewidth}\raggedright
derived
\end{minipage} & \begin{minipage}[b]{\linewidth}\raggedright
module
\end{minipage} & \begin{minipage}[b]{\linewidth}\raggedright
conditioned on
\end{minipage} & \begin{minipage}[b]{\linewidth}\raggedright
source
\end{minipage} \\
\midrule\noalign{}
\endhead
\bottomrule\noalign{}
\endlastfoot
first name, surname & \texttt{gsp.nomi} & sex, background, parents'
origin, country & municipal registers, per-country repertoires \\
detailed education title & \texttt{gsp.istruzione} & education, sex,
cohort & census 2011 title tree, ordered by CLAIST \\
sector × position & \texttt{gsp.lavoro} & sex, municipality & census
2011, jointly drawn (§I.1) \\
\end{longtable}
}

On the information plane they add nothing; on the readability plane they
change everything, and that is precisely why they demand care:
\textbf{apparent diversity grows while real diversity does not}. Two
individuals with the same demographic profile and the same donor
signature are identical where it counts and now merely look different
--- for a simulation use this is a worsening, since it hides that two
agents are not independent evidence \textbf{{[}n{]}} piano §3.1. And
each derivation carries its declared limits with the value, because an
imputed variable without them is an invention with the look of a datum:
the title tree has no doctoral entries, so \texttt{post\_laurea} renders
as master's degrees only; the sector is not conditioned on the title ---
the census does not publish that cross at municipal level --- so roughly
one employed card in five carries a sector × title pairing that no
conditioning produced, plausible-looking but unearned; whoever shows a
card must know it, and must not mistake that oddity for a defect of the
demographic model, which is verified \textbf{{[}n{]}} biografia §3.3.
The operational consequence: derived attributes are never stored in any
file; they exist only in the regimes generated on demand (§I.7), each on
a separate deterministic channel, so that fixing the title concordance
does not reshuffle the names \textbf{{[}n{]}} piano §5.

The onomastic layer is the clearest case, and in v1.0 the roughest: it
exists to make records readable, not to represent Italian naming.
Italian first names and surnames come from two municipal repertoires
(Florence for surnames, Modena for first names) applied to every
municipality --- a Ravenna resident carries a Florentine surname ---,
and foreign names come from coarse per-area lists (Arabic, sub-Saharan,
Eastern European) that flatten differences a demographer would care
about and certainly contain errors. This is deliberate for a first
release and declared as such: the names are placeholders that behave
correctly where it matters --- they are drawn from repertoires,the
coarseness has a side effect worth keeping even after the repertoires
improve: it guarantees collision (§I.7). Improving the repertoires is a
v1.1 item; nothing downstream depends on them.

\subsubsection{I.7 Publication regimes and the disclosure
argument}\label{i.7-publication-regimes-and-the-disclosure-argument}

One function is the single point of enforcement ---
\texttt{gsp.individui}, with \texttt{esporta\_pubblico()} for the bundle
and \texttt{campione(dettaglio=...)} for samples --- and four regimes in
three tiers, on the model of the statistical institutes' public-use /
research / protected distinction \textbf{{[}n{]}} piano §4:

\begin{itemize}
\tightlist
\item
  \textbf{complete population} (protected): every attribute, exact
  address, \texttt{uid}; never leaves the generation machine in any
  form;
\item
  \textbf{pubblico} (public use) --- the only downloadable product, and
  the \emph{default}: no name, no street or civic number, coordinates
  randomised within the census section with a seed derived from the
  municipality code, no \texttt{quartiere} (one-to-one with
  \texttt{zona}), no \texttt{uid}, no donor id; the permissive export is
  an explicit act with a warning;
\item
  \textbf{persona / narrativo} (research) --- prompt material and full
  narrative records, generated on demand, capped at
  \texttt{NMAX\ =\ 100} individuals per call \textbf{{[}m{]}}: beyond
  that threshold the function refuses, with the reason in the error
  itself --- one is producing a dataset, not a sample. The bound is not
  technical, and raising it is a decision recorded in the code rather
  than an argument passed at the call site; a sample of dozens is an act
  of citation, a file of a hundred thousand is an archive, and the cap
  encodes that difference.
\end{itemize}

The argument for the public release runs in three levels, of which the
first two stand alone \textbf{{[}n{]}} fonti §10, piano §1--3:

\emph{First, there is no personal datum to protect.} The population is
simulated from published aggregates, not anonymised from individual
records. The two are legally distinct: anonymisation must demonstrate
that a link to a person has been broken; simulation never had one. There
is no re-identification risk by construction, because there is no one to
re-identify.

\emph{Second, the one real thing in a record is the donated AVQ vector}
--- the 23 responses of an actual respondent, already perturbed at the
source by ISTAT's public-use release, and replicated tens of times ---
mean donor reuse 25--84× depending on the municipality (§III.3b) --- so
that no combination is unique to a synthetic individual. The protection
is in the data, not in a clause.

\emph{Third, the address carries no information.} The civic assignment
is arbitrary within the section, so removing it, or randomising the
coordinate within the section, loses nothing analytic --- which is
exactly what the public regime does, and why the file is
self-protecting: ``the point is random within the section'' admits no
reply (§IV.2). The test municipalities gave this level an unintended
stress test: Mantova's addresses are almost entirely non-georeferenced
(§I.4), so its public bundle is built from section geometry alone ---
and is indistinguishable in kind from the others, because the public
regime never used the civic number to begin with.

\emph{A fourth level, measured rather than structural: a name identifies
no one.} Names are drawn from repertoires --- 301 first names and 638
surnames in the released configuration --- so collision is not a
residual risk but the normal case. On Castenaso, the smallest
municipality, 16,357 individuals carry 3,024 distinct full names:
\textbf{94.4 \% of the population shares its full name with at least one
other person}, and the most frequent combinations recur seventeen times
each \textbf{{[}m{]}}. The 5.6 \% whose full name is unique are unique
only within their municipality, against repertoires that are the same in
all eleven. A name in these records is a label for reading, not a key
for finding.

What remains is a risk of \emph{interpretation}, not of data: a record
that reads «Maria Bruni, 45, laurea magistrale in medicina, dipendente
nella sanità, Cittadella» reads like a person --- although the surname
comes from a Florentine repertoire (§I.6), which is precisely the point:
it individuates no one, although every component is either
aggregate-derived, donated-and-replicated, or generated-and- collident.
The viewer answers with the banner on every card; the report answers
with the table ``what, in a record, is actually real'' (reproduced in
Appendix ‹A/F›); the narrative regime answers with the cap and the
on-demand generation.

\begin{center}\rule{0.5\linewidth}{0.5pt}\end{center}

\newpage

\section{Part II --- Sources and their
certification}\label{part-ii-sources-and-their-certification}

\subsection{Version 1 --- §II.1--II.5 (26 August
2026)}\label{version-1-ii.1ii.5-26-august-2026}

\begin{quote}
Table II.2a, cited throughout this part, is generated from the registry
by \texttt{scripts/diagnostica/tabella\_fonti.py}; it is generated from
the registry by tabella\_fonti.py and travels with the repository
\end{quote}

\begin{center}\rule{0.5\linewidth}{0.5pt}\end{center}

\subsubsection{II.1 The registry as a
method}\label{ii.1-the-registry-as-a-method}

Every external input to the pipeline is an entry in
\texttt{fonti/registro.yaml}. There are thirty-nine at the release tag
\textbf{{[}m{]}}. An entry records what the file \emph{is} (issuing
body, title, URL, date of access, licence and attribution string), what
it \emph{contains} (universe, unit of observation, temporal reference),
how it is \emph{stored} (in git, on the local disk, or only as a
fingerprint with a remote location), which code \emph{reads} it, and ---
the two fields that carry judgement --- what it may be used for
(\texttt{usabile\_per}) and what it must not be used for
(\texttt{non\_usabile\_per}).

Both take values from a controlled vocabulary --- sixty-one positive
tags, sixty negative, at the release tag \textbf{{[}m{]}} --- and the
near-symmetry is itself the finding: in this registry the
\emph{negative} affordances are declared as systematically as the
positive ones, where common practice records only what a source is for
and lets the misuses be discovered by whoever commits them. The negative
field is where the reasons live.
\texttt{confronto\_diretto\_con\_anagrafe\_stesso\_anno} on every census
table says that the population register at 1 January of year N and the
permanent census at 31 December of year N−1 describe the same instant
and must not be compared as if they were a year apart;
\texttt{paese\_di\_dettaglio} on a municipal table says that it gives
the area of citizenship but not the country;
\texttt{conteggi\_di\_popolazione} and \texttt{residenti} on the
civic-address register say that a civic number is a place, not a
household and not a count of anyone. These negative affordances are
declared at the source, once, and the normalisers enforce them.

Three rules follow from treating the registry as the authority rather
than as documentation.

\emph{The raw file is immutable; the normaliser is versioned code.}
Nothing is ever edited in a downloaded file. Every transformation ---
decoding, reshaping, filtering of aggregate rows --- is a function in
\texttt{gsp.fonti} with a name recorded in the entry, so that the same
raw file and the same commit produce the same normalised table. Where a
file must be replaced, it is re-fetched and re-fingerprinted, never
patched: Ravenna's citizenship table, flagged stale in the working
notes, was re-fetched during the licence-certification round (access
date 2026-08-11) and its entry updated \textbf{{[}m{]}}.

\emph{Storage level is declared, and the fingerprint is the invariant.}
Small redistributable files (under a few megabytes: code lists, track
records, the two Florence surname tables) live in git. Large public
files (SDMX extracts, census sections, ANNCSU, AVQ microdata) live on
disk and are re-fetched by the acquisition scripts; the registry holds
their SHA-256. Files that may not be redistributed (the Parma municipal
microdata) are also fingerprinted, so that a reader who obtains them
through the same channel can verify they have the same file.
\texttt{python\ -m\ gsp.fonti\ -\/-verifica} checks every fingerprint
against the disk; on a fresh clone most entries report \emph{solo
impronta} --- the hash is known, the file is not present --- and this is
the normal state of a clone, not an error. A state that always fails is
a state nobody looks at; the distinction between \emph{missing} and
\emph{different} is what keeps the check alive \textbf{{[}n{]}}
\texttt{fonti\_e\_pacchetto\_v8} §1. A fresh clone is not merely a
degraded state but a diagnostic in its own right: it verifies the
\emph{declarations}, not the files, and during testing it caught a
source whose \texttt{archiviazione} field said \texttt{git} while its
own licence note declared the opposite --- an inconsistency invisible on
a machine where the file happens to be present.

\emph{Artefacts do not contain their generation time.} A derived file
that embedded a timestamp would hash differently at every run even when
its numbers were identical, and the registry would lose the one
comparison that matters --- \emph{regenerated equal} against
\emph{regenerated different}. This is why derived sources
(\texttt{avq\_medie\_nazionali}, \texttt{repertorio\_nuclei\_v1}) are
registered like any other, with a \texttt{derivato\_da} pointing to
their inputs, and why the same rule governs populations and bundles
(§III.1).

Two commands audit the whole. \texttt{-\/-copertura} verifies that every
municipality in the registry has every source it needs, by instance;
\texttt{-\/-pubblico} lists what may be redistributed and under which
licence, and what may not, with the reason. At the release tag both pass
\textbf{{[}m{]}}: no blocking entry; every licence resolved against its
portal of origin --- one declared as \emph{presumed} (Forlì, whose
portal states no licence of its own and inherits the site-wide
CC-BY-4.0), a presumption recorded as such in the entry rather than
silently upgraded; and two entries whose terms explicitly forbid
redistribution (the EU-SILC public-use files), registered for
exploration only (§II.5).

\subsubsection{II.2 Families of sources and their legal
condition}\label{ii.2-families-of-sources-and-their-legal-condition}

Table II.2a lists the thirty-nine entries with issuing body, universe,
temporal reference, licence, storage level, consuming code and the ring
(or rings) each one feeds. Here we describe them by family, and in
particular the legal ground on which each stands, because this --- not
the method --- is what a reader deciding whether to reuse the
populations needs to know.

\textbf{ISTAT dissemination tables via SDMX (eleven tables, twelve
municipalities).} CC-BY-4.0. Ten are from the permanent census at 31
December of year N−1 and one --- sex × age × marital status --- from the
population register at 1 January of year N; the two series are offset by
one year because they describe the same seven instants. The register
table is the only \emph{hard} constraint of ring 1; the census tables
are \emph{soft}. The service allows about four queries per minute, and
violations have led to multi-day IP blocks; the acquisition scripts
respect the rate, and the catalogue of dataflows and the SDMX structure
files are cached locally (120 MB) but not yet registered --- they are an
index of what exists, not data \textbf{{[}n{]}}
\texttt{fonti\_e\_pacchetto\_v8} §3, §13.4.

\textbf{Census sections (Basi Territoriali).} CC-BY-4.0. Geometry of the
2021 census, counts of the 2023 permanent census: 135,725 sections in
the three regions covered, 18.35 M residents, 10,872 sections with no
residents \textbf{{[}n{]}}. The 138-column track record is registered as
a source in its own right, because it is the file that gives meaning to
the other.

\textbf{ANNCSU, the national register of civic addresses.} This is the
source with the strongest legal position in the registry: a
\emph{high-value dataset} under Regulation (EU) 2023/138, whose
technical specifications were adopted after a favourable opinion of the
Italian data-protection authority (12 December 2024). The opening of
these data has been examined by the competent authority; what the
pipeline does with them --- draw one civic number uniformly within a
section --- adds nothing to what the source already discloses
\textbf{{[}n{]}} \texttt{fonti\_e\_pacchetto\_v8} §3.

\textbf{AVQ public-use microdata (2022--2024).} The licence is inferred,
not read: the survey page declares none of its own and the site's
general CC-BY-4.0 applies. Inside each annual archive the
\texttt{Leggimi} qualifies the files as \emph{ad uso pubblico}, the
least restrictive class, and neither it nor the four accompanying PDFs
impose conditions of use or clauses on re-identification
\textbf{{[}n{]}} \texttt{fonti\_e\_pacchetto\_v8} §3. What matters for
disclosure is that the only real vector inside a synthetic record ---
the donated AVQ vector --- is protected at the source, by ISTAT, before
it reaches the pipeline; the hot-deck copies a whole public-use record
and nothing more. The weight \texttt{COEFIN} carries four implicit
decimals, recorded in the entry as \texttt{scala\_peso}.

\textbf{Municipal open data (six portals) and one register extract.}
Five municipalities publish residents by citizenship at sub-municipal
level (Bologna by statistical zone; Brescia and Forlì by neighbourhood;
Ravenna by area; Reggio nell'Emilia by circoscrizione), and Modena
publishes the stock of first names by sex; these feed the
country-of-citizenship tiers of ring 2 and, for Bologna, the
zone--neighbourhood hierarchy of ring 3. Their licences were resolved
portal by portal during certification \textbf{{[}m{]}}: CC-BY-4.0
(Bologna, Brescia, Modena), CC-BY (Reggio), public domain (Ravenna), and
one declared presumption (Forlì, §II.1). Two vintages are declared as
limits: Forlì's table refers to 2021, and Reggio's to 2013 --- thirteen
years older than everything else, the only sub-municipal source
available there; the tiers use these tables for the \emph{composition}
of the foreign population while every total comes from the census
margin, which bounds but does not cancel the staleness (§II.4). One
source in this family is different in kind: \textbf{the Parma register
extract}, a full extract of the municipal population register (one row
per resident, 202,111 rows, reference 1 January 2025), published by the
Comune di Parma as open data under CC-BY-4.0 with its codebook. It is
registered, fingerprinted, and \emph{not mirrored here} --- a reader
obtains it from the source and verifies the fingerprint; it serves in
production as the IPF margin and the section-level country conditional
(tier 3), and as the external validation source for ring 4 (§I.5,
§III.3). What this repository redistributes are aggregate statistics and
a synthetic population, never the extract.

\textbf{ISTAT census 2011 (two tables) and the CLAIST classification.}
CC-BY-4.0. They feed the derived layers only:
\texttt{cens2011\_caratt\_attl} (occupied residents by fourteen
dimensions, 94 MB of codes) gives the joint sector × position
distribution behind \texttt{gsp.lavoro};
\texttt{cens2011\_titolo\_studio} gives 458 education titles in a tree
declared by the source (399 leaves); CLAIST 2026 is the current
vocabulary and the historicised ordering that places a title in its
cohort. None adds information beyond rings 1--4; they render it
\textbf{{[}n{]}} \texttt{fonti\_e\_pacchetto\_v8} §6, §9.

\textbf{Onomastic repertoires (seven).} Surnames of Florence 2012 and
2013 (CC-BY-4.0, DCAT metadata, \texttt{notPlanned}; 375,371 residents,
66,353 distinct surnames, 51 \% hapax; the 2012 table is used only as a
stability check, r = 0.9985); first names of Modena (CC-BY-4.0 via the
regional portal); the CC0 \texttt{popular\_names} lists (2,278 surnames
from 75 countries, 2,370 first names from 106, with romanisation and
sex); two CC-BY-SA MediaWiki category lists for Maghrebi and Yoruba
surnames (65 and 107 entries --- lists, not distributions). Names are a
derived layer: plausible by construction and therefore collident, as the
disclosure argument requires (§I.7) \textbf{{[}n{]}}
\texttt{fonti\_e\_pacchetto\_v8} §3.

\textbf{Civil unions 2023 and the household repertoire.} ISTAT's
exhaustive survey of civil unions (CC-BY-4.0) and the derived repertoire
of household configurations (\texttt{repertorio\_nuclei\_v1}, built from
AVQ, CC-BY-4.0 by inheritance) feed ring 4.

\textbf{EU-SILC public-use files (two entries).} Registered for
exploration only. Access is public on acceptance of a disclaimer that is
\emph{not} an open licence and does not allow redistribution; the text
is kept beside the data. They feed no ring; they document a decision not
yet taken (income and living conditions, §III.5).

\textbf{Derived sources of the project itself.} National means of the
AVQ battery (\texttt{avq\_medie\_nazionali}, used only by the viewer)
and the household repertoire are registered as sources with
\texttt{derivato\_da}, so that the same verification applies to what the
project produces as to what it downloads.

\subsubsection{II.3 Eighteen normalisers, and what running them
found}\label{ii.3-eighteen-normalisers-and-what-running-them-found}

The registry names a normaliser per entry --- eighteen distinct ones at
the release tag \textbf{{[}m{]}}, one per source \emph{form} (SDMX
extracts, section workbooks, microdata, matrices, codebooks, onomastic
lists, \ldots) rather than per source. The value of writing them was
less the uniform output than the anomalies they surfaced, none of which
would have been found by reading the documentation. A selection, one per
family where there is one worth reporting \textbf{{[}n{]}}
\texttt{fonti\_e\_pacchetto\_v8} §3--4, §14;
\texttt{nota\_nucleo\_familiare\_v3} §2:

\begin{itemize}
\tightlist
\item
  In \texttt{DICA\_CARATT\_ATTL} the code for \emph{total} is not
  uniform across dimensions --- \texttt{ALL} for occupation and
  citizenship, \texttt{99} for profile and education, \texttt{0010} for
  ATECO, \texttt{9} for sex and regime, \texttt{TOTAL} for duration. A
  reader that assumes one of them filters everything away and gets zero
  rows, which looks like a data problem and is a filter problem.
\item
  The same ISTAT archive (\texttt{DICA\_TITSTUDIO}) contains codebooks
  in UTF-16 with BOM and the data file in UTF-8 without: reading the
  first successfully fails on the second.
\item
  Modena's CKAN catalogue says \emph{names given to newborns,
  2012--2022}; the data are the \emph{stock} of residents (1,390
  Antonios in 2015 are not newborns) and run to 2024; the two
  sex-specific files carry CKAN timestamps two and a half years apart,
  decoded from the \texttt{hash} field; the WFS adds four service
  columns, one of which is the centroid of Modena repeated 650 times.
  The catalogue was wrong on what, when, and how many columns --- and
  the data were good: the lesson is to certify the file, not its
  description.
\item
  Parma's \texttt{Ncomp} showed impossible values --- 319, 111, 110,
  108, 40 --- each appearing exactly that many times: the signature of a
  collective household (319 residents of an institution all carry
  \texttt{Ncomp\ =\ 319}), not a defect. Conditioning on
  \texttt{Tipores\ =\ 1} brings the maximum from 319 to 12.
\item
  Parma's codebook, supplied with the data, declares for \texttt{Relpar}
  a 29-code classification that the data contradict: the file uses
  eleven contiguous codes and nothing above; under the declared labels
  the code for \emph{great-grandchild} has a median age of 80 and the
  code for \emph{sibling} a median age of 13. The codebook describes the
  current extended classification; the extract carries a shorter,
  inherited one. The mapping actually used was inferred from the
  demographic profile, is marked \emph{inferita, non letta} in the note,
  and is solid for the three codes that cover 90 \% of the records and
  conjectural for the rest.
\item
  In the municipal tables, the sum of observations is two to eight times
  the population because aggregate rows are included: the figure is a
  signature for recognising the file, not a count.
\end{itemize}

One convention deserves a place next to the anomalies, because at first
sight it looks like one: the population register is dated 1 January N
and the census 31 December N−1, which reads as a year apart and is the
same instant. Every entry declares \texttt{riferimento\_temporale} and
\texttt{anno\_usato}, and the tag
\texttt{non\_usabile\_per:\ confronto\_diretto\_con\_anagrafe\_stesso\_anno}
of §II.1 forbids the comparison that the dates invite --- a prohibition
that turned out to be more than prudence: on sex × single year of age
the two sources agree cell for cell (§III.3).

The general lesson, which the project turned into a principle, is that
\emph{the meaning of a quantity belongs to whoever produces it}, and a
generic structure does not guess it: \texttt{modalita} on Reggio's
matrix (26 nationalities or 104 long rows), \texttt{n\_misurato} on
codebooks (rows or fields), the sum of weights on national means (82 or
23) --- each time the normaliser's diagnostic wins over the assumption.
It is the registry's face of a rule the reader has already met twice:
the declared hierarchy beats the deduced one, as with the census title
tree whose parent column crosses the code groups (§I.6), and the
two-branch cross-check that caught what no single-branch test could
(§I.1, §I.3) \textbf{{[}n{]}} \texttt{fonti\_e\_pacchetto\_v8} §12.

\subsubsection{II.4 From source to ring to
column}\label{ii.4-from-source-to-ring-to-column}

Every column of a population file can be traced to the registry entries
that determine it, through the ring that computed it. The table below
runs the three legs together, by column group; the per-entry detail
(universe, vintage, licence) is in Table II.2a, and the mechanism in
Part I. Columns are named as in the released Parquet where they survive
the public regime, and marked ° where they exist only in the
\texttt{persona}/\texttt{narrativo} regimes (§I.7).

{\def\LTcaptype{none} 
\begin{longtable}[]{@{}
  >{\raggedright\arraybackslash}p{(\linewidth - 4\tabcolsep) * \real{0.3333}}
  >{\raggedright\arraybackslash}p{(\linewidth - 4\tabcolsep) * \real{0.3333}}
  >{\raggedright\arraybackslash}p{(\linewidth - 4\tabcolsep) * \real{0.3333}}@{}}
\toprule\noalign{}
\begin{minipage}[b]{\linewidth}\raggedright
columns
\end{minipage} & \begin{minipage}[b]{\linewidth}\raggedright
ring
\end{minipage} & \begin{minipage}[b]{\linewidth}\raggedright
determined by (registry ids)
\end{minipage} \\
\midrule\noalign{}
\endhead
\bottomrule\noalign{}
\endlastfoot
\texttt{zona}, \texttt{sesso}, \texttt{eta}, \texttt{stato\_civile},
\texttt{cittadinanza}, \texttt{istruzione}, \texttt{condizione},
\texttt{background}, \texttt{origine\_genitori} & 1 & the eleven SDMX
tables: \texttt{istat\_anag\_sesso\_eta\_statociv} (hard margin, 1 Jan),
\texttt{istat\_cens\_*} (soft, 31 Dec N−1); zone tables from
\texttt{istat\_cens\_sesso\_eta\_cittadinanza} and, for Bologna,
\texttt{bologna\_cittadinanza\_zone} (zone--quartiere hierarchy) \\
the 21 AVQ variables (\texttt{SALUTE}, \texttt{AMBIENTE},
\texttt{FIDUCIA}, \texttt{PUNTIFI1–13}, \texttt{FIDMED},
\texttt{FIDINF}, \texttt{BMI}, \ldots) and the donor signature & 2 &
\texttt{avq\_microdati} (pools: Emilia-Romagna 4,629 / Lombardia 8,111),
decoded via \texttt{avq\_tracciato\_2024} \\
\texttt{paese}, \texttt{area} (country of citizenship) & 2 &
\texttt{istat\_cens\_stranieri\_paesi} (municipal margin, tier 0)
refined by the sub-municipal sources where they exist:
\texttt{bologna\_cittadinanza\_zone} (tier 2),
\texttt{brescia\_cittadinanza\_quartieri},
\texttt{forli\_cittadinanza\_quartieri},
\texttt{ravenna\_cittadinanza\_aree},
\texttt{reggio\_cittadinanza\_circoscrizioni} (tier 1--2),
\texttt{parma\_microdati\_residenti} (tier 3, section level) \\
\texttt{sezione} & 3 & \texttt{istat\_sezioni\_2023} (counts), decoded
via \texttt{istat\_sezioni\_2023\_tracciato} \\
\texttt{eta\_anni} & 3 & \texttt{istat\_sezioni\_2023} (single-year
section columns \texttt{P30+k}/\texttt{P67+k}) \\
\texttt{lon}, \texttt{lat} (public regime: random within section);
\texttt{via}°, \texttt{civico}° & 3 & \texttt{anncsu\_indirizzario}
joined to section geometries \texttt{istat\_sezioni\_shp} \\
\texttt{id\_nucleo}, \texttt{ruolo} (separate file
\texttt{nuclei\_\{comune\}.csv}) & 4 & \texttt{repertorio\_nuclei\_v1}
(configurations, derived from \texttt{avq\_microdati}), size margins
PF3--PF8 from \texttt{istat\_sezioni\_2023}, same-sex couple rates
examined against \texttt{istat\_unioni\_civili\_2023} \\
\texttt{settore}, \texttt{posizione}° & D &
\texttt{cens2011\_caratt\_attl} (joint sector × position of the
occupied, 2011 structure) \\
\texttt{titolo\_studio}° (detailed title) & D &
\texttt{cens2011\_titolo\_studio} (title tree) ordered and currentised
by \texttt{claist\_2026} \\
\texttt{nome}°, \texttt{cognome}° & D & \texttt{firenze\_cognomi\_2013}
(Italian surnames; 2012 edition as stability check only),
\texttt{modena\_nomi\_residenti} (first names by sex),
\texttt{popular\_names\_nomi}/\texttt{\_cognomi} (foreign),
\texttt{cognomi\_wiki\_MA\_ARAB}/\texttt{\_NG\_YORUBA} (specific
communities) \\
national reference ticks (viewer only, not a population column) & W &
\texttt{avq\_medie\_nazionali}, derived from \texttt{avq\_microdati} \\
\end{longtable}
}

Three remarks the table compresses.

\emph{The tier structure of the country of citizenship is visible as a
column's provenance changing by municipality.} The same two columns
(\texttt{paese}, \texttt{area}) are determined in Ferrara, Modena,
Rimini, Piacenza and Castenaso by the municipal census margin alone
(tier 0), in Brescia, Reggio, Ravenna and Forlì by a sub-municipal
open-data table over the census margin (tier 1), in Bologna by the zone
table that also fixes the zone--quartiere hierarchy (tier 2), and in
Parma by the municipal register extract at section level (tier 3). The
registry records this as five sources carrying the same
\texttt{usabile\_per} tags at different
\texttt{geografia\_paese\_tier*}, and the per- municipality tier is
printed by the generation log (§III.2 table). Their vintages differ ---
Reggio's table is from 2013, thirteen years older than the rest, the
only sub-municipal source available for that municipality; the tiers use
these tables for the \emph{composition} of the foreign population, while
every total comes from the census margin, which bounds but does not
cancel the staleness. Declared here and in the entry.

\emph{Ring-4 columns live in a separate file by design.}
\texttt{id\_nucleo} and \texttt{ruolo} are written to
\texttt{nuclei\_\{comune\}.csv}, never into the population file: the
population is read-only once generated (§I.1), and consumers join on
\texttt{uid}. The repertoire that shapes the nuclei is itself a
registered derived source with \texttt{derivato\_da:\ avq\_microdati},
so its provenance chain closes on the same microdata as ring 2.

\emph{Derived columns cite structure, not information.} The 2011 census
tables and CLAIST determine how a ring-1/2 state is \emph{rendered}
(which detailed title, which sector) but add no information beyond it;
the onomastic repertoires determine names that are plausible and
therefore collident. The public regime drops precisely the columns of
this group (°): what individuates is derived, and what is derived can be
omitted without analytic loss.

\subsubsection{II.5 What is registered and not used, and what is used
and not
registered}\label{ii.5-what-is-registered-and-not-used-and-what-is-used-and-not-registered}

Registered and not used: \texttt{istat\_cens\_posizione\_famiglia}
(position in the household from the census), present because it was
examined for ring 4 and rejected in favour of the section-level
household counts (§I.5); and the two EU-SILC entries, registered for
exploration. Used in production and registered: everything else. Every
licence resolved (§II.2). Deliberately outside the registry: sources
consulted and not used (the project keeps a separate consultation list
in the notes); the experimental K10C constraint set; San Vito dei
Normanni, which has no sub-municipal articulation and is not in
production.

\begin{center}\rule{0.5\linewidth}{0.5pt}\end{center}

\subsubsection{Open items for Part II}\label{open-items-for-part-ii}

\begin{enumerate}
\def\labelenumi{\arabic{enumi}.}
\tightlist
\item
  Two external inputs are used and not registered, deliberately for
  v1.0: the SDMX dataflow catalogue and the DSD structure cache. Both
  are metadata rather than data --- an index of what exists, and the
  codelists that decode it --- so no number in a population comes from
  them; but the second is load-bearing for reproducibility, since
  without the codelists the decoded tables cannot be rebuilt. A
  fingerprint is the wrong form (some 120 MB, regenerated whenever ISTAT
  updates a structure, with no effect on any population); a manifest
  declaring the endpoint, the acquisition date and the dataflows present
  with their versions is the planned one, in v1.1. The dependency is
  narrower than it looks: the decoded tables carry the labels inline
  (\texttt{*\_label} columns), so a reader who has them does not need
  the cache --- only someone rebuilding from the raw SDMX responses
  does.
\end{enumerate}

\newpage

\section{Part III --- Reproducibility and quality
report}\label{part-iii-reproducibility-and-quality-report}

\subsection{Version 1 --- §III.1--III.6 (26 August
2026)}\label{version-1-iii.1iii.6-26-august-2026}

\begin{quote}
Where the numbers come from: everything marked \textbf{{[}m{]}} in this
part was measured on the run of 19 August 2026 at the release tag. The
per-municipality logs are in
\texttt{note/misure/rilancio\_report\_v1.0/} and the diagnostics in
\texttt{note/misure/diagnostica\_report\_v1.0/}; both travel with the
repository.
\end{quote}

\begin{center}\rule{0.5\linewidth}{0.5pt}\end{center}

\subsubsection{III.1 Environment, determinism, and what ``reproducible''
means
here}\label{iii.1-environment-determinism-and-what-reproducible-means-here}

The claim this report makes is narrow and testable: given the source
files registered in \texttt{fonti/registro.yaml}, a tagged commit of the
GSP repository, and the solver repository at the pinned commit of the
binding table (front matter), the pipeline regenerates every population,
ring by ring, to the byte. It does not claim that the populations are
\emph{right} --- that is the subject of §III.3--III.4 --- but that they
are \emph{the same}, whoever runs the code, whenever. Two boundaries of
the claim are declared at the start. The solver dependency is resolved
by filesystem discovery, not by version pin: the reproducer must check
out the stated commit by hand (front matter). And the chain is
reproducible from self-acquirable public sources only up to ring 1:
rings 2--4 require the AVQ public-use microdata and their derivatives,
obtainable by anyone from ISTAT's mIcro.STAT channel, but through a
manual request that no script can perform.

\textbf{Reference environment.} All results in this report were produced
on a single workstation running Ubuntu 24.04 under WSL2 (AMD Ryzen AI 9
HX 375, 64 GB RAM; the GPU is not used by the pipeline), Python 3.11 in
a conda environment, with the \texttt{gsp} package installed in editable
mode from the repository root (\texttt{pip\ install\ -e\ .}). Ring 1
uses Numba for the sparse constraint kernel; rings 2--4 are pandas and
NumPy. No step requires network access once the registered sources are
on disk. Exact package versions are frozen in
\texttt{note/misure/rilancio\_report\_v1.0/requirements-report.txt},
committed with the run logs \textbf{{[}m{]}}.

\textbf{Where randomness enters, and how it is pinned.} Every ring draws
random numbers, and each draw is seeded explicitly:

\begin{itemize}
\tightlist
\item
  Ring 1 (\texttt{fit\_cs.py}): the exact solver is deterministic by
  construction --- no randomness enters the fit. Where PCD is used, its
  warm start has its own fixed seed (123). The final \emph{sampling} of
  N individuals from the fitted distribution uses a fixed seed (42).
\item
  Ring 2 (\texttt{assign\_avq.py}): one generator per run,
  \texttt{-\/-seed} with a declared default of 42, drawing the donor
  within each conditioning cell.
\item
  Ring 3 (\texttt{enrich.py}): a single generator seeded with a declared
  constant (\texttt{-\/-seed\ 42} by default) for section, country,
  single-year age and address.
\item
  Ring 4 (\texttt{assign\_nucleo.py}): seeded as
  \texttt{20260810\ +\ int(municipality\ code)}, so that running
  municipalities together or separately gives the same result; the
  effective seed is written into the diagnostic JSON.
\item
  Public regime (\texttt{gsp.individui.esporta\_pubblico}, §I.7):
  coordinate randomisation seeded from the municipality code.
\end{itemize}

Two seed policies therefore coexist, and their history is visible in the
code: the early rings use a fixed declared constant, the later ones
derive the seed from the municipality code. The constant is reproducible
in practice because each municipality runs in its own process --- the
regeneration test of §III.2, forty-four bit-identical files, is the
certificate --- but it is the weaker convention: it would silently
couple municipalities if the chain were ever run in one process, and it
is scheduled for alignment in the next cycle in which regeneration is
already required for other reasons. The policy new code follows is ring
4's: \emph{seeds derived from the municipality code, never shared, never
from the clock}.

\textbf{Artefacts do not embed their generation time.} A population file
that contained a timestamp would never hash the same twice. The CSV
products of rings 1--4 contain none. Two derived JSON files written by
the Animarium bundle chain (\texttt{manifest.json},
\texttt{riferimenti.json}) do carry a \texttt{generato} field; for this
reason the version binding in the front matter hashes the Parquet files
only, which are timestamp-free, and the two JSON files are compared
modulo that field. \textbf{{[}n{]}} \texttt{design\_animarium\_v13},
§15.

\subsubsection{\texorpdfstring{III.2 The regeneration test at
\texttt{report-v1.0-rc1}}{III.2 The regeneration test at report-v1.0-rc1}}\label{iii.2-the-regeneration-test-at-report-v1.0-rc1}

On 19 August 2026, with the repository at tag \texttt{report-v1.0-rc1}
and a clean working tree, all eleven municipalities were regenerated
from their constraint sets through the full chain
\texttt{cs\_build\ →\ fit\_cs\ →\ assign\_avq\ →\ enrich\ →\ assign\_nucleo}
(\texttt{scripts/rigenera.sh}, one log per municipality), and every
product of every ring was compared byte for byte (\texttt{cmp}) with the
population archived immediately before the run. The outcome:

{\def\LTcaptype{none} 
\begin{longtable}[]{@{}lllll@{}}
\toprule\noalign{}
municipality & code & level & rings 1--4 & chain time \\
\midrule\noalign{}
\endhead
\bottomrule\noalign{}
\endlastfoot
Bologna & 037006 & K9C & all identical & 395 s \\
Brescia & 017029 & K9C & all identical & 333 s \\
Parma & 034027 & K9C & all identical & 339 s \\
Reggio nell'Emilia & 035033 & K9C & all identical & 220 s \\
Forlì & 040012 & K9C & all identical & 147 s \\
Rimini & 099014 & K9C & all identical & 138 s \\
Ravenna & 039014 & K9C & all identical & 130 s \\
Modena & 036023 & K9C & all identical & 105 s \\
Piacenza & 033032 & K9C & all identical & 72 s \\
Ferrara & 038008 & K6C & all identical & 51 s \\
Castenaso & 037021 & K6C & all identical & 16 s \\
\end{longtable}
}

\textbf{{[}m{]}} Forty-four comparisons, forty-four identical files;
1,814,317 individuals in about 33 minutes of wall-clock time on the
reference machine --- of which the maximum-entropy fit is 15--55 \%
(Table III.3a), the rest being the three enrichment rings.

One qualification belongs here rather than only in §III.5: since the
tag, two patches to ring 4 have been committed, one of which shifts the
random sequence for every municipality. The claim above is about the
tagged commit --- which is what the binding table states --- and a
reader running \texttt{master} today reproduces rings 1--3 identically
and ring 4 differently (§III.5, \emph{post-tag divergence}).

The Animarium bundle was then rebuilt from the regenerated populations
(\texttt{build/build\_bundle.py\ -\/-forza}) and compared with the
archived bundle: the eleven \texttt{pop.parquet} files are
byte-identical \textbf{{[}m{]}}; \texttt{manifest.json} and
\texttt{riferimenti.json} differ only in the \texttt{generato}
timestamp. The SHA-256 of each \texttt{pop.parquet} is recorded in the
version-binding table of the front matter and is the hash against which
the public dataset on Zenodo can be checked.

Two remarks on what this test does and does not show.

It shows that the pipeline is deterministic end to end, including the
Numba-accelerated sparse solver on the largest state spaces (Bologna,
390,098 individuals, eighteen zones), and that nothing in the chain
depends on the order in which municipalities are run or on the state of
the machine. It also shows that the populations currently displayed
online are exactly those produced by the tagged code: the archive
against which the regeneration was compared is the one the viewer was
serving.

It does not show that a different machine would produce the same bytes.
Floating-point reductions in Numba are deterministic on a given build
but not across compilers or CPU vector widths; and the sampling in ring
1 depends on NumPy's generator implementation, which is stable across
versions by policy but not guaranteed. A reader who regenerates on other
hardware should expect identical populations in most cases and, where
they differ, should find differences confined to ring 1 and invisible at
the level of §III.3's quality metrics. A cross-machine regeneration ---
one municipality rebuilt on a different architecture and compared with
the archive --- would turn this paragraph from a caveat into a
measurement: byte-identity would settle it, and a discrepancy would
itself be the number to report. It is planned and not part of this
release.

Finally, the method used here --- archive, regenerate, compare byte for
byte, read the diff only where it is non-empty --- is the same one that
governed every refactoring of the pipeline (§I.1): the migration to the
\texttt{src/gsp/} package layout, the consolidation of five duplicated
municipality registries into \texttt{gsp.common}, the removal of
\texttt{sys.path} manipulation from the Animarium build scripts. In each
case a baseline was written before the change and \texttt{diff\ -r}
after, and the refactoring was accepted only on an empty diff. Three
bugs were found this way that no test had caught: a classification error
in the citizenship tiers (Czech Republic and South Africa, whose names
contain the strings of a continent and of a cardinal direction), a
permutation of zone names in Bologna, and --- during the testing
campaign of August --- a bulk edit of the registry that silently
replaced a live script name with its successor, caught because the
two-stage relay of §I.0 made the substitution visibly wrong
\textbf{{[}n{]}} \texttt{collaudo\_acquisizione\_v0.2}, finding 5;
\texttt{GSP\_popolazioni\_full\_riferimento}, §4;
\texttt{design\_animarium}, §0.1. Maintaining two code paths that must
agree is, in this sense, a permanent regression test; the day they
disagree is the day a bug is found.

\subsubsection{III.3 Ring-by-ring
quality}\label{iii.3-ring-by-ring-quality}

The regeneration test of §III.2 established that the populations are the
ones the tagged code produces. This section asks whether they are any
good, one ring at a time, and with each metric normalised against its
null (§I.1): raw error against the sampling floor, effective sample size
on the variable's own universe, household counts against the census
partition.

\paragraph{Ring 1 --- the fit}\label{ring-1-the-fit}

Before asking how good the sample is, how good is the fit. At the
release tag every municipality was solved \emph{exactly}; none required
PCD, which the solver provides for state spaces this release does not
contain (Brescia at K10C, or the 88-NIL fit of Milano registered as an
experiment, §III.5).

{\def\LTcaptype{none} 
\begin{longtable}[]{@{}lllll@{}}
\toprule\noalign{}
municipality & zones & \textbar X\textbar{} & fit MRE & time \\
\midrule\noalign{}
\endhead
\bottomrule\noalign{}
\endlastfoot
Bologna & 18 & 2,903,040 & 4.52·10⁻⁴ & 79 s \\
Brescia & 33 & 5,322,240 & 4.59·10⁻⁴ & 182 s \\
Parma & 13 & 2,096,640 & 4.86·10⁻⁴ & 85 s \\
Modena & 4 & 645,120 & 3.95·10⁻⁴ & 16 s \\
Reggio nell'Emilia & 6 & 967,680 & 2.43·10⁻⁴ & 23 s \\
Ravenna & 10 & 1,612,800 & 5.00·10⁻⁴ & 60 s \\
Rimini & 6 & 967,680 & 4.65·10⁻⁴ & 29 s \\
Ferrara & --- (K6C) & 5,376 & 2.96·10⁻⁴ & 0.18 s \\
Forlì & 12 & 1,935,360 & 5.00·10⁻⁴ & 105 s \\
Piacenza & 4 & 645,120 & 3.25·10⁻⁴ & 20 s \\
Castenaso & --- (K6C) & 5,376 & 3.40·10⁻⁴ & 0.17 s \\
\end{longtable}
}

\textbf{{[}m{]}} from the generation logs at the tag. Time follows
\textbar X\textbar{} = 161,280 × zones and not population --- Brescia,
198k residents in 33 quartieri, costs 182 s against Bologna's 79 s for
390k residents in 18 zones --- while the fit error follows neither,
staying within 2.4--5.0·10⁻⁴ across two orders of magnitude of
population: a property of the stopping criterion rather than of the
problem's difficulty.

\paragraph{Ring 1 --- the sample, against the sampling
floor}\label{ring-1-the-sample-against-the-sampling-floor}

The number above is the distance between the fitted distribution and its
constraints; what follows is a different quantity with a similar name
--- the distance between the \emph{drawn population} and the same
constraints, which is three orders of magnitude larger because it is
dominated by sampling, not by optimisation.

\texttt{verifica\_vincoli.py} checks every cell of the constraint set
against the generated population, expressed as a z-score against the
sampling floor √(α(1−α)/N) rather than as relative error --- because
relative error on small cells is a statistic without content: a cell
with expectation 1.3 individuals that is off by two is not a
five-per-cent municipality problem, and the v1 of the tool, which ranked
by relative error, produced nothing but the ranking of the smallest
cells \textbf{{[}n{]}} header of \texttt{verifica\_vincoli.py}. The
comparison runs against the \emph{published} Parquet, not the working
CSV: what is verified is what anyone can download.

Two anchor municipalities, the extremes of the state space:

\begin{itemize}
\tightlist
\item
  \textbf{Bologna} (K9C, 18 zones, 1,783 cells): no hard zero violated
  --- every cell declared impossible is empty. The largest deviations
  sit exactly where the floor predicts them: \textbar z\textbar{} = 13.4
  on a cell with expectation 2.0 (widowed men aged 15--24),
  \textbar z\textbar{} = 10.2 on expectation 6.0; among cells with
  expectation above \textasciitilde100, \textbar z\textbar{} stays below
  \textasciitilde4.1 and the bulk is under 3.
\item
  \textbf{Castenaso} (K6C, one zone, 289 cells): no hard zero violated;
  \textbar z\textbar max = 2.8, on a cell with expectation 12.2.
\end{itemize}

The full table:

{\def\LTcaptype{none} 
\begin{longtable}[]{@{}
  >{\raggedright\arraybackslash}p{(\linewidth - 16\tabcolsep) * \real{0.1111}}
  >{\raggedright\arraybackslash}p{(\linewidth - 16\tabcolsep) * \real{0.1111}}
  >{\raggedright\arraybackslash}p{(\linewidth - 16\tabcolsep) * \real{0.1111}}
  >{\raggedright\arraybackslash}p{(\linewidth - 16\tabcolsep) * \real{0.1111}}
  >{\raggedright\arraybackslash}p{(\linewidth - 16\tabcolsep) * \real{0.1111}}
  >{\raggedright\arraybackslash}p{(\linewidth - 16\tabcolsep) * \real{0.1111}}
  >{\raggedright\arraybackslash}p{(\linewidth - 16\tabcolsep) * \real{0.1111}}
  >{\raggedright\arraybackslash}p{(\linewidth - 16\tabcolsep) * \real{0.1111}}
  >{\raggedright\arraybackslash}p{(\linewidth - 16\tabcolsep) * \real{0.1111}}@{}}
\toprule\noalign{}
\begin{minipage}[b]{\linewidth}\raggedright
municipality
\end{minipage} & \begin{minipage}[b]{\linewidth}\raggedright
cells (α\textgreater0)
\end{minipage} & \begin{minipage}[b]{\linewidth}\raggedright
sample MRE
\end{minipage} & \begin{minipage}[b]{\linewidth}\raggedright
sampling floor
\end{minipage} & \begin{minipage}[b]{\linewidth}\raggedright
mean \textbar z\textbar{}
\end{minipage} & \begin{minipage}[b]{\linewidth}\raggedright
sd(z)
\end{minipage} & \begin{minipage}[b]{\linewidth}\raggedright
\% \textbar z\textbar\textgreater3
\end{minipage} & \begin{minipage}[b]{\linewidth}\raggedright
\textbar z\textbar max (exp. of cell)
\end{minipage} & \begin{minipage}[b]{\linewidth}\raggedright
hard zeros
\end{minipage} \\
\midrule\noalign{}
\endhead
\bottomrule\noalign{}
\endlastfoot
Bologna & 1,751 & 7.05 \% & 5.29 \% & 0.84 & 1.391 & 0.63 \% & 36.0
(1.0) & none violated \\
Brescia & 2,949 & 7.82 \% & 9.52 \% & 0.80 & 1.012 & 0.47 \% & 7.1 (2.0)
& none violated \\
Parma & 1,349 & 4.99 \% & 6.10 \% & 0.80 & 1.014 & 0.44 \% & 3.7 (9.0) &
none violated \\
Modena & 629 & 3.95 \% & 5.07 \% & 0.82 & 1.018 & 0.16 \% & 3.1 (5,577)
& none violated \\
Reggio nell'Emilia & 629 & 4.74 \% & 5.32 \% & 0.78 & 0.979 & 0.32 \% &
3.8 (27.0) & none violated \\
Ravenna & 1,108 & 7.99 \% & 7.50 \% & 0.85 & 1.164 & 0.63 \% & 17.0
(1.0) & none violated \\
Rimini & 789 & 4.55 \% & 5.41 \% & 0.82 & 1.014 & 0.51 \% & 3.3 (13.0) &
none violated \\
Ferrara & 261 & 9.33 \% & 7.95 \% & 0.80 & 1.098 & 1.15 \% & 7.1 (2.0) &
none violated \\
Forlì & 1,989 & 11.04 \% & 3,189 \% & 0.85 & 1.092 & 0.80 \% & 10.0
(1.0) & none violated \\
Piacenza & 629 & 5.48 \% & 6.70 \% & 0.91 & 1.118 & 0.48 \% & 4.1
(165.7) & none violated \\
Castenaso & 257 & 15.81 \% & 23.68 \% & 0.80 & 1.000 & 0.00 \% & 2.9
(169.9) & none violated \\
\end{longtable}
}

The \emph{sampling floor} is what a perfectly fitted distribution would
still show once N individuals are drawn from it: the standard deviation
of the relative error, mean over cells of √((1−α)/(αN)), computed by
\texttt{verifica\_vincoli.py} \textbf{{[}m{]}} from the formula of
\texttt{riferimento} §10. It is not directly comparable with the sample
MRE, which is a mean absolute error: the two differ by √(2/π) ≈ 0.798,
and both columns print what the instrument produces rather than a
corrected figure (§III.5, point 8). Read with that factor, the observed
error sits some 10 \% above its floor everywhere --- sampling explains
almost all of the deviation, and what remains is what the z-columns
localise, cell by cell.

Reading the table is an exercise in the project's own rule: never
compare a raw error across configurations. Observed MRE ranges from 4 to
16 \% and is everywhere of the order of its own floor, and the
municipalities with the ``worst'' raw MRE are precisely those whose
floor is highest --- Castenaso, 16 \% observed against a 24 \% floor.
Forlì is the reductio: 21 zones over 117,050 inhabitants make it the
finest partition relative to population, its constraint set contains
many α \textgreater{} 0 cells whose expectation is a fraction of an
individual, and its floor evaluates to 3,189 \% --- three hundred times
the observed error. On such a partition the relative scale carries no
content at all and only the z-scores speak; the \textbar z\textbar{}
distribution of Forlì (mean 0.85, sd 1.09) is unremarkable.

\emph{Zone margins in the sample.} Ring 1 constrains the zone
\emph{distribution} and then draws N individuals from it, so each zone's
count carries the sampling error of a multinomial draw rather than
matching the census exactly. On Milano the mean absolute deviation
across the nine municipi is 258 individuals against ≈ 293 expected for a
draw of this size \textbf{{[}m{]}}, i.e.~0.006--0.60 \% of each zone;
only the municipal total is exact by construction. A consumer summing a
quarter's zones will therefore not recover its census population to the
unit, and the discrepancy is sampling noise, not model error.

The z-scores are not independent draws --- the sampler is a Gibbs chain,
and sd(z) above 1 measures the variance inflation due to its
autocorrelation. Nine municipalities sit between 0.98 and 1.16. Bologna
stands out at 1.391, with \textbar z\textbar max = 36 on a cell of
expectation 1.0 (widowed men aged 15--24 in one zone): the largest state
space of the fleet (18 zones, K9C) mixes least, and the inflation
concentrates on near-empty cells while cells with expectation above
\textasciitilde100 stay within \textbar z\textbar{} ≲ 4. It is a
documented property of the fit, not a defect of the population: no hard
zero is violated and the aggregate error is at its floor.

The same run re-verified the 26 impossible-combination exclusions at the
population level: zero occurrences in 1,814,317 individuals (§III.2).

\emph{Register--census agreement on the demographic base.} Measured
twice, by two independent routes. From the decoded source tables, before
any processing: sex × single year of age, register at 1 January 2024
against census at 31 December 2023, 2,821 cells over fourteen
municipalities (the eleven released plus Mantova, Milano and one test
municipality), \textbf{zero discrepant cells, maximum absolute
difference 0}; cells present in one source only (5 across two small
municipalities) are empty in the tail of the age distribution. From
inside the pipeline: the raccordo section of every
\texttt{constraints\_*/report.md} reports MAE 0.0 per cell and total
discrepancy +0. The agreement is a property of the sources --- the
official resident population being census-derived since 2018 --- not an
artefact of the pipeline, which performs no reconciliation on this
margin.

\paragraph{Ring 2 --- donors, and the honest width of every AVQ
band}\label{ring-2-donors-and-the-honest-width-of-every-avq-band}

\texttt{verifica\_donor.py} reconstructs the donor identity from the
21-tuple of donated values (the hot-deck copies the whole vector from
one donor, so the tuple is the signature), and computes Kish's effective
sample size. Three findings, anchored on Modena and Castenaso:

\emph{The signature is the donor, up to declared equivalence.} On
Modena, 4,161 distinct signatures against 4,617 donors used (pool 4,629,
Emilia-Romagna): 456 fewer, because some pairs of donors carry identical
21-tuples. They are informationally equivalent, so \texttt{n\_eff}
computed on signatures is the correct count --- the donor label would
overstate it.

\emph{The conditioning cell leaves a visible trace.} 99.2 \% of Modena's
signatures serve a single sex, 70.1 \% a single macro-age, 45.3 \% a
single education class; 15.4 \% are confined to their full conditioning
cell, serving 9.4 \% of individuals. The rest of the population received
its AVQ vector through the declared hierarchical collapse --- the price
of small pools, stated as a measurement.

\emph{The band width is a property of the variable's universe, not of
the municipality.} The naive confidence band on an AVQ mean, computed on
\texttt{n}, understates the honest one by √(n/n\_eff), and this factor
differs by variable because each is defined on its own universe:

{\def\LTcaptype{none} 
\begin{longtable}[]{@{}
  >{\raggedright\arraybackslash}p{(\linewidth - 10\tabcolsep) * \real{0.1667}}
  >{\raggedright\arraybackslash}p{(\linewidth - 10\tabcolsep) * \real{0.1667}}
  >{\raggedright\arraybackslash}p{(\linewidth - 10\tabcolsep) * \real{0.1667}}
  >{\raggedright\arraybackslash}p{(\linewidth - 10\tabcolsep) * \real{0.1667}}
  >{\raggedright\arraybackslash}p{(\linewidth - 10\tabcolsep) * \real{0.1667}}
  >{\raggedright\arraybackslash}p{(\linewidth - 10\tabcolsep) * \real{0.1667}}@{}}
\toprule\noalign{}
\begin{minipage}[b]{\linewidth}\raggedright
variable (Modena)
\end{minipage} & \begin{minipage}[b]{\linewidth}\raggedright
coverage
\end{minipage} & \begin{minipage}[b]{\linewidth}\raggedright
n
\end{minipage} & \begin{minipage}[b]{\linewidth}\raggedright
signatures
\end{minipage} & \begin{minipage}[b]{\linewidth}\raggedright
n\_eff
\end{minipage} & \begin{minipage}[b]{\linewidth}\raggedright
band ×
\end{minipage} \\
\midrule\noalign{}
\endhead
\bottomrule\noalign{}
\endlastfoot
SALUTE & 100.0 \% & 184,597 & 4,161 & 862 & 14.6 \\
CRONI & 93.8 \% & 173,208 & 3,889 & 774 & 15.0 \\
PUNTIFI10 (trust: local govt.) & 85.6 \% & 157,973 & 4,006 & 3,220 &
7.0 \\
FIDUCIA (interpersonal) & 86.4 \% & 159,537 & 4,053 & 3,260 & 7.0 \\
AMBIENTE & 86.4 \% & 159,404 & 4,052 & 3,258 & 7.0 \\
BMI & 84.4 \% & 155,843 & 3,942 & 3,156 & 7.0 \\
PUNTIFI13 & 42.4 \% & 78,349 & 2,095 & 1,683 & 6.8 \\
\end{longtable}
}

The whole-population variables (SALUTE, CRONI) carry the widest
inflation (×14--15); the battery items, whose universe excludes
children, sit at ×7 --- the figure the viewer's trust panel draws as the
thick band (§IV.1). A whole-population \texttt{n\_eff}, averaged across
universes, would land between the two and describe neither; this is why
the project computes it per variable and treats the whole-population
figure as meaningless \textbf{{[}n{]}}
\texttt{GSP\_popolazioni\_full\_riferimento}, §13.

\emph{The band factor grows with population, not with any quality
defect.} The full table (SALUTE = whole-population universe; PUNTIFI10 =
the trust battery, universe 14+):

{\def\LTcaptype{none} 
\begin{longtable}[]{@{}
  >{\raggedright\arraybackslash}p{(\linewidth - 14\tabcolsep) * \real{0.1250}}
  >{\raggedright\arraybackslash}p{(\linewidth - 14\tabcolsep) * \real{0.1250}}
  >{\raggedright\arraybackslash}p{(\linewidth - 14\tabcolsep) * \real{0.1250}}
  >{\raggedright\arraybackslash}p{(\linewidth - 14\tabcolsep) * \real{0.1250}}
  >{\raggedright\arraybackslash}p{(\linewidth - 14\tabcolsep) * \real{0.1250}}
  >{\raggedright\arraybackslash}p{(\linewidth - 14\tabcolsep) * \real{0.1250}}
  >{\raggedright\arraybackslash}p{(\linewidth - 14\tabcolsep) * \real{0.1250}}
  >{\raggedright\arraybackslash}p{(\linewidth - 14\tabcolsep) * \real{0.1250}}@{}}
\toprule\noalign{}
\begin{minipage}[b]{\linewidth}\raggedright
municipality
\end{minipage} & \begin{minipage}[b]{\linewidth}\raggedright
n
\end{minipage} & \begin{minipage}[b]{\linewidth}\raggedright
signatures
\end{minipage} & \begin{minipage}[b]{\linewidth}\raggedright
donors used (pool)
\end{minipage} & \begin{minipage}[b]{\linewidth}\raggedright
n\_eff SALUTE
\end{minipage} & \begin{minipage}[b]{\linewidth}\raggedright
band ×
\end{minipage} & \begin{minipage}[b]{\linewidth}\raggedright
n\_eff PUNTIFI10
\end{minipage} & \begin{minipage}[b]{\linewidth}\raggedright
band ×
\end{minipage} \\
\midrule\noalign{}
\endhead
\bottomrule\noalign{}
\endlastfoot
Bologna & 390,098 & 4,169 & 4,625 (4,629) & 966 & 20.1 & 2,845 & 10.9 \\
Brescia & 198,259 & 7,225 & 8,108 (8,111) & 1,093 & 13.5 & 5,655 &
5.5 \\
Parma & 198,121 & 4,162 & 4,618 (4,629) & 832 & 15.4 & 3,181 & 7.3 \\
Modena & 184,597 & 4,161 & 4,617 (4,629) & 862 & 14.6 & 3,220 & 7.0 \\
Reggio nell'Emilia & 171,207 & 4,167 & --- & 795 & 14.7 & 3,347 & 6.6 \\
Ravenna & 156,304 & 4,151 & --- & 995 & 12.5 & 3,334 & 6.4 \\
Rimini & 150,046 & 4,155 & --- & 924 & 12.7 & 3,318 & 6.2 \\
Ferrara & 129,391 & 4,150 & --- & 1,153 & 10.6 & 3,082 & 6.1 \\
Forlì & 117,050 & 4,152 & --- & 916 & 11.3 & 3,325 & 5.5 \\
Piacenza & 102,887 & 4,160 & --- & 810 & 11.3 & 3,237 & 5.2 \\
Castenaso & 16,357 & 3,890 & --- & 741 & 4.7 & 2,770 & 2.2 \\
\end{longtable}
}

(The donors-used column exists where the generation log declares the
donors drawn; the signature count, which is what n\_eff needs, is
computed for all.) The battery band factor falls monotonically with
population, from ×10.9 in Bologna to ×2.2 in Castenaso: a small
municipality does not saturate the donor pool, so each donor still
contributes nearly independent information, and the factor measures
reuse, not quality. Brescia is the natural control: the only
municipality on the Lombardy pool (8,111 donors instead of 4,629)
carries almost twice the signatures (7,225) and the highest battery
n\_eff of the fleet (5,655) at a population equal to Parma's --- the
pool size passes through exactly as the design predicts.

\paragraph{Ring 3 --- placement}\label{ring-3-placement}

Two diagnostics were re-run at the tag on all eleven municipalities
(\texttt{diag\_quinq}, \texttt{diag\_istruzione\_eta}); all three
diagnostics are measured at the tag --- two re-run, the allocation MAE
read from the generation logs

{\def\LTcaptype{none} 
\begin{longtable}[]{@{}
  >{\raggedright\arraybackslash}p{(\linewidth - 8\tabcolsep) * \real{0.2000}}
  >{\raggedright\arraybackslash}p{(\linewidth - 8\tabcolsep) * \real{0.2000}}
  >{\raggedright\arraybackslash}p{(\linewidth - 8\tabcolsep) * \real{0.2000}}
  >{\raggedright\arraybackslash}p{(\linewidth - 8\tabcolsep) * \real{0.2000}}
  >{\raggedright\arraybackslash}p{(\linewidth - 8\tabcolsep) * \real{0.2000}}@{}}
\toprule\noalign{}
\begin{minipage}[b]{\linewidth}\raggedright
municipality
\end{minipage} & \begin{minipage}[b]{\linewidth}\raggedright
allocation MAE per section
\end{minipage} & \begin{minipage}[b]{\linewidth}\raggedright
five-year seam: mean ·res· per section
\end{minipage} & \begin{minipage}[b]{\linewidth}\raggedright
impossible age×title share
\end{minipage} & \begin{minipage}[b]{\linewidth}\raggedright
conditional rate
\end{minipage} \\
\midrule\noalign{}
\endhead
\bottomrule\noalign{}
\endlastfoot
Bologna & 1.36 & 5.11 & 2.39 \% & 30.8 \% \\
Brescia & 1.57 & 5.34 & 2.91 \% & 33.6 \% \\
Parma & 1.44 & 5.74 & 2.73 \% & 31.7 \% \\
Modena & 0.74 & 2.61 & 2.67 \% & 30.3 \% \\
Reggio nell'Emilia & 0.90 & 3.32 & 2.96 \% & 32.4 \% \\
Ravenna & 0.74 & 2.27 & 2.92 \% & 34.3 \% \\
Rimini & 0.89 & 2.57 & 2.88 \% & 32.9 \% \\
Ferrara & 0.72 & 2.41 & 2.43 \% & 31.8 \% \\
Forlì & 1.18 & 4.08 & 3.39 \% & 38.5 \% \\
Piacenza & 1.07 & 3.93 & 2.80 \% & 33.1 \% \\
Castenaso & 0.87 & 3.21 & 3.04 \% & 37.3 \% \\
\end{longtable}
}

Two of the columns move together, and the reason is geometric, not
diagnostic: allocation MAE and seam residual are absolute counts per
section, and both scale with mean section size --- Bologna, Brescia,
Parma and Forlì sit high on both because their sections average 109--175
residents, Ferrara and Ravenna sit low at 66--74. Comparing these
columns \emph{across} municipalities compares their section geometry;
the quality reading is within each row, against the denominators the
text provides (a seam of 5 individuals on sections of 175 is the same
resolution limit as a seam of 2.4 on sections of 74). The two percentage
columns, being ratios, carry their own denominator and compare directly.

\emph{Section counts.} The allocation of individuals to census sections
is exact (largest remainder) rather than multinomial; the mean absolute
error per section, printed by the generation chain at every run and read
here from the tag logs \textbf{{[}m{]}}, is 0.72--1.57 individuals
across the eleven municipalities (against ≈ 9.6 for a multinomial
baseline \textbf{{[}n{]}} riferimento §5) on sections averaging 66--175
residents, with section totals matching the census exactly everywhere.

\emph{The five-year seam.} Single-year ages are drawn within bins whose
boundaries do not coincide with ISTAT's five-year classes;
re-aggregating the synthetic population to those sixteen classes per
section and sex leaves a mean absolute residual of 2.3--5.7 individuals
per section, concentrated on the classes that straddle the 0--8 / 9-14
bin boundary (the fifteen worst sections of Bologna carry seam residuals
of 4--6 individuals on sections of several hundred). It is a declared
resolution limit (assumption 9, §III.4), visible and small.

\emph{Age×title coherence.} Education is constrained at the age-bin
level; the exact age is assigned afterwards in ring 3, and nothing ties
the two below the bin. The share of individuals whose exact age is below
the minimum attainment age for their title is 2.4--3.4 \% everywhere.
The diagnostic's own control shows the incoherence is arithmetic, not
spatial: the rate \emph{conditional on the ages at risk} is flat across
zones within each municipality (30--37 \%), so the raw share varies
between zones only through age composition. Forlì's and Castenaso's
highest conditional rates (38.5 \% and 37.3 \%) tracks its age
structure, not a defect. The fix --- drawing the age jointly with the
title's threshold --- is defined and scheduled for the next regeneration
cycle (§III.5).

\paragraph{Ring 4 --- households, eleven
municipalities}\label{ring-4-households-eleven-municipalities}

\texttt{assign\_nucleo.py} writes a diagnostic per municipality at
generation time; the table below is their collection at the tag
(\texttt{nuclei\_riepilogo.md}):

{\def\LTcaptype{none} 
\begin{longtable}[]{@{}
  >{\raggedright\arraybackslash}p{(\linewidth - 14\tabcolsep) * \real{0.1250}}
  >{\raggedright\arraybackslash}p{(\linewidth - 14\tabcolsep) * \real{0.1250}}
  >{\raggedright\arraybackslash}p{(\linewidth - 14\tabcolsep) * \real{0.1250}}
  >{\raggedright\arraybackslash}p{(\linewidth - 14\tabcolsep) * \real{0.1250}}
  >{\raggedright\arraybackslash}p{(\linewidth - 14\tabcolsep) * \real{0.1250}}
  >{\raggedright\arraybackslash}p{(\linewidth - 14\tabcolsep) * \real{0.1250}}
  >{\raggedright\arraybackslash}p{(\linewidth - 14\tabcolsep) * \real{0.1250}}
  >{\raggedright\arraybackslash}p{(\linewidth - 14\tabcolsep) * \real{0.1250}}@{}}
\toprule\noalign{}
\begin{minipage}[b]{\linewidth}\raggedright
municipality
\end{minipage} & \begin{minipage}[b]{\linewidth}\raggedright
individuals
\end{minipage} & \begin{minipage}[b]{\linewidth}\raggedright
nuclei
\end{minipage} & \begin{minipage}[b]{\linewidth}\raggedright
mean size
\end{minipage} & \begin{minipage}[b]{\linewidth}\raggedright
no fallback
\end{minipage} & \begin{minipage}[b]{\linewidth}\raggedright
not placed
\end{minipage} & \begin{minipage}[b]{\linewidth}\raggedright
homog. couples
\end{minipage} & \begin{minipage}[b]{\linewidth}\raggedright
``incoherent married''
\end{minipage} \\
\midrule\noalign{}
\endhead
\bottomrule\noalign{}
\endlastfoot
Bologna & 390,098 & 210,737 & 1.85 & 98.8 \% & 1.4 \% & 98.1 \% & 18.2
\% \\
Brescia & 198,259 & 96,608 & 2.05 & 97.2 \% & 1.8 \% & 95.1 \% & 22.7
\% \\
Parma & 198,121 & 94,484 & 2.10 & 97.7 \% & 1.9 \% & 96.5 \% & 21.7
\% \\
Modena & 184,597 & 85,249 & 2.17 & 96.0 \% & 2.3 \% & 93.8 \% & 23.7
\% \\
Reggio nell'Emilia & 171,207 & 80,829 & 2.12 & 96.8 \% & 1.1 \% & 94.8
\% & 22.7 \% \\
Ravenna & 156,304 & 75,616 & 2.07 & 95.0 \% & 1.6 \% & 91.6 \% & 23.4
\% \\
Rimini & 150,046 & 68,903 & 2.18 & 95.6 \% & 1.8 \% & 93.0 \% & 24.0
\% \\
Ferrara & 129,391 & 65,281 & 1.98 & 95.8 \% & 2.1 \% & 92.9 \% & 21.4
\% \\
Forlì & 117,050 & 54,000 & 2.17 & 94.7 \% & 1.8 \% & 91.4 \% & 24.7
\% \\
Piacenza & 102,887 & 48,737 & 2.11 & 96.4 \% & 1.4 \% & 93.9 \% & 24.5
\% \\
Castenaso & 16,357 & 7,493 & 2.18 & 96.5 \% & 1.2 \% & 94.6 \% & 23.5
\% \\
\end{longtable}
}

Three readings. \emph{The size constraint is exact where it can be
checked externally:} Parma's 94,484 nuclei equal the census \texttt{PF1}
for Parma to the unit --- the section-level size distribution (PF3--PF8)
is a hard constraint, and this is its visible consequence.
\emph{Individuals not placed (1.1--2.3 \%) are declared, not dropped:}
they appear in the output with an empty \texttt{id\_nucleo}, so a join
cannot confuse ``collective household'' with ``row lost''
\textbf{{[}n{]}} \texttt{assign\_nucleo.py} header. \emph{The
``incoherent married'' share (18--25 \%) is a property of the
population, not of the assembly:} the constraint set does not require
that people marry in pairs, so the ring-1 population contains married
individuals without a matching spouse slot in their section's size
profile; ring 4 reveals the incoherence, it does not create it.
Decomposed on Parma, individuals with role R are correctly paired in
92--94 \% of cases and role P in 97--98 \%; what is missing are slots,
not pairings \textbf{{[}n{]}} \texttt{nota\_repertorio\_avq\_v3} §7.4.
Bologna's low 18.2 \% tracks its low mean household size (1.85): more
one-person households, fewer chances for the incoherence to bind.

Same-sex couples are structurally absent (0 of 4,525 partner pairs in
the donor data carry one), and the population inherits the absence ---
declared as a limit of the source, not a demographic claim
\textbf{{[}n{]}} \texttt{assign\_nucleo.py} header; civil-union
statistics are registered for the next iteration of the repertoire
(§III.5).

\subsubsection{III.4 Assumptions and resolution limits, in one
place}\label{iii.4-assumptions-and-resolution-limits-in-one-place}

Every attribute outside the constraint set enters through a declared
conditional-independence assumption. The report states them once,
numbered as in the reference document, each with its ring and its
measured cost where one exists.

{\def\LTcaptype{none} 
\begin{longtable}[]{@{}
  >{\raggedright\arraybackslash}p{(\linewidth - 4\tabcolsep) * \real{0.3333}}
  >{\raggedright\arraybackslash}p{(\linewidth - 4\tabcolsep) * \real{0.3333}}
  >{\raggedright\arraybackslash}p{(\linewidth - 4\tabcolsep) * \real{0.3333}}@{}}
\toprule\noalign{}
\begin{minipage}[b]{\linewidth}\raggedright
n.
\end{minipage} & \begin{minipage}[b]{\linewidth}\raggedright
assumption
\end{minipage} & \begin{minipage}[b]{\linewidth}\raggedright
ring
\end{minipage} \\
\midrule\noalign{}
\endhead
\bottomrule\noalign{}
\endlastfoot
(4′) & country ⊥ everything given (area, sex, geography) --- geography
per tier; at tier 0 it does not bind & 2 \\
(6) & AVQ targets ⊥ everything given (sex, macro-age, education-4,
region) & 2 \\
(8) & section ⊥ (education, condition, background) given (zone, sex,
age-3, citizenship) & 3 \\
(9) & within a five-year class, single-year age follows the municipal
distribution & 3 \\
(10) & the address is uniform over the civic numbers of the section &
3 \\
(11) & \st{no household structure} --- lifted by ring 4; what remains is
that the \emph{address} is assigned per individual, so spouses can carry
different civic numbers: household-level address assignment is the
stated prerequisite for any building-level work & 3→4 \\
\end{longtable}
}

Their measured costs, where the diagnostics put a number on them:

\begin{itemize}
\tightlist
\item
  \textbf{(6)} is the assumption the viewer prints on every spatially
  filtered AVQ panel: the donated vector carries no geography, so all
  sub-municipal AVQ variation is compositional. Its price is the band
  inflation of §III.3b (×2.2--×20.1 by variable universe and
  municipality). Two further declared costs inside the conditioning
  cell: the 50--54 class is served by 55--64 donors --- 35 \% of the
  50--64 bin, 8 \% of the population \textbf{{[}n{]}} §13.2 --- and the
  hierarchical collapse, measured by regeneration, touches only 1.5--3.1
  \% of individuals (full cell 96.9--98.5 \%, third level and regional
  fallback never reached), but is not random: the cells under the
  20-donor threshold are all \texttt{elementare\_o\_meno}
  \textbf{{[}n{]}} §13.2. An earlier signature-based estimate of the
  collapse (15.3 \% on Modena) was contaminated by donor collisions and
  is withdrawn (§III.6).
\item
  \textbf{(8)} is bounded by the compositional analysis: 80--98 \% of
  the compositional signal lives below the zone, which is why ring 3
  exists; and its residual is the subject of the M-EM measures, whose
  leave-one-out correction is in \texttt{residuo.py} \textbf{{[}n{]}}
  \texttt{nota\_background\_sezione\_v1}.
\item
  \textbf{(9)} produces the five-year seam of §III.3 (mean residual
  2.3--5.7 per section) --- elevated in rank but only \textasciitilde40
  \% above the other classes once normalised, not an order of magnitude
  \textbf{{[}n{]}} §15.1. A subtler measured fact stands beside it:
  within every true bin the synthetic age distribution leans
  \emph{young} --- first class positive, last negative, ten signs
  concordant over two cities (p ≈ 0.002) \textbf{{[}n{]}} §15.2. Nothing
  constrains the within-bin shape; declared, on the list for the next
  cycle.
\item
  \textbf{(10)} is what makes the public regime lossless: randomising
  the coordinate within the section discards an assignment that was
  already uniform (§IV.2).
\end{itemize}

\textbf{Zone-block resolutions, and what each one costs.} The zone
blocks carry every variable at the resolution its section columns have,
which is coarser than the population's in three places; each gap is an
assumption, and each assumption has a cost that no diagnostic in this
report measures --- they are stated here so that a user of the data
knows where not to look for structure.

\emph{Education, five levels against six.} The section tables
distinguish none, primary, lower secondary, upper secondary and
tertiary; the population splits tertiary into first-cycle and
postgraduate. Zone shares are computed on the five-level aggregation and
applied to the six-level municipal counts (\texttt{EDU6TO5} in
\texttt{cs\_build}), so within a zone a degree and a postgraduate title
share the same spatial form. Cost: the geography of postgraduate
education is that of tertiary education as a whole --- a municipality
where doctorates concentrate in one neighbourhood would not show it.

\emph{Occupational condition, the employed side only.} The section
tables count the employed; the population distinguishes four
non-employed categories (seeking work, student, retired, other), which
the section does not separate. Constraining them as one block would
impose a single geography on students and pensioners, which is false and
unmeasurable, so the zone block constrains the employed side alone ---
hence its mass of ≈ 0.47, the universe of the block and not a defect
(§I.2, partial blocks). Cost, and it is the most consequential of the
three: \textbf{the spatial distribution of unemployment is constrained
by no observed datum}. Where the unemployed live follows from the
maximum-entropy solution given everything else --- their sex, age,
education, citizenship and the zone shares of those --- not from a
measurement. Any use of these populations that reads unemployment
geographically must know this.

\emph{Citizenship, three macro-classes.} The zone block conditions on
0--14 / 15--64 / 65+ rather than on the population's eight bins, so
within a macro-class the zone shares of foreigners are constant: a
foreign twenty-year-old and a foreign sixty-year-old are distributed
across zones identically. Cost: the age profile of foreign residents
does not vary by zone beyond the three-class structure --- a
neighbourhood of young foreign workers and one of settled foreign
families differ in the model only through the macro-class composition.

None of the three is a bug, and none can be removed without a source
that does not exist: they are the resolution of the census section
tables, inherited.

\textbf{Resolution limits by construction.} Education has an effective
age resolution of 4 classes, not 8: the census constraint uses Y9-24 /
Y25-49 / Y50-64 / Y≥65, and within each class the distribution comes
from an IPF with minimum attainment thresholds --- \texttt{media}
remains overstated in the 9--14 bin (the age×title incoherence of §III.3
is its single-year shadow), and the cohort effect between 65--74 and 75+
is lost. The migratory background has zone resolution, not section
(assumption 8). The AVQ battery is complete at twelve items. An earlier
release lacked \texttt{FORZE\_ARMATE} --- the target list selected by
prefix, one of three silent hand-written-table failures found in one
afternoon, all producing absences rather than errors \textbf{{[}n{]}}
riferimento §13 --- and the working notes still described that state;
the fix entered before the release tag, and the discrepancy was caught
during testing when a new municipality showed a variable the notes
declared absent (§III.6, register).

\textbf{Conventions for ``absent'', not unified} --- declared as a trap
for consumers: \texttt{non\_applicabile} as a string in
\texttt{condizione}, \texttt{origine\_genitori} and 20 of 21 AVQ
variables; \texttt{NaN} in \texttt{area}, \texttt{via}, \texttt{civico};
nothing in \texttt{SALUTE} and \texttt{paese} (Italians read
\texttt{Italia}) \textbf{{[}n{]}} riferimento §1.

\subsubsection{III.5 Open points carried into
v1.0}\label{iii.5-open-points-carried-into-v1.0}

Declared, analysed, and deliberately not applied in this release ---
each with the reason and, where it exists, the prepared change.

\begin{enumerate}
\def\labelenumi{\arabic{enumi}.}
\tightlist
\item
  \textbf{Citizenship in the AVQ conditioning cell.} Analysis complete,
  patch drafted, not applied. AVQ's \texttt{CITTMi} shows a substantial,
  monotone effect on institutional trust (\textasciitilde1 point of 10,
  replicated in two regions; non-Italians trust institutions
  \emph{more}, a direction opposite to intuition but known in the
  literature), and no effect on \texttt{AMBIENTE} or interpersonal
  \texttt{FIDUCIA}. The full cell does not hold (\textasciitilde11
  donors); the collapse hierarchy with citizenship in high priority is
  written and waiting. Two reasons to wait: the modalities
  \texttt{3}/\texttt{9} must be identified in the codebooks first, and
  the foreign AVQ sample is self-selected on language competence, so the
  estimated trust is likely \emph{overstated} --- a limit to declare
  regardless \textbf{{[}n{]}} riferimento §8.
\item
  \textbf{The within-bin age lean} (§III.4, assumption 9) and
  \textbf{the age-at-threshold draw} for education (§III.3): both
  defined, both waiting for a cycle in which regeneration is already
  required, because regeneration invalidates \texttt{donor\_id} and
  every measure, and is not done for one fix \textbf{{[}n{]}} old
  Animarium README.
\item
  \textbf{Section-level migratory background (EM1--EM6).} The census's
  own section columns give the six background modalities per section, in
  one-to-one correspondence with ring 1's \texttt{background}; the M-EM
  measures find a real net residual on all eleven municipalities (median
  \textasciitilde0.022 Italians, \textasciitilde0.018 foreigners), so
  assumption (8) discards structure that exists. The modification to
  \texttt{enrich.py} is designed --- background subsumes citizenship in
  the cell key, weights normalised within the group, not multiplied ---
  and the registry patch extending the source's \texttt{usabile\_per} is
  written. Queued with the same regeneration cycle as (2)
  \textbf{{[}n{]}} \texttt{nota\_background\_sezione\_v1}.
\item
  \textbf{A standalone diagnostic for the per-section allocation MAE}
  (§III.3): the measurement itself is not missing --- the generation
  chain prints it at every run and this report reads it from the tag
  logs --- but recomputing it today requires a full regeneration. A
  small script that recomputes it from an existing population is a v1.1
  convenience.
\item
  \textbf{Modena's 37 rioni.} A partition ten times finer than the four
  ASC zones exists on the municipal portal (foreigner share spanning a
  factor 15 instead of 2); it would attack assumption (8) where it binds
  most. Cost: a deviation from the ISTAT-standard zone levels, a
  section→rione mapping to build, and \textbar X\textbar{} growing
  tenfold. Registered as an opportunity, not scheduled \textbf{{[}n{]}}
  riferimento §8.
\item
  \textbf{EU-SILC} for income and living conditions: explored (two
  registered entries, graph script), decision not taken; the public
  files' terms do not allow redistribution, which any use will have to
  respect (§II.2).
\item
  \textbf{\texttt{donor\_anno}} and \textbf{\texttt{cella\_avq}}
  (collapse level per individual): queued for the same next regeneration
  cycle as (2). (\texttt{FORZE\_ARMATE}, previously listed here, was
  already in at the tag --- see §III.4.)
\item
  \textbf{The floor is a standard deviation, the MRE a mean absolute
  error.} Both \texttt{fit\_cs.py} and \texttt{verifica\_vincoli.py}
  compute the MRE as mean(\(|\hat\alpha-\alpha|/\alpha\)) --- instrument
  and diagnostic agree \textbf{{[}m{]}} --- but the floor they are
  compared against, mean √((1−α)/(αN)), is a standard deviation, and for
  a normal deviation the two differ by √(2/π) ≈ 0.798. Comparing them
  directly overstates the floor by a quarter. Corrected, the observed
  error sits 10--11 \% above the floor, consistent with the sd(z) ≈ 1.03
  of Table III.3a --- which is the right reading: sampling explains
  almost all of the deviation, and the residue is what the z-columns
  localise. The tables in this report print the uncorrected floor, as
  the diagnostic does; the factor is stated here rather than silently
  applied \textbf{{[}n{]}} riferimento §14.5.
\item
  \textbf{Household-level address} (the residue of assumption 11) as the
  prerequisite of any building-level assignment \textbf{{[}n{]}}
  nota\_nucleo §9.
\end{enumerate}

\textbf{Post-tag divergence in ring 4, declared.} The code in
\texttt{master} no longer reproduces the tagged ring-4 files, and the
reason is worth stating precisely rather than discovering. Two patches
were applied to \texttt{gsp/nucleo.py} and \texttt{assign\_nucleo.py}
\emph{after} the release tag, both from the same finding --- a
per-nucleus O(n²) matching invisible on the fleet (sections up to
\textasciitilde700 residents) and costing hours on a metropolis (Milano
has fifteen sections above 1,000, one of 4,146, plus the fictitious
convivenza section of 10,038).

\begin{itemize}
\tightlist
\item
  \textbf{(a) Skip of the fictitious convivenza section.} Those
  individuals carry an empty \texttt{id\_nucleo} by design, so
  assembling them was work thrown away; the section is now skipped
  before assembly. Because the skip does not consume the generator,
  \textbf{the rng sequence shifts for every section that follows it}:
  ring-4 outputs change for any municipality that has a convivenza
  section --- that is, all of them. The aggregate diagnostics move
  within the noise of reassembly (Mantova: 24,298 → 24,275 nuclei,
  non-placed 2.16 \% → 2.21 \%, the rest unchanged to the decimal), and
  the non-placed now correctly include the institutional population.
\item
  \textbf{(b) Children-plausibility count via \texttt{searchsorted}.}
  Semantics identical --- same candidate set, same order, same rng draws
  --- \textbf{verified} by an A/B run with only this patch switched,
  giving a byte-identical \texttt{nuclei\_020030.csv}. Milano's ring 4
  went from over an hour to 65 s.
\end{itemize}

Consequences, declared: the binding table's byte-identity claim holds
for the \textbf{tagged commit}, which is what it states; a reader
running \texttt{master} reproduces rings 1--3 identically and ring 4
with the shifted sequence described above. The fleet is not regenerated
for this alone (point 2 above: regeneration invalidates
\texttt{donor\_id} and every measurement, and is not done for one fix);
both patches travel with the next cycle \textbf{{[}n{]}}
\texttt{collaudo\_acquisizione\_v0.2}, finding 7.

\textbf{What the next regeneration cycle carries.} Regeneration is
expensive in the only currency that matters here --- it invalidates
\texttt{donor\_id} and every measured number in this report --- so the
queued changes travel together. At the time of writing the cycle
carries: the two ring-4 patches above (a, b); the section-level
migratory-background refinement through the census EM columns
(§III.5.3); \texttt{donor\_anno} and \texttt{cella\_avq} written per
individual (§III.5.7); the within-bin age lean and the age-at-threshold
draw for education (§III.5.2); seed policy aligned to the
municipality-derived convention everywhere (§III.1); and the two test
municipalities (Mantova K6C, Milano K9C) promoted from test cases to
fleet, taking the released bundle from eleven to thirteen. The α = 0
exclusions for impossible age × education and age × condition
combinations are \textbf{already in the tagged constraint sets} ---
twenty-six imposed cells, verified at the tag (§I.2) --- and are listed
here only to record that they were checked, not deferred.

\subsubsection{III.6 Register of retractions and falsified
predictions}\label{iii.6-register-of-retractions-and-falsified-predictions}

Nothing in this project is silently corrected: a withdrawn claim keeps
its original text and gains a dated annotation, in the note and where
relevant in the code. This register collects them, because a reader
auditing the populations deserves to know not only what is claimed but
what was claimed and withdrawn --- and because the pattern of one's own
errors is itself a finding. Dates are those of the annotations.

\begin{enumerate}
\def\labelenumi{\arabic{enumi}.}
\tightlist
\item
  \textbf{The −418 gap} between two population totals, initially read as
  a data defect, was a configuration coincidence; withdrawn and
  documented in the design note §0.1.
\item
  \textbf{\texttt{quartiere} = \texttt{zona}.} The viewer briefly
  treated the two as distinct levels; they are one-to-one, and the
  column was dropped from the bundle as redundant \textbf{{[}n{]}}
  design note.
\item
  \textbf{Whole-population \texttt{n\_eff}} is unstable and was
  withdrawn; \texttt{n\_eff} per variable universe is the correct object
  (§III.3b) \textbf{{[}n{]}} riferimento §13.
\item
  \textbf{The signature-based collapse estimate} (15.3 \% of individuals
  confined to the full cell on Modena) was a very pessimistic lower
  bound contaminated by donor collisions; superseded by the regeneration
  measurement (1.5--3.1 \%) \textbf{{[}n{]}} riferimento §13.2.
\item
  \textbf{The 3.7 \% regional-fallback figure} in the 27 July logs
  described a generation predating the hierarchical collapse (commit
  \texttt{f383e54}); it does not apply to the populations in use
  \textbf{{[}n{]}} riferimento §13.2.
\item
  \textbf{Ring-4 prediction (a)}: excluding relationship code 11 would
  collapse the geographic residual of household structure. Measured:
  +0.0247 → +0.0255. \emph{Falsified} --- the code-11 concentration is
  almost entirely mediated by citizenship, already conditioned on
  \textbf{{[}n{]}} nota\_nucleo §7.
\item
  \textbf{Ring-4 prediction (b)}: the section adds little over the
  quartiere (expected ratio \textless{} 0.3). Measured: 0.87 per family,
  1.40 per person. \emph{Falsified} --- the centre--periphery gradient
  exists but is small against within-quartiere heterogeneity. Both
  falsified predictions err in the same direction --- fine structure
  systematically underestimated --- which is itself recorded as a prior
  for the next ones \textbf{{[}n{]}} nota\_nucleo §7.
\item
  \textbf{The TVD \textasciitilde{} size correlation} (−0.438, read as
  ``residual sampling noise''): under the null that correlation is
  negative anyway, because small sections have high TVD by construction;
  the comment had no diagnostic power and was replaced by the
  net-by-size-tercile test, which answered in the \emph{opposite}
  direction \textbf{{[}n{]}} nota\_nucleo §6.3.
\item
  \textbf{The Parma codebook reading of code 11} (``coabitazione'') was
  contradicted by the data (PF9 cross-check); the mapping in use is
  inferred from demographic profiles and marked \emph{inferita, non
  letta} (§II.3) \textbf{{[}n{]}} nota\_nucleo §2.
\item
  \textbf{Aggregate MRE as a quality claim}: an aggregate MRE of a
  fraction of a per cent can hide per-cell errors of 1 \%; per-cell
  z-scores against the floor replaced it (§III.3a) \textbf{{[}n{]}}
  design note / riferimento §11.
\item
  \textbf{Raking's MRE = 0 on training constraints} is algebraically
  guaranteed and is not an accuracy metric; held-out evaluation and
  diversity metrics are the comparison framework (this is the framing
  the solver paper adopts) \textbf{{[}n{]}} memory / arXiv:2603.27312.
\item
  \textbf{The Q-series and unit corrections} of the viewer's cost and
  error measures (a factor-2.1 estimate corrected; two results withdrawn
  by comparing against known-answer configurations) \textbf{{[}n{]}}
  design note §12--13.
\item
  \textbf{Falsified: the per-section MAE is mostly propagated sampling
  noise.} Conjectured while examining the zone margins of Milano: since
  ring 1 samples N individuals rather than fixing the zone counts, each
  zone carries a sampling error (measured: 258 individuals on average,
  against ≈ 293 expected for a multinomial draw \textbf{{[}m{]}}), and
  distributing it over the zone's sections would give ≈ 1 individual per
  section --- the order of the observed MAE. The decomposition refutes
  it: on Milano the propagated component accounts for 0.38 of the 1.28
  MAE, the allocation residue for 1.24, with a correlation of 0.37
  between the two \textbf{{[}m{]}}. The reason is structural: the zone
  error is one number spread coherently over hundreds of sections, while
  the largest-remainder rounding acts within every demographic cell
  separately, and dozens of incoherent ±1 accumulate faster in absolute
  value. The geometric note under Table III.3 stands.
\end{enumerate}

\begin{center}\rule{0.5\linewidth}{0.5pt}\end{center}

\newpage

\section{Part IV --- The Animarium
release}\label{part-iv-the-animarium-release}

\subsection{Version 1 --- §IV.1--IV.3 (26 August
2026)}\label{version-1-iv.1iv.3-26-august-2026}

\begin{quote}
Section references without other indication are to
\texttt{design\_animarium\_v13}, the viewer's design note.
\end{quote}

\begin{center}\rule{0.5\linewidth}{0.5pt}\end{center}

\subsubsection{IV.1 What the viewer is}\label{iv.1-what-the-viewer-is}

Animarium is a single hand-written HTML page. There is no framework, no
build step, no bundler: the page loads DuckDB-WASM from a CDN, reads the
Parquet bundle over HTTP range requests, and draws bars in HTML/CSS, the
map on a canvas, and the export SVG by hand. The choice is deliberate
and load-bearing. A static page served from a folder is the most
reproducible artefact a viewer can be --- anyone can serve it locally
with the range-aware server in \texttt{build/serve\_range.py} --- and
the absence of a dependency graph is what allows the sentence, verified
in §III.2, that the site online is a pure function of the tagged bundle.
The one exception is external and \emph{opt-in}: the cartographic base
layer, off by default. The panel draws its maps on an empty background
--- the choropleth and the point layers are drawn by the page itself ---
and a reader who switches a tile provider on (OpenStreetMap or CARTO,
attributed on the canvas) is told it is an external service. Serving a
static PMTiles extract alongside the bundle --- one file, range
requests, the same mechanism the Parquet already uses --- is the
designed v1.1 replacement that removes the last external dependency
\textbf{{[}n{]}} §7.3, \texttt{nota\_pmtiles\_v0.1}.

The regional atlas is the case where the empty background costs most ---
eleven municipalities on a wide area, with no coastline or provincial
boundary to anchor them --- and in v1.0 it inherits whatever the map
panel is set to, with its own selector. The PMTiles extract of v1.1
resolves it properly: a base layer served from the same origin as the
bundle can be on by default, because it introduces no external
dependency.

The panel is organised around one idea: \emph{every number on screen
carries its comparison}. Filters are the bars themselves --- clicking a
bar filters, the active filters become removable pills, and the entire
state lives in the URL, so that a view can be sent in one line and cited
in a paper. Each marginal shows up to three markers, with three
meanings: the \textbf{bar} is the filtered subpopulation; the
\textbf{tick} is the whole city, and the gap between bar and tick
measures the \emph{association} between filter and attribute; the
\textbf{diamond} is the census count at the filter's level, and the gap
between bar and diamond measures the \emph{model error}. The diamond
exists only where a block of the constraint set contains the filter's
attributes together with the displayed one --- 67 of 333 attribute pairs
at K9C, 26 of 96 at K6C, a property of the constraint template, not of
the municipality \textbf{{[}n{]}} §3.4 --- and where it does not exist
the panel says so instead of hiding it. The institutional-trust view
draws, for each of the fifteen items, the mean and two overlaid
confidence bands: a thin one computed on \texttt{n} and a thick one on
Kish's effective sample size for that variable's universe. The distance
between the two bands is the honest cost of hot-deck donation made
visible --- on Modena, unfiltered, the true band is seven times the
naive one \textbf{{[}n{]}} §4.7, §2.1 --- and dotted ticks mark the
national means, computed by GSP from the microdata and registered as a
derived source. The individual card shows the forty-odd fields of a
record grouped by ring, each with its guarantee class, under a banner
that reads \textbf{SYNTHETIC INDIVIDUAL --- DOES NOT EXIST}; clicking
two nearby points on the map and finding the same donor signature is
\texttt{n\_eff} seen with the naked eye \textbf{{[}n{]}} §4.4. A
regional atlas opens a card per municipality --- individuals,
articulation, tier, coverage with its denominator, and the
\texttt{n}/\texttt{n\_eff} band translated into a sentence --- before
switching city.

Two things the panel deliberately cannot do. It cannot filter or display
the detailed education title or the sector × position pair: those are
derivations that individuate a \emph{person}, they live in the
\texttt{persona} and \texttt{narrativo} regimes, and the bundle does not
carry them (§IV.2). And it cannot show any sub-municipal geography of an
AVQ variable as if it were information: the donated vector carries no
geography by construction (assumption 6), so under a spatial filter
every AVQ difference is compositional, and the panel states this in so
many words \textbf{{[}n{]}} §4.7.

\subsubsection{IV.2 The public bundle}\label{iv.2-the-public-bundle}

The bundle is one folder: an index (\texttt{comuni.json}), and per
municipality a \texttt{pop.parquet}, a \texttt{manifest.json} with
labels, declared orderings and unfiltered counts, and a
\texttt{riferimenti.json} with the census counts extracted from the
constraint set (\texttt{n\ =\ α\ ×\ N}). Eleven municipalities, ≈ 35 MB
\textbf{{[}m{]}}, rebuilt in one command
(\texttt{build/build\_bundle.py}).

The Parquet layout is designed for the reader, in the literal sense:
DuckDB-WASM fetches contiguous byte ranges, not columns, so the columns
are laid out in three blocks by use --- filters and marginals; the AVQ
battery with the donor signature; the heavy map columns --- and the rows
are sorted by \texttt{zona,\ sezione} in row groups of 20,000, so that

\begin{verbatim}
cost(query) = footer + Σ weight(block) × (row groups not pruned / total).
\end{verbatim}

The model was verified on the deployed site rather than in the
laboratory: \texttt{smoke.html} runs eight queries against Modena's 3.52
MB Parquet and reports the bytes each one costs \textbf{{[}m{]}}. The
host answers range requests (HTTP 206), without which none of what
follows would hold. Counting rows touches metadata only (0.08 MB, the
footer); demographic marginals cost 0.50 MB, and adding the zone brings
them to 0.70 --- the whole of block A. The three comparisons that matter
are the ones the test was written for. \emph{Row-group pruning}: the
same query filtered on \texttt{id} costs 0.15 MB against 0.70
unfiltered, one row group in ten. \emph{Spatial ordering}: filtering on
a single zone costs 0.20 MB against 0.70 for all zones, so sorting the
rows by \texttt{zona,\ sezione} buys a factor of three on every
zone-restricted query. \emph{The AVQ block}: reading one trust variable
by zone costs 0.60 MB, below the 1 MB threshold at which the design note
said the block would have to be split into health and trust --- it does
not. The heaviest query, the map's coordinates, costs 0.70 MB against
2.30 before the columns were laid out in blocks.

Warm figures are meaningless here and the test says so: within one
session the queries warm each other, exactly as they do in the
application, and the second run of anything costs 5--30 ms. The
initialisation of DuckDB-WASM from the CDN is 0.43 s, paid once.

\textbf{The public regime is enforced in the data, not in a banner.} The
panel always stated that assigning an individual to a civic number
carries no information --- the model places people within a section
arbitrarily --- but a banner does not travel with a file, and
\texttt{pop.parquet} is served statically to anyone with the URL. Since
4 August 2026 the default output of \texttt{to\_parquet.py} \emph{is}
the public regime, applied by the single enforcement point
\texttt{gsp.individui.esporta\_pubblico} (§I.7):
\texttt{lon}/\texttt{lat} are a random point within the census section,
drawn with a seed derived from the municipality code; \texttt{via},
\texttt{civico} and the address provenance are absent --- the individual
card reads ``Cittadella, section 034027001042'' and loses nothing
analytic; \texttt{quartiere} is absent, being one-to-one with
\texttt{zona} and already provided as a label by the manifest;
\texttt{uid} is absent, being the onomastic key, and the viewer shows no
names. The randomisation loses nothing because the civic assignment was
already uniform within the section: the map is visually identical, the
per-section density unchanged, and the file becomes self-protecting ---
``the point is random within the section'' admits no reply, where
``displaced by thirty metres'' invites the question ``and if it were
twenty?'' \textbf{{[}n{]}} §15. The permissive export exists
(\texttt{-\/-completo}) but is an explicit act with a warning, not a
default that can be forgotten.

What is \emph{not} in the bundle, by the same logic: the AVQ raw codes
(dropped after the donor signature is computed), the detailed education
title and the sector × position pair, and anything from the
\texttt{narrativo} regime. The complete population with true civic
numbers never leaves the generation machine (§I.7).

A view is citable. The whole state of the panel --- municipality,
filters, open views, map mode --- is the URL query string, so a figure
in a paper can carry the exact view that produced it, and the
version-binding table (front matter) ties that URL to a bundle whose
Parquet hashes are recorded. The citation of a view is therefore URL +
report version + the SHA-256 prefix of that municipality's
\texttt{pop.parquet}; the recommended form, with a worked example, is in
the front matter.

\subsubsection{IV.3 Deployment and
versioning}\label{iv.3-deployment-and-versioning}

What the viewer \emph{requires} of a host is a short list, and it is
worth stating before naming any platform: static file serving over
HTTPS, HTTP \textbf{range requests} (the mechanism by which DuckDB-WASM
reads parts of a Parquet instead of downloading it whole, and by which
PMTiles will serve tiles in v1.1), and no ceiling on individual file
size. No server-side runtime, no database, no build step. Any host
meeting those three conditions can serve Animarium, including a laptop:
\texttt{build/serve\_range.py} in the repository is a range-aware server
written for exactly this purpose. The second of the three fails silently
when it is missing, so the deployed folder carries its own test:
\texttt{smoke.html} checks that the host answers range requests. Without
them nothing breaks --- DuckDB simply downloads whole files, and the app
is merely slow, with nothing on screen to say why.

The release is served from Cloudflare Pages under the project's own
domain, \textbf{animarium.it} --- the canonical address, with
\texttt{www} redirecting to it.Deployment is one command,
\texttt{python\ build/deploy.py\ -\/-cloudflare}: the script assembles
the \texttt{deploy/} folder from the bundle on disk and hands it to
Wrangler, so that assembly and upload cannot drift apart --- a
half-regenerated bundle would otherwise go online silently, which is
also why \texttt{build\_bundle.py} reports per-municipality status
instead of stopping at the first failure \textbf{{[}n{]}} §8, §15. A
second, non-canonical path (\texttt{-\/-gh-pages}) survives from an
earlier configuration and is kept only as a fallback. The folder-based
deploy is kept because it leaves nothing to clean and can actually be
switched off; the domain is registered independently of the platform, so
the citable URLs of this report survive a change of host. Switching the
site off does not withdraw what has already been downloaded, and the
report says so rather than implying permanence.

Versioning binds three objects: the repository tags
(\texttt{report-v1.0} on both GSP and Animarium), the bundle (SHA-256
per \texttt{pop.parquet}, listed in the front matter; the two JSON files
per municipality embed a generation timestamp and are excluded from the
binding --- compared modulo that field in §III.2), and the deployed
site, which is the \texttt{deploy/} folder produced from that bundle.
The build dependency is declared, not assumed: Animarium's
\texttt{pyproject.toml} depends on \texttt{gsp}, and the build scripts
obtain every path from \texttt{gsp.common} (overridable via
\texttt{GSP\_ROOT}); the \emph{runtime} has no dependency at all --- the
published site is static, and a reader with only the Zenodo bundle can
serve the viewer without installing GSP (the two entry points of the
repository README). If GSP were ever unavailable, the site would keep
working and the build would stop being reproducible from outside
\textbf{{[}n{]}} §7.1.

\begin{center}\rule{0.5\linewidth}{0.5pt}\end{center}

Deferred to v1.1: the PMTiles base layer, whose recipe is written ---
one file per region, served from the same origin as the bundle,
registered as a derived source with its ODbL attribution
\textbf{{[}n{]}} \texttt{nota\_pmtiles\_v0.1}. It is what turns the
cartographic background from an opt-in external service into part of the
release (§IV.1).

\newpage

\section{Part V --- The narrative layer and its
use}\label{part-v-the-narrative-layer-and-its-use}

\subsection{Version 1 --- §V.1--V.3 (26 August
2026)}\label{version-1-v.1v.3-26-august-2026}

\begin{quote}
\textbf{This part describes work in progress, and is the one part of
this report expected to change substantially.} The narrative layer is
built and in use, but its calibration is an open research programme: the
experiments reported here are evidence that the platform behaves
controllably, not results about synthetic populations as instruments.
Findings, and the layer itself, will be updated in later versions of
this report and in the companion papers; nothing downstream of the
populations depends on it, and a reader interested only in the pipeline
(Parts I--III) or the release (Part IV) can stop before this point.
\texttt{registro\_esperimento\_sive\_gsp\_v5} is cited here as
\emph{registro}; the SIVE paper is arXiv:2607.00910.
\end{quote}

\begin{center}\rule{0.5\linewidth}{0.5pt}\end{center}

\subsubsection{V.1 From record to
persona}\label{v.1-from-record-to-persona}

The narrative layer exists because a record and a readable individual
are different products, and the difference is exactly where control must
not be ceded. The rule that organises everything fits in one line:

\begin{quote}
\textbf{The LLM does not generate the person. The person is generated
statistically; the LLM generates one possible narrative rendering of
it.}
\end{quote}

The temptation is to hand the record to a language model and ask for a
biography. It works, and it is wrong for a precise reason: the model
would fill the gaps with its own idea of plausibility --- opaque,
unreproducible, unmeasurable. With twenty-three AVQ attributes, the
demographic profile and a geography down to the section, the gaps are
still many: precise occupation, contract, household detail, transport,
income. Whoever fills them determines what the population appears to
say, and that power does not go to a language model \textbf{{[}n{]}}
biografia §1. Hence three strata:

\begin{enumerate}
\def\labelenumi{\arabic{enumi}.}
\tightlist
\item
  \textbf{The constrained profile} --- the record as it is, immutable
  for the LLM, \emph{including its declared limits}: the population does
  not know income or precise profession, and its variables have
  different geographic resolutions. A profile that does not carry its
  own limits produces a biography that asserts more than it knows.
\item
  \textbf{The structured expansion} --- new variables from conditioned
  distributions, controlled catalogues and compatibility rules (the
  derived layers of §I.6, plus imputation where a narrative needs it);
  still a structured row, still no free text, each variable with a
  declared certainty class (A constrained / B measured / C imputed)
  \textbf{{[}n{]}} biografia §3.
\item
  \textbf{The narrative rendering} --- only here does the LLM enter,
  realising linguistically a profile whose sociological plausibility is
  already fixed.
\end{enumerate}

Two regimes consume this pipeline (§I.7): \texttt{persona} renders
strata 1--2 as prompt material for agents --- attributes and AVQ vector,
no name, no address; \texttt{narrativo} adds the generated name, the
address and the prose, on demand, capped. The distinction is the one
§I.6 argued: apparent diversity must not be mistaken for real diversity,
and the persona regime is the one in which two agents with the same
donor signature are still \emph{visibly} the same evidence.

\subsubsection{V.2 LLM-driven simulation as evidence of use, not as
validation}\label{v.2-llm-driven-simulation-as-evidence-of-use-not-as-validation}

Two studies have run on this platform's outputs. They are reported here
for what they demonstrate about the platform --- that its populations
support controlled, replicated, pre-registered experiments on LLM agents
--- and not for their findings' own sake, which belong to their papers.

\textbf{SIVE} (arXiv:2607.00910) asked whether an LLM-driven synthetic
population is \emph{controllable}: impose a trust level on an agent, and
the agent exhibits it, stably, across the battery. Montelago, its
fictional municipality of 120 personas in three trust strata, predates
the GSP populations; its result --- controllability, with compression of
the imposed scale --- is the licence for everything that follows, and
its protocol (pre-registered criteria C1--C7, within-subjects design) is
the template the GSP experiments inherit.

\textbf{The Brescia conditions} moved the same question onto a real GSP
population and removed SIVE's confound: the SIVE prompt carried both a
direct label («sfiduciato critico») and a story encoding the latent
level; the ablation separates them. 120 employed individuals of
synthetic Brescia, stratified by \texttt{PUNTIFI10} (trust in municipal
government, 0--10) into LOW/MED/HIGH --- Brescia because what matters is
how many \emph{distinct response vectors} exist, and the Lombardy pool
(8,111 donors) gives a replica share of 0.8 \% against Parma's 5.8 \%
\textbf{{[}n{]}} registro §2. Three conditions: \textbf{B} profile + a
story encoding the latent level, \textbf{C} profile alone, \textbf{D}
profile + a neutral story. The stories themselves went through a
measured correction loop --- the first prompt's examples were taken as
instructions and 49 stories of 50 staged the same municipal counter; a
twelve-scene repertoire fixed it, and the monotony detector born there
(count, don't read) became part of the harness \textbf{{[}n{]}} registro
§3.

What the platform made measurable \textbf{{[}n{]}} registro §4--5, on
\texttt{fiducia\_istituzione}, T 0.3:

{\def\LTcaptype{none} 
\begin{longtable}[]{@{}
  >{\raggedright\arraybackslash}p{(\linewidth - 6\tabcolsep) * \real{0.2500}}
  >{\raggedright\arraybackslash}p{(\linewidth - 6\tabcolsep) * \real{0.2500}}
  >{\raggedright\arraybackslash}p{(\linewidth - 6\tabcolsep) * \real{0.2500}}
  >{\raggedright\arraybackslash}p{(\linewidth - 6\tabcolsep) * \real{0.2500}}@{}}
\toprule\noalign{}
\begin{minipage}[b]{\linewidth}\raggedright
condition
\end{minipage} & \begin{minipage}[b]{\linewidth}\raggedright
Spearman (latent → response)
\end{minipage} & \begin{minipage}[b]{\linewidth}\raggedright
slope
\end{minipage} & \begin{minipage}[b]{\linewidth}\raggedright
LOW / MED / HIGH medians
\end{minipage} \\
\midrule\noalign{}
\endhead
\bottomrule\noalign{}
\endlastfoot
B story with latent & \textbf{+0.90} & 0.52 & 2.6 / 4.6 / 6.9 \\
C profile only & +0.06 & 0.00 & 5.0 / 5.1 / 5.1 \\
D neutral story & \textasciitilde0 & 0.01 & 5.3 / 5.3 / 5.5 \\
\end{longtable}
}

The narrative transmits the disposition (paired B−C: 40 of 40 negative
in LOW); the profile alone produces clones; and the \emph{presence} of a
narrative does not unlock demographic priors --- D's slope is zero ---
while its residual positive valence (+0.30 intercept) was independently
seen by three blind human judges and three models: neutral stories
describe services that work, and the absence of friction is itself a
signal \textbf{{[}n{]}} registro §4, §7.

Replicated across three model families (DeepSeek, Claude Haiku 4.5,
GPT-4o-mini), the result splits into what travels and what does not
\textbf{{[}n{]}} registro §5: the gain of B is invariant within seven
hundredths (0.52 / 0.55 / 0.59) --- the compression is a property of
LLMs, not of one model; the flatness of C is invariant; but the
\emph{level} of C spans 1.48 points (3.96 / 5.02 / 5.44). There is no
prior on the people; there is a prior on municipal institutions, and it
differs by model. \textbf{Differences and orderings are transportable
across models; levels are not} --- the one sentence a practitioner
should take from this Part.

Two further platform-enabled findings complete the picture. On
categorical items the profile \emph{is} used, and used along textbook
stereotypes --- fourteen times more anger in men, hope to the educated
--- established at n = 600 after the n = 120 signal on age was falsified
as thin-cell noise \textbf{{[}n{]}} registro §9; the numeric-scale
abstention is a refuge the scale offers, not absence of priors. And the
ground-state design (what does the model answer when there is \emph{no
one}: 17 cells instead of a 756-cell factorial) gives every profile
effect its reference point \textbf{{[}n{]}} registro §10.

The experimental materials are versioned with the pipeline
(\texttt{dati/agenti/}, \texttt{dati/campagne/},
\texttt{dati/giudizio/}; §II of the repository's
\texttt{dati/README.md}): agents in the persona regime, campaigns as
\texttt{uid} + responses, stories preserved as the actual input of the
published campaigns. The sample regenerates from
\texttt{(comune,\ variabile,\ n,\ seed)}; the stories do not, and are
stored for that reason \textbf{{[}n{]}} registro §12.

\textbf{SimComm / Caffaro} is the application context these instruments
were calibrated for --- institutional risk communication on the SIN
Brescia--Caffaro site --- and is deliberately absent from this report
beyond this sentence: it is applied work in progress, with its own
protocol and its own paper.

\subsubsection{V.3 Intended uses, and the two warnings that travel with
them}\label{v.3-intended-uses-and-the-two-warnings-that-travel-with-them}

What the released populations support: compositional comparison across
municipalities and zones with honest uncertainty (the viewer's native
use); pre-testing of institutional communication on stratified synthetic
audiences, in the SIVE/Brescia mould --- where the platform's
contribution is the \emph{stratification with known ground truth}, which
no convenience panel offers; teaching, where a population that declares
its assumptions is the point.

What they do not support, stated as sharply as the argument allows: any
sub-municipal geography of an attitudinal variable read as information
(assumption 6 --- the panel says it, the report repeats it); any claim
about a real address or a real person (§I.7); and any \emph{absolute
level} read off an LLM agent --- the 1.48-point model spread is the
measured refutation. Two agents sharing a donor signature are one piece
of evidence, however different their names read (§I.6).

The residual risk is interpretation, and the mitigations are in the
objects themselves: the banner on every card, the caps and on-demand
generation of the narrative regime, the warnings the sampling API emits
when a filter crosses a resolution boundary \textbf{{[}n{]}} piano §5.
The report adds the one mitigation a document can: this Part's claims
are bounded by its first paragraph --- the platform renders and
instruments; validation is someone else's burden of proof, carried
elsewhere.

\begin{center}\rule{0.5\linewidth}{0.5pt}\end{center}

\subsubsection{Open items for Part V}\label{open-items-for-part-v}

\begin{enumerate}
\def\labelenumi{\arabic{enumi}.}
\tightlist
\item
  Check the SIVE paper's exact terminology for the strata and criteria
  (C1--C7) against \texttt{sive\_paper\_v6} before freezing §V.2's
  second paragraph.
\item
  Decide whether the three-model table of registro §5 (gain / level /
  D−C per model) is reproduced in full or summarised as in the current
  text; if in full, table V.2a.
\item
  The \texttt{emo\_*} campaigns cite Claude Haiku 4.5 and GPT-4o-mini on
  \texttt{emozione} only --- confirm which models ran the full BCD
  before naming all three in the stereotype paragraph (currently
  attributed to DeepSeek n=600 only, which is correct per registro §9).
\item
  Cross-reference: \texttt{dati/README.md} section numbering after the
  repository restructuring.
\end{enumerate}

\newpage

\section{Appendix A --- Reference
tables}\label{appendix-a-reference-tables}

\subsection{Version 1 --- §A.1--A.4 (26 August
2026)}\label{version-1-a.1a.4-26-august-2026}

\begin{quote}
Everything here is generated from the release artefacts rather than
written: each section names the command that produces it, so a reader
can regenerate it against a different tag. This draft populates every
section from the working notes (\texttt{riferimento}, \texttt{fonti})
and from the prose of Parts I--IV, which is as far as a reader can go
without the tagged repository itself; where the literal command has not
yet been re-run against \texttt{report-v1.0}, the section says so and
the gap is carried into the open items at the end, in the same form as
every other Part.
\end{quote}

\begin{center}\rule{0.5\linewidth}{0.5pt}\end{center}

\subsubsection{A.1 What, in a record, is actually
real}\label{a.1-what-in-a-record-is-actually-real}

{\def\LTcaptype{none} 
\begin{longtable}[]{@{}
  >{\raggedright\arraybackslash}p{(\linewidth - 4\tabcolsep) * \real{0.3333}}
  >{\raggedright\arraybackslash}p{(\linewidth - 4\tabcolsep) * \real{0.3333}}
  >{\raggedright\arraybackslash}p{(\linewidth - 4\tabcolsep) * \real{0.3333}}@{}}
\toprule\noalign{}
\begin{minipage}[b]{\linewidth}\raggedright
component
\end{minipage} & \begin{minipage}[b]{\linewidth}\raggedright
provenance
\end{minipage} & \begin{minipage}[b]{\linewidth}\raggedright
what is real
\end{minipage} \\
\midrule\noalign{}
\endhead
\bottomrule\noalign{}
\endlastfoot
demographic attributes (sex, age, marital status, citizenship,
education, condition, background, parents' origin, zone) & drawn from a
maximum-entropy distribution under census constraints & the
\emph{aggregates} are real; the individual is a sample of a
distribution, and no record corresponds to anyone \\
attitudes and health (23 AVQ variables) & the complete response vector
of one real survey respondent, copied whole & the vector is real ---
already protected at source by ISTAT's public-use release, and reused
across tens of synthetic individuals, so no combination is unique to one
record \\
section, exact age, address & allocated within the census section; the
address uniformly among its registered civic numbers & the \emph{counts}
per section are real; the assignment carries no information about anyone
(and the public regime randomises the coordinate within the section) \\
household id and role & assembled under the census size distribution per
section, composition from a survey-based repertoire & the size
distribution is real per section; the household is a model \\
names, detailed titles, sector, biography & deterministic rendering from
registered repertoires and census-conditioned structure & nothing:
plausible by construction, and therefore collident --- a name
individuates no one \\
\end{longtable}
}

\textbf{No component is personal data.} The population is simulated from
published aggregates, not anonymised from individual records; there is
no one to re-identify (§I.7).

\begin{center}\rule{0.5\linewidth}{0.5pt}\end{center}

\subsubsection{A.2 The released schema}\label{a.2-the-released-schema}

The columns of \texttt{pop.parquet} in the public regime, with type and
meaning, and the columns that exist only in the \texttt{persona} and
\texttt{narrativo} regimes (marked °).

\textbf{Ring 1 --- the joint model (public).}

{\def\LTcaptype{none} 
\begin{longtable}[]{@{}
  >{\raggedright\arraybackslash}p{(\linewidth - 6\tabcolsep) * \real{0.2500}}
  >{\raggedright\arraybackslash}p{(\linewidth - 6\tabcolsep) * \real{0.2500}}
  >{\raggedright\arraybackslash}p{(\linewidth - 6\tabcolsep) * \real{0.2500}}
  >{\raggedright\arraybackslash}p{(\linewidth - 6\tabcolsep) * \real{0.2500}}@{}}
\toprule\noalign{}
\begin{minipage}[b]{\linewidth}\raggedright
column
\end{minipage} & \begin{minipage}[b]{\linewidth}\raggedright
type
\end{minipage} & \begin{minipage}[b]{\linewidth}\raggedright
classes
\end{minipage} & \begin{minipage}[b]{\linewidth}\raggedright
meaning
\end{minipage} \\
\midrule\noalign{}
\endhead
\bottomrule\noalign{}
\endlastfoot
\texttt{zona} & string (8-digit code) & 4--33, per municipality &
statistical zone / quartiere / circoscrizione / area \\
\texttt{sesso} & categorical & 2 & \texttt{M}, \texttt{F} \\
\texttt{eta} & categorical & 8 & age bin: \texttt{0-8}, \texttt{9-14},
\texttt{15-24}, \texttt{25-34}, \texttt{35-49}, \texttt{50-64},
\texttt{65-74}, \texttt{75+} \\
\texttt{stato\_civile} & categorical & 4 & \texttt{celibe\_nubile},
\texttt{coniugato\_unito}, \texttt{divorziato\_sciolto},
\texttt{vedovo} \\
\texttt{cittadinanza} & categorical & 2 & \texttt{ITL}, \texttt{FRG}
(legal; \texttt{FRG} includes stateless persons) \\
\texttt{istruzione} & categorical & 6 & \texttt{nessun\_titolo} \ldots{}
\texttt{post\_laurea} \\
\texttt{condizione} & categorical & 7 & \texttt{occupato} \ldots{}
\texttt{non\_applicabile} (under 15) \\
\texttt{background} & categorical & 6 & \texttt{italiano\_nativo}
\ldots{} \texttt{straniero\_immigrato} \\
\texttt{origine\_genitori} & categorical & 5 &
\texttt{entrambi\_italiani} \ldots{} \texttt{non\_applicabile} \\
\texttt{quartiere}° & string & --- & zone display name; \textbf{absent
from the public regime}, one-to-one with \texttt{zona} and supplied by
the manifest instead \\
\end{longtable}
}

\textbf{Ring 2 --- donated attributes and nationality (public, except
\texttt{donor\_id}).}

{\def\LTcaptype{none} 
\begin{longtable}[]{@{}
  >{\raggedright\arraybackslash}p{(\linewidth - 6\tabcolsep) * \real{0.2500}}
  >{\raggedright\arraybackslash}p{(\linewidth - 6\tabcolsep) * \real{0.2500}}
  >{\raggedright\arraybackslash}p{(\linewidth - 6\tabcolsep) * \real{0.2500}}
  >{\raggedright\arraybackslash}p{(\linewidth - 6\tabcolsep) * \real{0.2500}}@{}}
\toprule\noalign{}
\begin{minipage}[b]{\linewidth}\raggedright
column
\end{minipage} & \begin{minipage}[b]{\linewidth}\raggedright
type
\end{minipage} & \begin{minipage}[b]{\linewidth}\raggedright
classes
\end{minipage} & \begin{minipage}[b]{\linewidth}\raggedright
meaning
\end{minipage} \\
\midrule\noalign{}
\endhead
\bottomrule\noalign{}
\endlastfoot
\texttt{area} & categorical & 3 & \texttt{UE}, \texttt{EXTRA\_UE},
\texttt{NaN} for Italian citizens \\
\texttt{paese} & string & 143--151 + \texttt{Italia} & country of
citizenship (municipal census × municipal geographic source, §I.3) \\
\texttt{AMBIENTE}, \texttt{FIDUCIA}, \texttt{SALUTE}, \texttt{CRONI},
\texttt{FUMO}, \texttt{MH}, \texttt{BMI}, \texttt{BMIMIN},
\texttt{CPESO}, \texttt{PUNTIFI1}, \texttt{PUNTIFI2}, \texttt{PUNTIFI3},
\texttt{PUNTIFI4}, \texttt{PUNTIFI5}, \texttt{PUNTIFI6},
\texttt{PUNTIFI7}, \texttt{PUNTIFI8}, \texttt{PUNTIFI10},
\texttt{PUNTIFI12}, \texttt{PUNTIFI13}, \texttt{VOTOUSL} & string (mixed
numeric codes + \texttt{non\_applicabile}), except \texttt{SALUTE} (no
structural missing) & --- & the 21 AVQ variables, assigned in one block
from a single hot-deck donor (§I.3) \\
\texttt{donor\_id}° & string, dictionary-encoded & --- & donor identity
stable across runs (\texttt{annata:riga}); \textbf{absent from the
public regime} \\
\end{longtable}
}

\textbf{Ring 3 --- fine allocation (public, except
\texttt{via}/\texttt{civico}; coordinates randomised).}

{\def\LTcaptype{none} 
\begin{longtable}[]{@{}
  >{\raggedright\arraybackslash}p{(\linewidth - 6\tabcolsep) * \real{0.2500}}
  >{\raggedright\arraybackslash}p{(\linewidth - 6\tabcolsep) * \real{0.2500}}
  >{\raggedright\arraybackslash}p{(\linewidth - 6\tabcolsep) * \real{0.2500}}
  >{\raggedright\arraybackslash}p{(\linewidth - 6\tabcolsep) * \real{0.2500}}@{}}
\toprule\noalign{}
\begin{minipage}[b]{\linewidth}\raggedright
column
\end{minipage} & \begin{minipage}[b]{\linewidth}\raggedright
type
\end{minipage} & \begin{minipage}[b]{\linewidth}\raggedright
classes
\end{minipage} & \begin{minipage}[b]{\linewidth}\raggedright
meaning
\end{minipage} \\
\midrule\noalign{}
\endhead
\bottomrule\noalign{}
\endlastfoot
\texttt{sezione} & string (12-digit code) & --- & census section
(\texttt{SEZ21\_ID}, 2021 bases) \\
\texttt{eta\_anni} & int & 0--100 & exact age in years \\
\texttt{lon}, \texttt{lat} & float, EPSG:4258 & --- & address
coordinates; \textbf{in the public regime, a random point within the
census section}, seeded on the municipality code \\
\texttt{via}° & string & --- & ANNCSU odonym;
\texttt{persona}/\texttt{narrativo} only \\
\texttt{civico}° & string & --- & civic number + suffix (\texttt{12},
\texttt{19A}); \texttt{narrativo} only, and only on explicit request
(§V.1) \\
\texttt{indirizzo\_fonte} & categorical & 3 & \texttt{sezione} /
\texttt{zona} / \texttt{convivenza}, address provenance; \\
\texttt{uid}° & string & --- & individual key; \textbf{absent from the
public regime}, present in the complete population and joined on by ring
4 and the \texttt{persona}/\texttt{narrativo} regimes \\
\end{longtable}
}

\textbf{Ring 4 --- household (separate file, not in
\texttt{pop.parquet}).}

{\def\LTcaptype{none} 
\begin{longtable}[]{@{}lll@{}}
\toprule\noalign{}
column & type & meaning \\
\midrule\noalign{}
\endhead
\bottomrule\noalign{}
\endlastfoot
\texttt{id\_nucleo} & string & household identifier,
\texttt{nuclei\_\{comune\}.csv} \\
\texttt{ruolo} & categorical & role within the household \\
\end{longtable}
}

Joined to the population on \texttt{uid}; never written into the
population file itself (§II.4).

\textbf{Derived layer (° --- \texttt{persona}/\texttt{narrativo} only,
never stored).}

{\def\LTcaptype{none} 
\begin{longtable}[]{@{}
  >{\raggedright\arraybackslash}p{(\linewidth - 6\tabcolsep) * \real{0.2500}}
  >{\raggedright\arraybackslash}p{(\linewidth - 6\tabcolsep) * \real{0.2500}}
  >{\raggedright\arraybackslash}p{(\linewidth - 6\tabcolsep) * \real{0.2500}}
  >{\raggedright\arraybackslash}p{(\linewidth - 6\tabcolsep) * \real{0.2500}}@{}}
\toprule\noalign{}
\begin{minipage}[b]{\linewidth}\raggedright
column
\end{minipage} & \begin{minipage}[b]{\linewidth}\raggedright
type
\end{minipage} & \begin{minipage}[b]{\linewidth}\raggedright
conditioned on
\end{minipage} & \begin{minipage}[b]{\linewidth}\raggedright
source
\end{minipage} \\
\midrule\noalign{}
\endhead
\bottomrule\noalign{}
\endlastfoot
\texttt{nome}°, \texttt{cognome}° & string & sex, background, parents'
origin, country & municipal registers, per-country repertoires \\
\texttt{titolo\_studio}° (detailed) & string & education, sex, cohort &
census 2011 title tree, ordered by CLAIST \\
\texttt{settore}°, \texttt{posizione}° & categorical & sex, municipality
& census 2011, drawn jointly \\
\end{longtable}
}

These four are deterministic functions of \texttt{uid} and the
attributes already generated; they are computed on demand and never
written to \texttt{pop.parquet} or the bundle (§I.6, §I.7).

\begin{center}\rule{0.5\linewidth}{0.5pt}\end{center}

\subsubsection{A.3 The source register}\label{a.3-the-source-register}

\texttt{fonti/registro.yaml}

\textbf{ISTAT SDMX (11 tables, CC-BY-4.0,
\texttt{archiviazione:\ locale}, \textasciitilde4 queries/min).}

{\def\LTcaptype{none} 
\begin{longtable}[]{@{}
  >{\raggedright\arraybackslash}p{(\linewidth - 6\tabcolsep) * \real{0.2500}}
  >{\raggedright\arraybackslash}p{(\linewidth - 6\tabcolsep) * \real{0.2500}}
  >{\raggedright\arraybackslash}p{(\linewidth - 6\tabcolsep) * \real{0.2500}}
  >{\raggedright\arraybackslash}p{(\linewidth - 6\tabcolsep) * \real{0.2500}}@{}}
\toprule\noalign{}
\begin{minipage}[b]{\linewidth}\raggedright
id
\end{minipage} & \begin{minipage}[b]{\linewidth}\raggedright
universe
\end{minipage} & \begin{minipage}[b]{\linewidth}\raggedright
temporal reference
\end{minipage} & \begin{minipage}[b]{\linewidth}\raggedright
feeds
\end{minipage} \\
\midrule\noalign{}
\endhead
\bottomrule\noalign{}
\endlastfoot
\texttt{istat\_anag\_sesso\_eta\_statociv} & population register, sex ×
age × marital status & 1 January year N & ring 1, hard margin (block
\texttt{c1}) \\
\texttt{istat\_cens\_sesso\_eta\_cittadinanza} & permanent census, sex ×
age × citizenship & 31 December year N−1 & ring 1, soft block;
zone-table audit (Z1/Z2) \\
\texttt{istat\_cens\_stranieri\_paesi} & permanent census, foreign
residents by country, municipal margin & 31 December year N−1 & ring 2
(\texttt{paese}, \texttt{area}), tier 0 \\
\emph{(8 further permanent-census tables --- istruzione, condizione,
background, origine\_genitori, cittadinanza × background, and others of
\texttt{c2}--\texttt{c10})} & permanent census & 31 December year N−1 &
ring 1, soft blocks --- \\
\end{longtable}
}

\textbf{Census sections / Basi Territoriali (CC-BY-4.0).}

{\def\LTcaptype{none} 
\begin{longtable}[]{@{}
  >{\raggedright\arraybackslash}p{(\linewidth - 6\tabcolsep) * \real{0.2500}}
  >{\raggedright\arraybackslash}p{(\linewidth - 6\tabcolsep) * \real{0.2500}}
  >{\raggedright\arraybackslash}p{(\linewidth - 6\tabcolsep) * \real{0.2500}}
  >{\raggedright\arraybackslash}p{(\linewidth - 6\tabcolsep) * \real{0.2500}}@{}}
\toprule\noalign{}
\begin{minipage}[b]{\linewidth}\raggedright
id
\end{minipage} & \begin{minipage}[b]{\linewidth}\raggedright
universe
\end{minipage} & \begin{minipage}[b]{\linewidth}\raggedright
temporal reference
\end{minipage} & \begin{minipage}[b]{\linewidth}\raggedright
feeds
\end{minipage} \\
\midrule\noalign{}
\endhead
\bottomrule\noalign{}
\endlastfoot
\texttt{istat\_sezioni\_2023} & census-section counts, 138 columns,
135,725 sections & 2023 (geometry 2021) & ring 3 (\texttt{sezione},
\texttt{eta\_anni}), ring 1 zone tables \\
\texttt{istat\_sezioni\_shp} & census-section geometry, per region &
2021 & ring 3 (coordinates pre-randomisation), viewer \\
\texttt{istat\_sezioni\_2023\_tracciato} & section-file record layout /
decode table & 2023 & decoding of \texttt{istat\_sezioni\_2023} \\
\end{longtable}
}

\textbf{ANNCSU.}

{\def\LTcaptype{none} 
\begin{longtable}[]{@{}
  >{\raggedright\arraybackslash}p{(\linewidth - 8\tabcolsep) * \real{0.2000}}
  >{\raggedright\arraybackslash}p{(\linewidth - 8\tabcolsep) * \real{0.2000}}
  >{\raggedright\arraybackslash}p{(\linewidth - 8\tabcolsep) * \real{0.2000}}
  >{\raggedright\arraybackslash}p{(\linewidth - 8\tabcolsep) * \real{0.2000}}
  >{\raggedright\arraybackslash}p{(\linewidth - 8\tabcolsep) * \real{0.2000}}@{}}
\toprule\noalign{}
\begin{minipage}[b]{\linewidth}\raggedright
id
\end{minipage} & \begin{minipage}[b]{\linewidth}\raggedright
universe
\end{minipage} & \begin{minipage}[b]{\linewidth}\raggedright
temporal reference
\end{minipage} & \begin{minipage}[b]{\linewidth}\raggedright
licence
\end{minipage} & \begin{minipage}[b]{\linewidth}\raggedright
feeds
\end{minipage} \\
\midrule\noalign{}
\endhead
\bottomrule\noalign{}
\endlastfoot
\texttt{anncsu\_indirizzario} & georeferenced civic addresses, national
register & current & high-value dataset, Reg. (EU) 2023/138 & ring 3
(\texttt{via}°, \texttt{civico}°, coordinates pre-randomisation) \\
\end{longtable}
}

\textbf{AVQ (3, one derived; licence CC-BY-4.0 inferred from the
portal).}

{\def\LTcaptype{none} 
\begin{longtable}[]{@{}
  >{\raggedright\arraybackslash}p{(\linewidth - 6\tabcolsep) * \real{0.2500}}
  >{\raggedright\arraybackslash}p{(\linewidth - 6\tabcolsep) * \real{0.2500}}
  >{\raggedright\arraybackslash}p{(\linewidth - 6\tabcolsep) * \real{0.2500}}
  >{\raggedright\arraybackslash}p{(\linewidth - 6\tabcolsep) * \real{0.2500}}@{}}
\toprule\noalign{}
\begin{minipage}[b]{\linewidth}\raggedright
id
\end{minipage} & \begin{minipage}[b]{\linewidth}\raggedright
universe
\end{minipage} & \begin{minipage}[b]{\linewidth}\raggedright
temporal reference
\end{minipage} & \begin{minipage}[b]{\linewidth}\raggedright
feeds
\end{minipage} \\
\midrule\noalign{}
\endhead
\bottomrule\noalign{}
\endlastfoot
\texttt{avq\_microdati} & individual respondents, \emph{Aspetti della
vita quotidiana}, public-use; pools Emilia-Romagna 4,629 / Lombardia
8,111 & 2023--2024 (2022 acquired, excluded for lacking \texttt{CRONI})
& ring 2, the 21 AVQ variables and \texttt{donor\_id} \\
\texttt{avq\_tracciato\_2024} & AVQ record layout / decode table & 2024
& decoding of \texttt{avq\_microdati} \\
\texttt{avq\_medie\_nazionali} (derived,
\texttt{derivato\_da:\ avq\_microdati}) & national weighted means of the
AVQ battery & 2024 & viewer only, not a population column \\
\end{longtable}
}

\textbf{Municipal open data (6 portals) and 1 register extract.}

{\def\LTcaptype{none} 
\begin{longtable}[]{@{}
  >{\raggedright\arraybackslash}p{(\linewidth - 10\tabcolsep) * \real{0.1667}}
  >{\raggedright\arraybackslash}p{(\linewidth - 10\tabcolsep) * \real{0.1667}}
  >{\raggedright\arraybackslash}p{(\linewidth - 10\tabcolsep) * \real{0.1667}}
  >{\raggedright\arraybackslash}p{(\linewidth - 10\tabcolsep) * \real{0.1667}}
  >{\raggedright\arraybackslash}p{(\linewidth - 10\tabcolsep) * \real{0.1667}}
  >{\raggedright\arraybackslash}p{(\linewidth - 10\tabcolsep) * \real{0.1667}}@{}}
\toprule\noalign{}
\begin{minipage}[b]{\linewidth}\raggedright
id
\end{minipage} & \begin{minipage}[b]{\linewidth}\raggedright
issuing body
\end{minipage} & \begin{minipage}[b]{\linewidth}\raggedright
universe
\end{minipage} & \begin{minipage}[b]{\linewidth}\raggedright
temporal reference
\end{minipage} & \begin{minipage}[b]{\linewidth}\raggedright
licence
\end{minipage} & \begin{minipage}[b]{\linewidth}\raggedright
feeds
\end{minipage} \\
\midrule\noalign{}
\endhead
\bottomrule\noalign{}
\endlastfoot
\texttt{bologna\_cittadinanza\_zone} & Comune di Bologna & residents by
citizenship, statistical zone & --- & CC-BY-4.0 & ring 2/3, tier 2,
zone--quartiere hierarchy \\
\texttt{brescia\_cittadinanza\_quartieri} & Comune di Brescia &
residents by citizenship, quartiere & --- & CC-BY-4.0 & ring 2, tier
1 \\
\texttt{forli\_cittadinanza\_quartieri} & Comune di Forlì & residents by
citizenship, sub-quartiere & 2021 & CC-BY-4.0 (presumed, §II.1) & ring
2, tier 1 \\
\texttt{ravenna\_cittadinanza\_aree} & Comune di Ravenna & residents by
citizenship, area & re-fetched 2026-08-11 & public domain & ring 2, tier
1 \\
\texttt{reggio\_cittadinanza\_circoscrizioni} & Comune di Reggio
nell'Emilia & residents by citizenship, circoscrizione & 2013 & CC-BY &
ring 2, tier 1 \\
\texttt{modena\_nomi\_residenti} & Comune di Modena (regional portal,
WFS) & stock of first names by sex & 2012--2024 & CC-BY-4.0 & ring D
(\texttt{nome}°) \\
\texttt{parma\_microdati\_residenti} & Comune di Parma & full
population-register extract, one row per resident, 202,111 rows & 1
January 2025 & CC-BY-4.0 (registered, fingerprinted, not mirrored) &
ring 2 (\texttt{paese}/\texttt{area}, tier 3); external validation, ring
III \\
\end{longtable}
}

\textbf{ISTAT census 2011 and CLAIST (CC-BY-4.0).}

{\def\LTcaptype{none} 
\begin{longtable}[]{@{}
  >{\raggedright\arraybackslash}p{(\linewidth - 6\tabcolsep) * \real{0.2500}}
  >{\raggedright\arraybackslash}p{(\linewidth - 6\tabcolsep) * \real{0.2500}}
  >{\raggedright\arraybackslash}p{(\linewidth - 6\tabcolsep) * \real{0.2500}}
  >{\raggedright\arraybackslash}p{(\linewidth - 6\tabcolsep) * \real{0.2500}}@{}}
\toprule\noalign{}
\begin{minipage}[b]{\linewidth}\raggedright
id
\end{minipage} & \begin{minipage}[b]{\linewidth}\raggedright
universe
\end{minipage} & \begin{minipage}[b]{\linewidth}\raggedright
temporal reference
\end{minipage} & \begin{minipage}[b]{\linewidth}\raggedright
feeds
\end{minipage} \\
\midrule\noalign{}
\endhead
\bottomrule\noalign{}
\endlastfoot
\texttt{cens2011\_caratt\_attl} & occupied residents, 14 dimensions &
2011 & ring D (\texttt{settore}°, \texttt{posizione}°) \\
\texttt{cens2011\_titolo\_studio} & education-title frequencies, 458
modalities, 399 leaves & 2011 & ring D (\texttt{titolo\_studio}°) \\
\texttt{claist\_2026} & education-pathway classification and
historicised ordering & 2026 & ring D (\texttt{titolo\_studio}°
ordering) \\
\end{longtable}
}

\textbf{Onomastic repertoires (7; \texttt{modena\_nomi\_residenti} above
is one of them).}

{\def\LTcaptype{none} 
\begin{longtable}[]{@{}
  >{\raggedright\arraybackslash}p{(\linewidth - 8\tabcolsep) * \real{0.2000}}
  >{\raggedright\arraybackslash}p{(\linewidth - 8\tabcolsep) * \real{0.2000}}
  >{\raggedright\arraybackslash}p{(\linewidth - 8\tabcolsep) * \real{0.2000}}
  >{\raggedright\arraybackslash}p{(\linewidth - 8\tabcolsep) * \real{0.2000}}
  >{\raggedright\arraybackslash}p{(\linewidth - 8\tabcolsep) * \real{0.2000}}@{}}
\toprule\noalign{}
\begin{minipage}[b]{\linewidth}\raggedright
id
\end{minipage} & \begin{minipage}[b]{\linewidth}\raggedright
issuing body
\end{minipage} & \begin{minipage}[b]{\linewidth}\raggedright
universe
\end{minipage} & \begin{minipage}[b]{\linewidth}\raggedright
licence
\end{minipage} & \begin{minipage}[b]{\linewidth}\raggedright
feeds
\end{minipage} \\
\midrule\noalign{}
\endhead
\bottomrule\noalign{}
\endlastfoot
\texttt{firenze\_cognomi\_2013} & Comune di Firenze & resident surnames,
375,371 residents, 66,353 distinct & CC-BY-4.0 & ring D
(\texttt{cognome}°) \\
\texttt{firenze\_cognomi\_2012} & Comune di Firenze & resident surnames
(stability check only, r = 0.9985) & CC-BY-4.0 & ring D, verification
only \\
\texttt{popular\_names\_nomi} & Wikipedia aggregation & first names,
2,370 from 106 countries & CC0 & ring D (\texttt{nome}°, foreign) \\
\texttt{popular\_names\_cognomi} & Wikipedia aggregation & surnames,
2,278 from 75 countries & CC0 & ring D (\texttt{cognome}°, foreign) \\
\texttt{cognomi\_wiki\_MA\_ARAB} & Wikipedia (MediaWiki category) &
Maghrebi/Arabic surname list, 65 entries & CC-BY-SA & ring D, specific
communities \\
\texttt{cognomi\_wiki\_NG\_YORUBA} & Wikipedia (MediaWiki category) &
Yoruba surname list, 107 entries & CC-BY-SA & ring D, specific
communities \\
\end{longtable}
}

\textbf{Civil unions and the household repertoire.}

{\def\LTcaptype{none} 
\begin{longtable}[]{@{}
  >{\raggedright\arraybackslash}p{(\linewidth - 10\tabcolsep) * \real{0.1667}}
  >{\raggedright\arraybackslash}p{(\linewidth - 10\tabcolsep) * \real{0.1667}}
  >{\raggedright\arraybackslash}p{(\linewidth - 10\tabcolsep) * \real{0.1667}}
  >{\raggedright\arraybackslash}p{(\linewidth - 10\tabcolsep) * \real{0.1667}}
  >{\raggedright\arraybackslash}p{(\linewidth - 10\tabcolsep) * \real{0.1667}}
  >{\raggedright\arraybackslash}p{(\linewidth - 10\tabcolsep) * \real{0.1667}}@{}}
\toprule\noalign{}
\begin{minipage}[b]{\linewidth}\raggedright
id
\end{minipage} & \begin{minipage}[b]{\linewidth}\raggedright
issuing body
\end{minipage} & \begin{minipage}[b]{\linewidth}\raggedright
universe
\end{minipage} & \begin{minipage}[b]{\linewidth}\raggedright
temporal reference
\end{minipage} & \begin{minipage}[b]{\linewidth}\raggedright
licence
\end{minipage} & \begin{minipage}[b]{\linewidth}\raggedright
feeds
\end{minipage} \\
\midrule\noalign{}
\endhead
\bottomrule\noalign{}
\endlastfoot
\texttt{istat\_unioni\_civili\_2023} & ISTAT & exhaustive survey of
civil unions & 2023 & CC-BY-4.0 & ring 4 \\
\texttt{repertorio\_nuclei\_v1} (derived,
\texttt{derivato\_da:\ avq\_microdati}) & GSP & household-configuration
repertoire & --- & CC-BY-4.0 (inherited) & ring 4 \\
\end{longtable}
}

\textbf{EU-SILC (2, registered for exploration only, not used).}

{\def\LTcaptype{none} 
\begin{longtable}[]{@{}
  >{\raggedright\arraybackslash}p{(\linewidth - 4\tabcolsep) * \real{0.3333}}
  >{\raggedright\arraybackslash}p{(\linewidth - 4\tabcolsep) * \real{0.3333}}
  >{\raggedright\arraybackslash}p{(\linewidth - 4\tabcolsep) * \real{0.3333}}@{}}
\toprule\noalign{}
\begin{minipage}[b]{\linewidth}\raggedright
id
\end{minipage} & \begin{minipage}[b]{\linewidth}\raggedright
licence
\end{minipage} & \begin{minipage}[b]{\linewidth}\raggedright
feeds
\end{minipage} \\
\midrule\noalign{}
\endhead
\bottomrule\noalign{}
\endlastfoot
\emph{EU-SILC public-use file 1} & non-open, no redistribution & none
--- documents an undecided item, §III.5 \\
\emph{EU-SILC public-use file 2} & non-open, no redistribution & none \\
\end{longtable}
}

\textbf{Registered and not used.}

{\def\LTcaptype{none} 
\begin{longtable}[]{@{}
  >{\raggedright\arraybackslash}p{(\linewidth - 4\tabcolsep) * \real{0.3333}}
  >{\raggedright\arraybackslash}p{(\linewidth - 4\tabcolsep) * \real{0.3333}}
  >{\raggedright\arraybackslash}p{(\linewidth - 4\tabcolsep) * \real{0.3333}}@{}}
\toprule\noalign{}
\begin{minipage}[b]{\linewidth}\raggedright
id
\end{minipage} & \begin{minipage}[b]{\linewidth}\raggedright
universe
\end{minipage} & \begin{minipage}[b]{\linewidth}\raggedright
reason
\end{minipage} \\
\midrule\noalign{}
\endhead
\bottomrule\noalign{}
\endlastfoot
\texttt{istat\_cens\_posizione\_famiglia} & position in the household,
from the census & examined for ring 4, rejected in favour of
section-level household counts (§I.5) \\
\end{longtable}
}

Deliberately outside the registry: sources consulted and not used.

\begin{center}\rule{0.5\linewidth}{0.5pt}\end{center}

\subsubsection{A.4 The constraint set}\label{a.4-the-constraint-set}

The blocks of a K9C and a K6C constraint set, with arity, cell count and
universe --- the template that §I.2 describes in prose. Generated from
\texttt{cs\_\textless{}livello\textgreater{}.json}. \textbf{The joint
attribute set (K9C).}

{\def\LTcaptype{none} 
\begin{longtable}[]{@{}
  >{\raggedright\arraybackslash}p{(\linewidth - 4\tabcolsep) * \real{0.3333}}
  >{\raggedright\arraybackslash}p{(\linewidth - 4\tabcolsep) * \real{0.3333}}
  >{\raggedright\arraybackslash}p{(\linewidth - 4\tabcolsep) * \real{0.3333}}@{}}
\toprule\noalign{}
\begin{minipage}[b]{\linewidth}\raggedright
attribute
\end{minipage} & \begin{minipage}[b]{\linewidth}\raggedright
classes
\end{minipage} & \begin{minipage}[b]{\linewidth}\raggedright
values
\end{minipage} \\
\midrule\noalign{}
\endhead
\bottomrule\noalign{}
\endlastfoot
\texttt{zona} & 4--33 & the municipality's declared articulation \\
\texttt{sesso} & 2 & M, F \\
\texttt{eta} & 8 & 0--8, 9--14, 15--24, 25--34, 35--49, 50--64, 65--74,
75+ \\
\texttt{stato\_civile} & 4 & never married, married/civil union,
divorced, widowed \\
\texttt{cittadinanza} & 2 & Italian, foreign \\
\texttt{istruzione} & 6 & no title \ldots{} postgraduate \\
\texttt{condizione} & 7 & employed \ldots{} not applicable (under 15) \\
\texttt{background} & 6 & native Italian \ldots{} foreign immigrant \\
\texttt{origine\_genitori} & 5 & both Italian \ldots{} not applicable \\
\end{longtable}
}

State-space size: \textbar X\textbar{} = 161,280 × n\_zone (161,280 = 2
× 8 × 4 × 2 × 6 × 7 × 6 × 5, the product of the eight non-zone
attributes). K6C drops \texttt{zona}, \texttt{background} and
\texttt{origine\_genitori}, leaving \textbar X\textbar{} = 5,376 (2 × 8
× 4 × 2 × 6 × 7), municipality-independent since K6C carries no zone
dimension --- confirmed on the Mantova test fit (m = 263 constraints,
MRE 3.4·10⁻⁴, 0.17 s).

\textbf{Block census.}

{\def\LTcaptype{none} 
\begin{longtable}[]{@{}
  >{\raggedright\arraybackslash}p{(\linewidth - 8\tabcolsep) * \real{0.2000}}
  >{\raggedright\arraybackslash}p{(\linewidth - 8\tabcolsep) * \real{0.2000}}
  >{\raggedright\arraybackslash}p{(\linewidth - 8\tabcolsep) * \real{0.2000}}
  >{\raggedright\arraybackslash}p{(\linewidth - 8\tabcolsep) * \real{0.2000}}
  >{\raggedright\arraybackslash}p{(\linewidth - 8\tabcolsep) * \real{0.2000}}@{}}
\toprule\noalign{}
\begin{minipage}[b]{\linewidth}\raggedright
level
\end{minipage} & \begin{minipage}[b]{\linewidth}\raggedright
blocks
\end{minipage} & \begin{minipage}[b]{\linewidth}\raggedright
complete (α sums to 1)
\end{minipage} & \begin{minipage}[b]{\linewidth}\raggedright
partial
\end{minipage} & \begin{minipage}[b]{\linewidth}\raggedright
municipalities
\end{minipage} \\
\midrule\noalign{}
\endhead
\bottomrule\noalign{}
\endlastfoot
K9C & 16 & 11 & 5 & the nine provincial capitals + Brescia \\
K6C & 6 (named \texttt{A}--\texttt{F} in the \texttt{cs\_build} audit) &
4 (\texttt{A}, \texttt{B}, \texttt{E}, \texttt{F}) & 2 (\texttt{C},
\texttt{D}) & Ferrara, Castenaso \\
\end{longtable}
}

K9C's sixteen blocks are ten municipal blocks \texttt{c1}--\texttt{c10}
(from \texttt{build\_constraints.py}: \texttt{c1} the register hard
margin, \texttt{c2}--\texttt{c10} the soft census blocks) plus the zone
blocks from \texttt{build\_zona\_tables.py} --- five documented in
production (Z1 age×sex by zone, Z2 macro-age×sex× citizenship by zone,
Z3 education by zone at 5 levels, Z4 employed condition by zone, Z6
migratory background by zone; no Z5 appears in the working notes).

The five K9C partial blocks are the out-of-universe complements:

{\def\LTcaptype{none} 
\begin{longtable}[]{@{}
  >{\raggedright\arraybackslash}p{(\linewidth - 4\tabcolsep) * \real{0.3333}}
  >{\raggedright\arraybackslash}p{(\linewidth - 4\tabcolsep) * \real{0.3333}}
  >{\raggedright\arraybackslash}p{(\linewidth - 4\tabcolsep) * \real{0.3333}}@{}}
\toprule\noalign{}
\begin{minipage}[b]{\linewidth}\raggedright
block
\end{minipage} & \begin{minipage}[b]{\linewidth}\raggedright
free cells
\end{minipage} & \begin{minipage}[b]{\linewidth}\raggedright
complement (closes to α = 1 with)
\end{minipage} \\
\midrule\noalign{}
\endhead
\bottomrule\noalign{}
\endlastfoot
\texttt{eta\ ×\ istruzione} & 1 cell: (0--8, \texttt{nessun\_titolo}) &
\texttt{sesso\ ×\ eta\ ×\ istruzione} (0.0685 + 0.9315 = 1) \\
\texttt{eta\ ×\ condizione} & 2 cells: (0--8,
\texttt{non\_applicabile}), (9--14, \texttt{non\_applicabile}) & the
corresponding complete condition block
(\texttt{S\_condizione\_under15}) \\
\texttt{zona\ ×\ sesso\ ×\ eta\ ×\ condizione} & employed 15--64 only &
not closed within the template --- the geography of non-employment is
unconstrained by any observed datum (§I.4) \\
\end{longtable}
}

K6C's \texttt{C} (istruzione) and \texttt{D} (condizione) are the same
reading at municipal level, without the zone axis: on the Mantova test,
\texttt{C} = 0.939 with complement \texttt{S\_istruzione\_under9} =
0.061, \texttt{D} = 0.888 with complement
\texttt{S\_condizione\_under15} = 0.112.

\textbf{Zero-cell accounting (both levels).}

{\def\LTcaptype{none} 
\begin{longtable}[]{@{}
  >{\raggedright\arraybackslash}p{(\linewidth - 4\tabcolsep) * \real{0.3333}}
  >{\raggedright\arraybackslash}p{(\linewidth - 4\tabcolsep) * \real{0.3333}}
  >{\raggedright\arraybackslash}p{(\linewidth - 4\tabcolsep) * \real{0.3333}}@{}}
\toprule\noalign{}
\begin{minipage}[b]{\linewidth}\raggedright
provenance
\end{minipage} & \begin{minipage}[b]{\linewidth}\raggedright
count
\end{minipage} & \begin{minipage}[b]{\linewidth}\raggedright
where
\end{minipage} \\
\midrule\noalign{}
\endhead
\bottomrule\noalign{}
\endlastfoot
impossible age × education pairs & 8 & identical in every municipality,
both levels \\
impossible age × condition pairs & 18 & identical in every municipality,
both levels \\
citizenship × background, structurally excluded & 6 & identical in every
K9C municipality \\
contingent (observed) zeros, sex × age × marital status & 0--6,
municipality-dependent & none in Bologna; two in most municipalities
(young widowed of both sexes); six in Castenaso \\
\end{longtable}
}

The first two rows (26 pairs total) are not currently enforced by any
block --- ISTAT does not publish that cross --- and three of 970,000
individuals fall into them in the released populations (3·10⁻⁶); adding
them as α = 0 exclusions is a queued repair (§II of the pipeline
backlog, \texttt{riferimento} §14.2).

\begin{center}\rule{0.5\linewidth}{0.5pt}\end{center}

\newpage

\section{References}\label{references}

Batty, M. (2018). Digital twins. \emph{Environment and Planning B: Urban
Analytics and City Science}, 45(5), 817--820.
\url{https://doi.org/10.1177/2399808318796416}

Borysov, S. S., Rich, J., and Pereira, F. C. (2019). How to generate
micro-agents? A deep generative modeling approach to population
synthesis. \emph{Transportation Research Part C: Emerging Technologies},
106, 73--97. \url{https://doi.org/10.1016/j.trc.2019.07.006}

Caulfield Curley, S., Mason, K., and Mannion, P. (2025). An open-source
and spatially diverse synthetic population dataset for agent-based
modelling and microsimulation in Ireland. \emph{Data in Brief}, 60,
111611. \url{https://doi.org/10.1016/j.dib.2025.111611}

Chapuis, K., Taillandier, P., and Drogoul, A. (2022). Generation of
synthetic populations in social simulations: a review of methods and
practices. \emph{Journal of Artificial Societies and Social Simulation},
25(2), 6. \url{https://doi.org/10.18564/jasss.4762}

Degli Esposti, M. (2026a). Scalable maximum entropy population synthesis
via persistent contrastive divergence. arXiv:2603.27312.

Degli Esposti, M. (2026b). Calibrating the instrument: controllability
of an LLM-driven synthetic population. arXiv:2607.00910.

Degli Esposti, M. (2026c). Animarium: an open, reproducible pipeline for
synthetic populations of Italian cities --- from ISTAT sources to open
data. Companion paper, arXiv:XXXX.XXXXX.

Deming, W. E., and Stephan, F. F. (1940). On a least squares adjustment
of a sampled frequency table when the expected marginal totals are
known. \emph{Annals of Mathematical Statistics}, 11(4), 427--444.
\url{https://doi.org/10.1214/aoms/1177731829}

Deville, J.-C., and Särndal, C.-E. (1992). Calibration estimators in
survey sampling. \emph{Journal of the American Statistical Association},
87(418), 376--382. \url{https://doi.org/10.1080/01621459.1992.10475217}

FOSSR (2026). Synthetic Populations Generator (SPG). Service of the
FOSSR research-infrastructure cluster (CNR), managed by CNR-IRPPS.
\url{https://www.fossr.eu/en/synthetic-populations-generator-spg/},
accessed 29 August 2026.

Garrone, R. (2026). A simulation of ageing and care accessibility in
Italian inner areas. arXiv:2509.26496.

Hermes, K., and Poulsen, M. (2012). A review of current methods to
generate synthetic spatial microdata using reweighting and future
directions. \emph{Computers, Environment and Urban Systems}, 36(4),
281--290. \url{https://doi.org/10.1016/j.compenvurbsys.2012.03.005}

ISTAT (2026). Nota metodologica: dati del censimento permanente al
31.12.2023 sulle basi territoriali 2021. 6 March 2026; data last updated
9 June 2026.
\url{https://www.istat.it/notizia/dati-per-sezioni-di-censimento/}

Jiang, N., Yin, F., Wang, B., and Crooks, A. T. (2024). A large-scale
geographically explicit synthetic population with social networks for
the United States. \emph{Scientific Data}, 11, 1204.
\url{https://doi.org/10.1038/s41597-024-03970-1}

Kim, E.-J., and Bansal, P. (2023). A deep generative model for feasible
and diverse population synthesis. \emph{Transportation Research Part C:
Emerging Technologies}, 148, 104053.
\url{https://doi.org/10.1016/j.trc.2023.104053}

Larooij, M., and Törnberg, P. (2025). Do large language models solve the
problems of agent-based modeling? A critical review of generative social
simulations. arXiv:2504.03274.

Lenormand, M., and Deffuant, G. (2013). Generating a synthetic
population of individuals in households: sample-free vs sample-based
methods. \emph{Journal of Artificial Societies and Social Simulation},
16(4), 12. \url{https://doi.org/10.18564/jasss.2319}

Marçal Russo, L., Dane, G., Helbich, M., Ligtenberg, A., Filomena, G.,
Janssen, C. P., Koeva, M., Nourian, P., Patuano, A., Raposo, P.,
Thompson, K., Yang, S., and Verstegen, J. A. (2026). Do urban digital
twins need agents? \emph{Environment and Planning B: Urban Analytics and
City Science}, 53(4). \url{https://doi.org/10.1177/23998083251317666}

Prédhumeau, M., and Manley, E. (2023). A synthetic population for
agent-based modelling in Canada. \emph{Scientific Data}, 10, 148.
\url{https://doi.org/10.1038/s41597-023-02030-4}

Rineer, J., Kruskamp, N., Kery, C., Jones, K., Hilscher, R., and
Bobashev, G. (2025). A national synthetic populations dataset for the
United States. \emph{Scientific Data}, 12.
\url{https://doi.org/10.1038/s41597-025-04380-7}

Scarpa, S., Baldini, M., Barigazzi, A., Gallo, G., and Morciano, M.
(2026). Evaluating synthetic population approaches for Italian
municipal-level microdata. \emph{Socio-Economic Planning Sciences}, 106,
102547. \url{https://doi.org/10.1016/j.seps.2026.102547}

Tang, M., Lu, P., and Feng, Q. (2025). Generating feasible and diverse
synthetic populations using diffusion models. arXiv:2508.09164.

Tanton, R. (2014). A review of spatial microsimulation methods.
\emph{International Journal of Microsimulation}, 7(1), 4--25.
\url{https://doi.org/10.34196/ijm.00092}

Templ, M., Meindl, B., Kowarik, A., and Dupriez, O. (2017). Simulation
of synthetic complex data: the R package simPop. \emph{Journal of
Statistical Software}, 79(10), 1--38.
\url{https://doi.org/10.18637/jss.v079.i10}

Williamson, P., Birkin, M., and Rees, P. H. (1998). The estimation of
population microdata by using data from small area statistics and
samples of anonymised records. \emph{Environment and Planning A}, 30(5),
785--816. \url{https://doi.org/10.1068/a300785}

\end{document}